\pdfoutput=1
\documentclass[nofootinbib,fleqn,twocolumn,superscriptaddress]{revtex4-1}
\usepackage{color}
\usepackage{aas_macros}
\usepackage{hyperref,breakurl}
\usepackage[export]{adjustbox}
\usepackage{bm}

\newcommand{\ims}{\textsc{im3shape}}
\newcommand{\ngm}{\textsc{ngmix}}

\newcommand{\hmpc}{\ensuremath{h^{-1}\mathrm{Mpc}}}

\newcommand{\bea}{\begin{eqnarray}}
\newcommand{\be}{\begin{equation}}
\newcommand{\ben}{\begin{enumerate}}
\newcommand{\bi}{\begin{itemize}}
\newcommand{\eea}{\end{eqnarray}}
\newcommand{\ee}{\end{equation}}
\newcommand{\ei}{\end{itemize}}
\newcommand{\een}{\end{enumerate}}

\newcommand{\eps}{\epsilon}
\newcommand{\mr}{\mathrm}

\newcommand{\bxi}{\ensuremath{\bm\xi}}

\begin{document}

\title{Cosmic Shear Measurements with DES Science Verification Data}

\date{\today}

\author{M.~R.~Becker}
\email[Corresponding author: ]{beckermr@stanford.edu}
\affiliation{Department of Physics, Stanford University, 382 Via Pueblo Mall, Stanford, CA 94305, USA}
\affiliation{Kavli Institute for Particle Astrophysics \& Cosmology, P. O. Box 2450, Stanford University, Stanford, CA 94305, USA}
\author{M.~A.~Troxel}
\affiliation{Jodrell Bank Center for Astrophysics, School of Physics and Astronomy, University of Manchester, Oxford Road, Manchester, M13 9PL, UK}
\author{N.~MacCrann}
\affiliation{Jodrell Bank Center for Astrophysics, School of Physics and Astronomy, University of Manchester, Oxford Road, Manchester, M13 9PL, UK}
\author{E.~Krause}
\affiliation{Kavli Institute for Particle Astrophysics \& Cosmology, P. O. Box 2450, Stanford University, Stanford, CA 94305, USA}
\author{T.~F.~Eifler}
\affiliation{Department of Physics and Astronomy, University of Pennsylvania, Philadelphia, PA 19104, USA}
\affiliation{Jet Propulsion Laboratory, California Institute of Technology, 4800 Oak Grove Dr., Pasadena, CA 91109, USA}
\author{O.~Friedrich}
\affiliation{Max Planck Institute for Extraterrestrial Physics, Giessenbachstrasse, 85748 Garching, Germany}
\affiliation{Universit\"ats-Sternwarte, Fakult\"at f\"ur Physik, Ludwig-Maximilians Universit\"at M\"unchen, Scheinerstr. 1, 81679 M\"unchen, Germany}
\author{A.~Nicola}
\affiliation{Department of Physics, ETH Zurich, Wolfgang-Pauli-Strasse 16, CH-8093 Zurich, Switzerland}
\author{A.~Refregier}
\affiliation{Department of Physics, ETH Zurich, Wolfgang-Pauli-Strasse 16, CH-8093 Zurich, Switzerland}
\author{A.~Amara}
\affiliation{Department of Physics, ETH Zurich, Wolfgang-Pauli-Strasse 16, CH-8093 Zurich, Switzerland}
\author{D.~Bacon}
\affiliation{Institute of Cosmology \& Gravitation, University of Portsmouth, Portsmouth, PO1 3FX, UK}
% ...
\collaboration{The DES Collaboration}
\noaffiliation

\label{firstpage}
\begin{abstract}
We present measurements of weak gravitational lensing cosmic shear two-point statistics using Dark Energy Survey Science Verification data. We demonstrate that our 
results are robust to the choice of shear measurement pipeline, either \ngm\ or \ims, and robust to the choice of two-point statistic, including both real and Fourier-space statistics. 
Our results pass a suite of null tests including tests for B-mode contamination and direct tests for any dependence of the two-point functions on a set of 16 observing 
conditions and galaxy properties, such as seeing, airmass, galaxy color, galaxy magnitude, etc. 
We furthermore use a large suite of simulations to compute the covariance matrix of the cosmic shear measurements and 
assign statistical significance to our null tests. We find that our covariance matrix is consistent with the halo model prediction, indicating that it has the appropriate level of halo sample variance. 
We compare the same jackknife procedure applied to the data and the simulations in order to search for additional sources of noise not captured by the simulations. We find no statistically 
significant extra sources of noise in the data. The overall detection significance with tomography for our highest source density catalog is 9.7$\sigma$. 
Cosmological constraints from the measurements in this work are presented in a companion paper \citep{deswlcosmo}.
\end{abstract}

\maketitle

\section{Introduction}\label{sec:intro}
Cosmic shear, the weak gravitational lensing of galaxies due to large-scale structure, is one of the most statistically powerful probes of Dark Energy, massive neutrinos, and potential modifications to General Relativity~\citep{detf,esoesa}. Due to its powerful potential as a cosmological probe, many ongoing and future surveys
(Kilo-Degree Survey: KiDS\footnote{\tt{http://kids.strw.leidenuniv.nl/}},  
Hyper Suprime-Cam survey: HSC\footnote{\tt{http://www.naoj.org/Projects/HSC/HSCProject.html}}, 
the Dark Energy Survey: DES\footnote{\tt{http://www.darkenergysurvey.org}}, 
the Large Synoptic Survey Telescope: LSST\footnote{\tt{http://www.lsst.org}}, 
Euclid\footnote{\tt{http://sci.esa.int/euclid}} and 
WFIRST\footnote{\tt{http://wfirst.gsfc.nasa.gov}})
will employ cosmic shear as one of their principle cosmological probes. 
Cosmic shear two-point measurements, in their simplest form, are made by correlating the shapes of many millions of galaxies 
as a function of their separation in angle. Additionally, if the galaxies can be separated as a function of redshift, then 
tomographic cosmic shear measurements can be made by cross-correlating galaxies at different redshifts, which can probe 
the evolution of large-scale structure. The galaxies themselves have intrinsic shapes that are an order of magnitude larger 
than the cosmic shear signal, which means that cosmic shear measurements involve extracting small correlations from a large, 
shape noise-dominated background. Competitive cosmological constraints from cosmic shear will require of order percent level or better 
measurements at all steps of the analysis, from shear measurement to the measurements of cosmic shear two-point functions 
(see, e.g., \citet{weinberg2013} or \citet{kilbinger2014} for a review).

Cosmic shear was first detected in 2000 \citep{Bacon:2000yp,Kaiser:2000if,Wittman:2000tc,van_Waerbeke:2000rm}. The most recent results have 
detected correlated shapes on scales from a few to 60 arcminutes from the Deep Lens Survey~\citep{jee2013}, the Sloan Digital Sky Survey 
\citep{lin2012,huff2014}, KiDS \citep{kuijken2015} and the Canada-France-Hawaii Legacy Survey~\citep{kilbinger13}, including in 6 redshift bins~\citep{heymans13}. Future cosmic shear 
measurements will be very high signal-to-noise and over much larger survey areas, yielding a wealth of cosmological information. 

Cosmic shear measurements are challenging for a variety of reasons. First and foremost, shear measurements are 
subject to biases that can arise from a number of sources. These biases are usually split into additive and multiplicative 
components. Sources of additive biases include inaccuracies in the modeling of the point spread function (PSF), inaccuracies in 
correcting for the effect of the PSF on galaxy images, astrometric errors, and contaminating flux from nearby galaxies. Multiplicative biases can arise from 
the effects of noise on the shear measurement process, incorrect estimates of the size of the PSF, and, for model-fitting methods, 
mismatches between an object's true underlying structure and the model employed in the shear measurement process. Additionally, 
many modern shear measurement methods require accurate estimates of the distribution of galaxy shapes and profiles in the absence 
of lensing to either serve as priors in the extraction of shapes from the data or to directly make corrections to the data. These priors can 
be estimated from high-resolution \textit{Hubble Space Telescope} imaging, but must be matched to the observational sample under consideration. 

Significant computational and scientific challenges in cosmic shear measurements remain, even in the presence of perfect shear measurements. 
The cosmic shear field is the result of lensing by the non-linearly evolved matter density field. Accurate predictions for the non-linear matter power 
spectrum, even just for pure dark matter models, are computationally expensive and are needed at every point in parameter space in order to extract 
cosmological parameters. Emulators, like the Coyote Universe \citep{heitmann2014}, have solved this problem for typical cosmologies and Dark 
Energy models, but neglect important physical effects, like galaxy formation, on the matter power spectrum. Additionally, some physical effects of 
galaxy formation break the assumption that galaxies are randomly oriented in the absence of lensing. These effects, called intrinsic alignments, can 
introduce correlations in the shapes of galaxies that are not due to lensing, complicating the interpretation of cosmic shear measurements 
\citep[see, e.g.,][]{Troxel20151,kirk2015}. Furthermore, even if the mean signal can be modeled properly, the covariance matrix of cosmic shear measurements is 
dominated by sample variance, requiring either extensive suites of numerical simulations or complicated halo model calculations. The (mis-)estimation 
of photometric redshifts (photo-$z$s) from imaging data is yet another important source of bias in the modeling of cosmic shear measurements. Finally, for precise 
cosmic shear measurements, lensing magnification, second-order lensing effects, and source selection effects will be important.

In this work, we present cosmic shear measurements from Dark Energy Survey (DES) Science Verification (SV) data (Gruendl et al. in preparation; 
Rykoff et al. in preparation) using the shear catalogs by \citet{jarvis2015}. We employ a combination of two shear estimation codes and two photometric 
redshift estimation codes, each of which takes a different approach to many of the issues described above. Additionally, we use a suite of ray-traced weak 
lensing simulations to compute the sample variance contributions to the covariance matrix of our measurements. We then present an extensive suite of 
tests of both the signals in the data and the covariance matrices. These tests include comparisons of the covariance matrices to halo model predictions, 
null tests of B-mode contamination, and null tests based on comparing the signal between halves of the source galaxy sample split by survey metadata, 
like seeing, depth, etc. Overall, we find no statistically significant contamination. This paper is closely related to three other papers, namely the presentation 
of the DES SV shear catalog \citep{jarvis2015}, the presentation of the DES SV photometric redshifts for weak lensing \citep{bonnett2015b}, and a companion 
paper that presents constraints on cosmological parameters using the measurements in this paper \citep{deswlcosmo}.

This work is organized as follows. In Section~\ref{sec:data}, we describe the DES SV shear catalogs and photometric redshifts. Then we 
describe the mock catalogs used in this work in Section~\ref{sec:mocks}. Next, in Section~\ref{sec:2ptcov}, we present our detections of 
cosmic shear with DES SV data and our real-space two-point function estimators. Appendix~\ref{app:altebstats} describes alternate two-point 
estimators besides the real-space correlation functions used for the bulk of this work. We discuss the estimation and validation of our covariance
 estimation in Section~\ref{sec:covmatrix}. Then, we describe our suite of null and consistency tests of our measurements in Section~\ref{sec:systematics}. 
 Finally, we conclude in Section~\ref{sec:conclusions}. The shear 
 correlation functions and simulation covariance matrices from this work are available as online supplementary material with this paper. 

\section{Data}\label{sec:data}
The DES SV data with weak lensing measurements consists of $139$ square degrees of five-band imaging with roughly 7 exposures per band on average 
\citep{DiehlDECam2012,flaugher2012,HonscheidControl2012,flaugher2015}. The depth of the data is somewhat shallower than the expected $\sim$10-exposure 
average depth of the DES five-year data. The basic reductions and co-add source detection were done with the DES data management (DESDM) 
system as described in \citet{desai2012} and Gruendl et al. (in preparation). We use the shear measurements from \citet{jarvis2015} performed on the 
DES SV Gold sample of galaxies (Rykoff et al., in preparation). For more information on the shear measurements and recommended cuts, we refer the 
reader to \citet{jarvis2015}. The shear measurement pipelines and photo-$z$s used in this work are described below for 
completeness. We use the ``reduced shear'' ellipticity definition \citep{schneider1995}. Finally, note that the two shear measurement codes 
used in this work are not identical, employing different cuts and different parts of the DES SV data. Thus they have different overall source number densities and photometric 
redshift distributions. These differences, which we expect to be smaller in future DES analyses (see \citet{jarvis2015}), 
have no effect on the major conclusions of this work and are in fact important in verifying the robustness of our results. 

\subsection{Shear Measurement Pipeline 1: \ngm}\label{sec:ngmix}
The \ngm\footnote{\url{https://github.com/esheldon/ngmix}} pipeline \citep{sheldon2014} uses sums of Gaussians to represent 
simple galaxy models \citep{hogg2013}. The model parameters of each object are sampled using Markov Chain Monte Carlo (MCMC) 
techniques applied to a full likelihood which forward models the galaxy and its convolution with the PSF. The total likelihood for each 
object is a product of the likelihoods of the individual images of each object. The $r$-, $i$- and $z$-bands are fit simultaneously with 
the same model shape, but different amplitudes. The samples of the likelihood are then used with the \verb+lensfit+ algorithm 
\citep{miller2007} to measure the shear of each object using a prior on the intrinsic distribution of shapes from the \verb+GREAT3+ 
\citep{great3hb} release of the COSMOS galaxy sample. The final effective source number density of the \ngm\ catalog is $\simeq6.1$ galaxies 
per square arcminute.\footnote{We use the following definitions of effective source density $n_{\rm eff}$ and the effective 
shape noise per component $\sigma_{\rm SN}$, which are appropriate for the two-point 
function estimators employed in this work. $n_{\rm eff}=(\sum_{i}w_{i} s_{i})^{2}/\left(\Omega\sum_{i}w_{i}^{2} s_{i}^{2}\right)$ 
and $\sigma^{2}_{SN}=\left(\sum_{i} w_{i}^{2}\left(e_{1}^{2}+e_{2}^{2}\right)\right)/\left(2\sum_{i} w_{i}^{2} s_{i}^{2}\right)$ 
where $w_{i}$ are the weights, $s_{i}$ are the sensitivities, $e_{i}$ are the shear components, $\Omega$ is the survey areas 
and the index $i$ runs over all of the galaxies.} 
Each source has an associated weight and we use the average sensitivity over both directions, as described in \citet{jarvis2015}

\subsection{Shear Measurement Pipeline 2: \ims}\label{sec:im3shape}
The \ims\footnote{\url{https://bitbucket.org/joezuntz/im3shape}} pipeline is built on the \ims~code described in \citet{zuntz2013},
with configuration and modifications for its application to DES SV data
described in \citet{jarvis2015}.  \ims~is a forward-modelling
maximum likelihood code that uses a Levenberg-Marquardt algorithm to fit (in the
configuration used here) two different models to galaxy images, one a de
Vaucouleurs bulge and the other an exponential disc, including the effect of the
PSF and pixelization.  The better-fitting model is then used to give an
ellipticity estimate. Maximum-likelihood parameter sets computed by \ims\ and similar codes have a bias we
refer to as \emph{noise bias} \citep{refregier2012,kacprzak2013}. This bias is removed using a
calibration scheme based on the work of \citet{kacprzak2013}.  The scheme is applied to an ensemble of galaxies using the 
mean bias calibration for the ensemble; different subsets of objects thus use different correction factors.
The final \ims\ catalog has an effective number density of
$\simeq4.1$ galaxies per square arcminute. Each source in \ims\ has a
weight, two additive noise bias corrections (one each for $e_1$ and $e_2$) and a
single multiplicative correction.

\subsection{Photometric Redshifts}\label{sec:photoz}
Based on an extensive comparison of four photo-$z$ methods' impacts on the two-point correlation function in \citet{bonnett2015b} 
and a comparison of a much larger set of photo-$z$ methods in \citet{sanchez2014}, we have selected \texttt{SkyNet} \citep{Bonnett2015,Graff2013} 
for our fiducial photo-$z$ tomography. Galaxies are split into tomographic bins of equal lensing 
weight for the \ngm\ catalog according to the mean of the photo-$z$ PDF for each galaxy produced from \texttt{SkyNet}. The resulting 
tomographic bin boundaries are then used for galaxies in both shear catalogs. For a given shear code, the redshift distribution of each tomographic 
bin is estimated from summing the redshift probability distributions of each individual galaxy according to their weights assigned by the shear code.
The relative agreement between the photo-$z$ estimates and its impact on the correlation function is discussed in more detail in \citet{bonnett2015b}.

\begin{figure*}
\begin{center}
\includegraphics[width=\columnwidth]{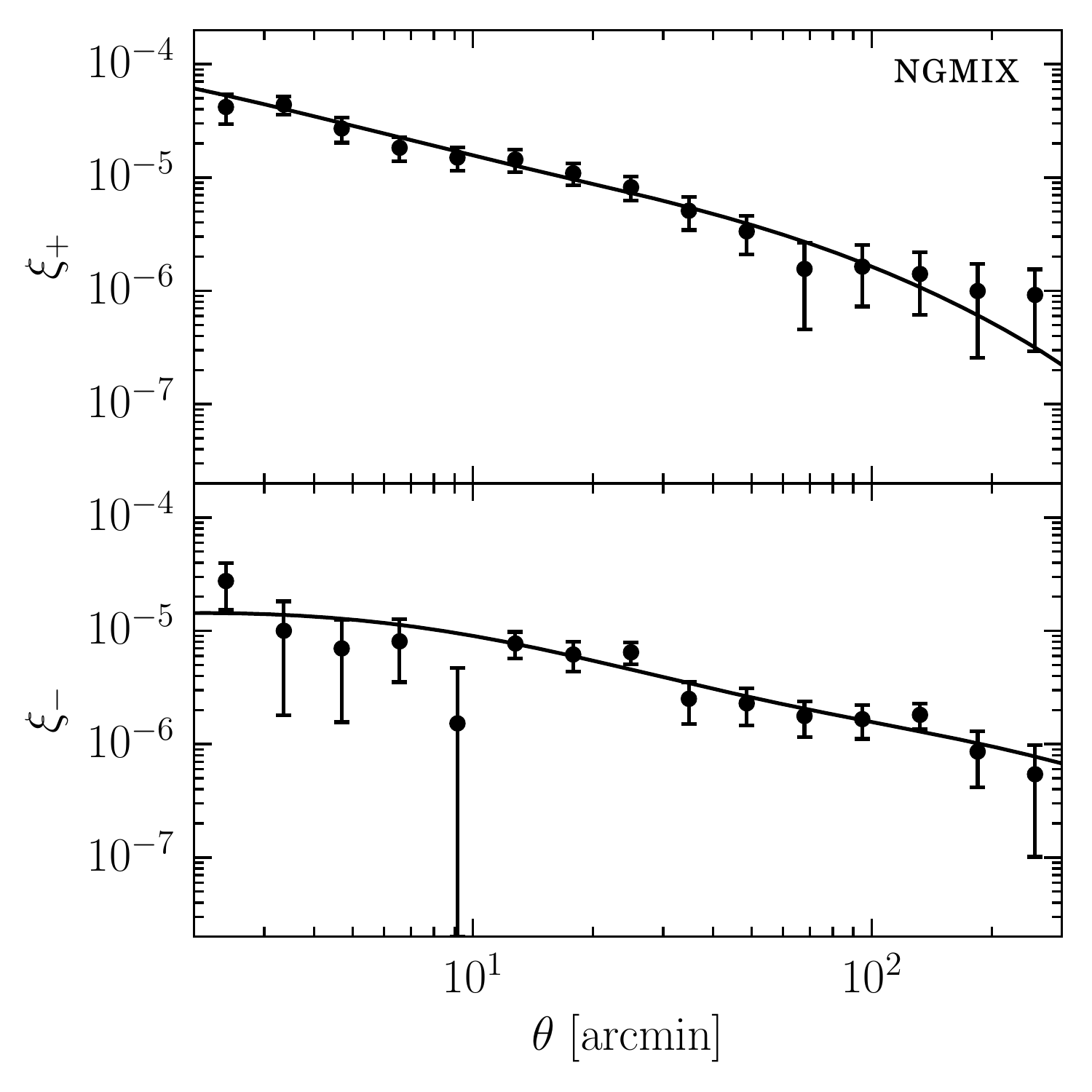}
\includegraphics[width=\columnwidth]{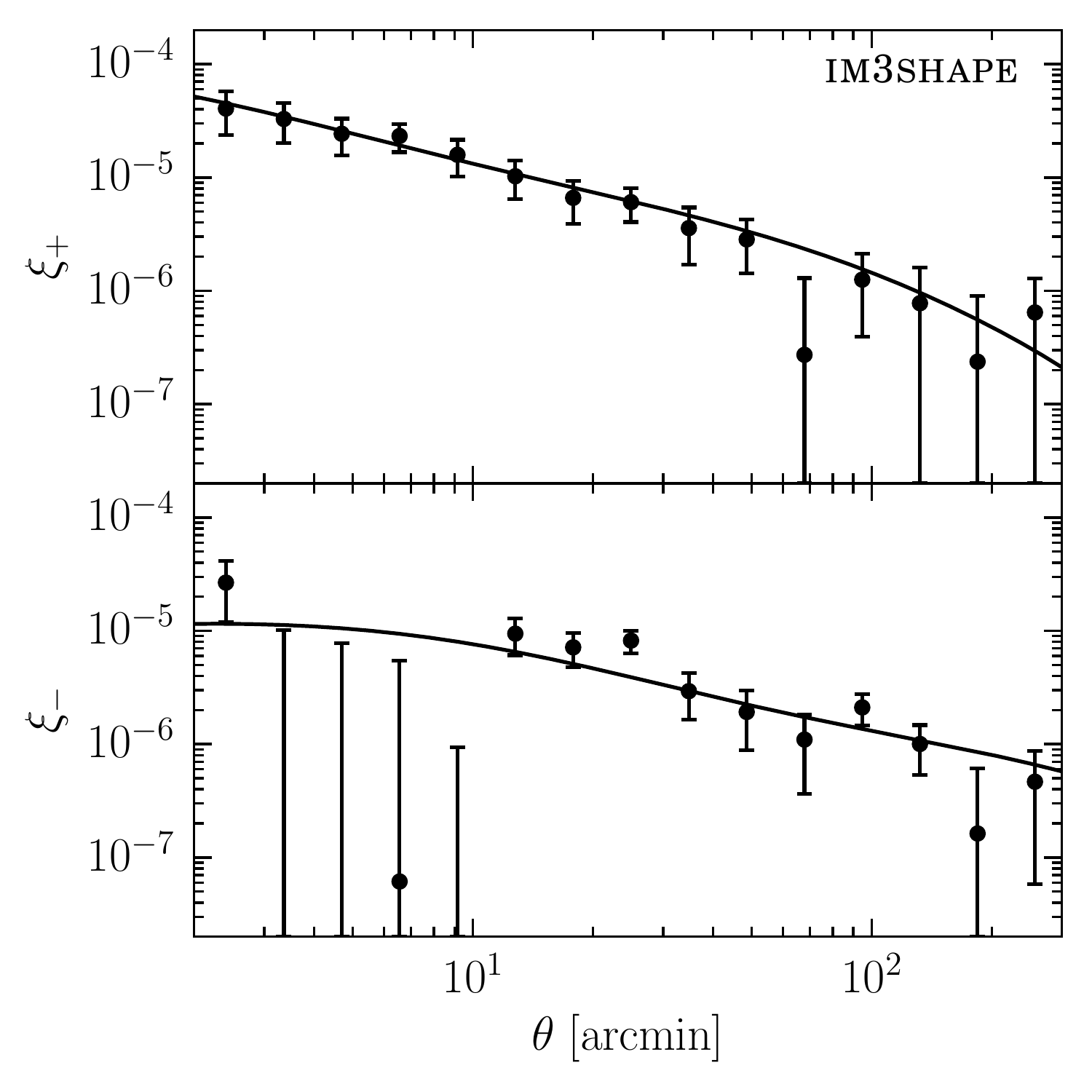}
\end{center}
\caption[]{The measured shear correlation functions $\xi_{+/-}$ for a single tomographic bin for the \ngm\ shape catalog (left) and \ims\ shape catalog (right). The single tomographic bin 
corresponds to redshift distribution shown in Figure \ref{fig:geomphotoz}, $z\approx0.3-1.3$. Note that the redshift distributions of the two catalogs are not identical, so that the shear correlation functions are not expected to match. A detailed comparison of the two catalogs is described in Section~\ref{sec:const}. Negative measurements are shown as 
upper limits. The error bars show the $1\sigma$ uncertainties from the mock catalogs with the appropriate level of shape noise for each shear pipeline. 
The black solid lines show the predictions from a flat, $\Lambda$CDM model described in Section~\ref{sec:mocks} --- not chosen to fit the data.
\label{fig:xinotomo}}
\end{figure*}

\begin{figure*}
\begin{center}
\includegraphics[width=\columnwidth]{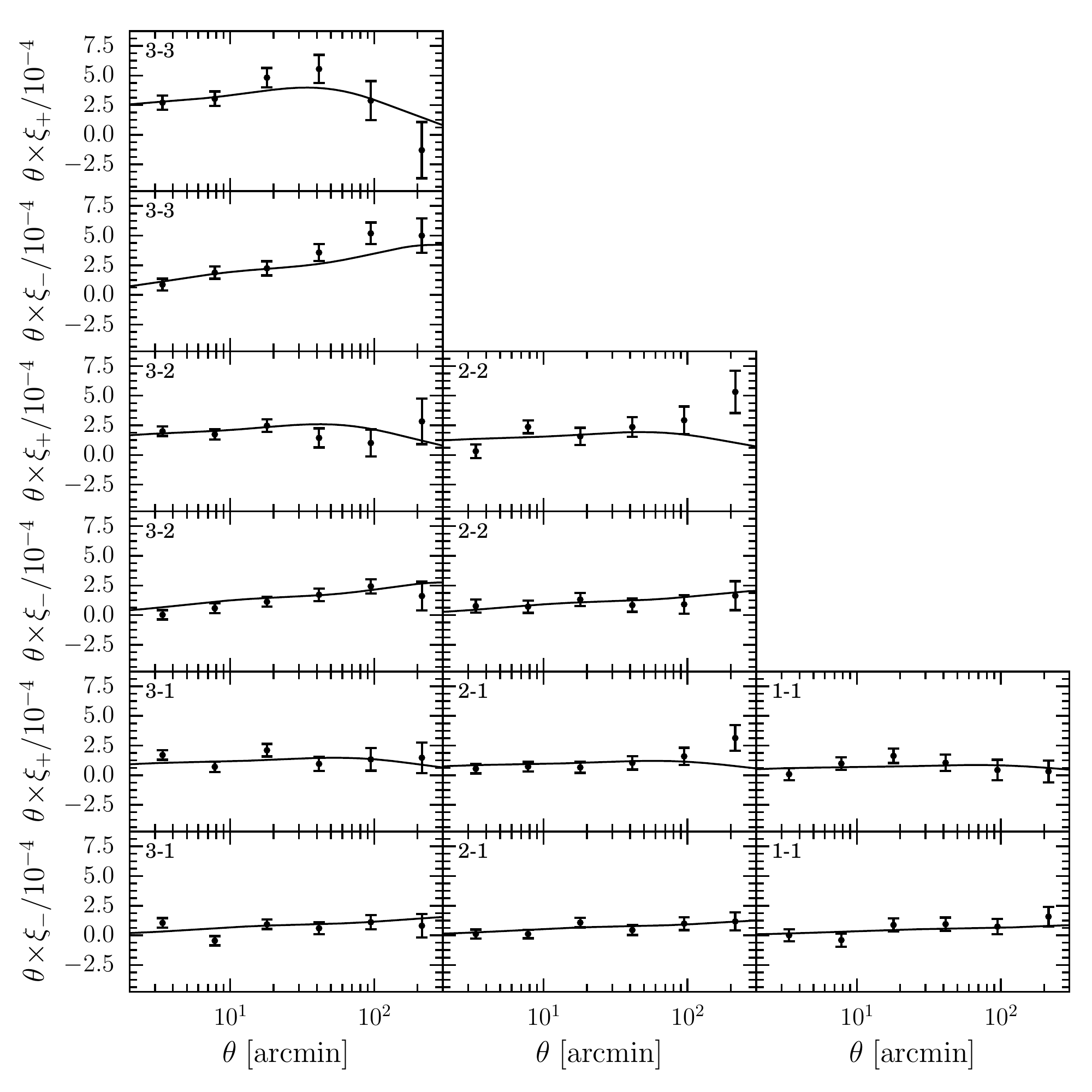}
\includegraphics[width=\columnwidth]{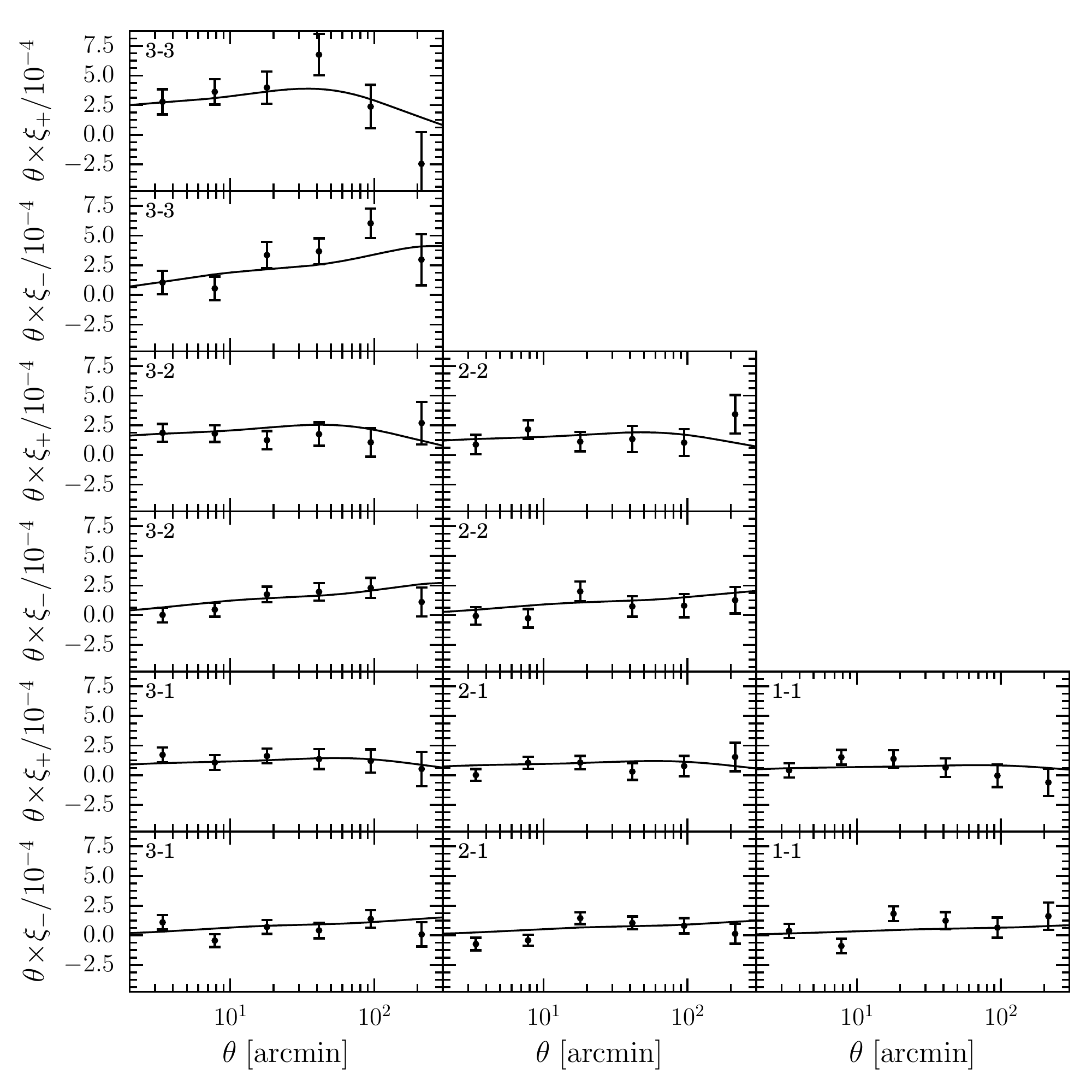}
\end{center}
\caption[]{The measured shear correlation functions $\xi_{+/-}$ times $\theta$ in six angular bins and three tomographic bins for the \ngm\ shape catalog (left) and \ims\ shape catalog (right). 
The tomographic bins correspond to those shown in Figure \ref{fig:geomphotoz}, $z\approx0.30-0.55,0.55-0.83,0.83-1.30$, 
and are labeled from 1 to 3, increasing with redshift. Thus, panel `3-2' shows the cross-correlation between the 
highest and middle redshift bins. The error bars show the $1\sigma$ uncertainties from the mock catalogs with the appropriate level of shape noise for each 
shear pipeline. As in Figure~\ref{fig:xinotomo}, the black solid lines show the predictions from our fiducial $\Lambda$CDM model --- not chosen to fit the data.  \label{fig:xitomo}}
\end{figure*}

\section{Mock Catalogs}\label{sec:mocks}
We use a set of 126 mock catalogs to compute the covariance matrix of the shear correlation functions, 
E/B-mode statistics, power spectra and null statistics described in the following sections. These mock catalogs are constructed 
from seven sets of simulations consisting of three N-body light cones pieced together along the line of sight. 
We use 1050 \hmpc, 2600 \hmpc\ and 4000 \hmpc\ boxes with $1400^3$, $2048^3$ and $2048^3$ 
particles respectively. We use a flat, $\Lambda$CDM model with $\Omega_{m}=0.286$, $\Omega_{\Lambda}=0.714$, 
$n_{s} =0.96$, $h=0.7$, $\Omega_{b}=0.047$, $w=-1$ and $\sigma_{8}=0.820$. The initial conditions are generated 
at redshift 49 with \verb+2LPTic+, a second-order Lagrangian perturbation theory initial conditions generator \citep{crocce2006}  
using linear power spectra from the \verb+CAMB+ Boltzmann code \citep{lewis2002}. The N-body evolution is computed with an efficient dark-matter-only version of the 
\texttt{Gadget-2} code \citep{springel2005}, \texttt{LGadget-2}. We have implemented our own on-the-fly 
light cone generator directly into the \texttt{LGadget-2} code (Busha et al. in preparation). 
We produce a full-sky light cone which formally replicates the N-body box eight times. However, each 
final simulation covers only one octant of the full-sky, $\simeq5,000$ square degrees, eliminating the replications. 
As the DES SV area with weak lensing measurements is only $139$ square degrees, we divide each 
simulation into 18 different pieces using the observed SV mask to construct 126 total mock catalogs. 
This procedure has the advantage of properly computing the halo sample variance contributions to the 
lensing covariance matrices due to the fact that each patch is embedded in the large-scale modes of the box. 

We place lensing sources randomly in angle with in the DES SV mask (see \citet{jarvis2015} for the details of the mask), 
and with the redshift distribution of the tomographic bins defined above. 
Then the weak lensing shear for each source is computed using the \verb+CALCLENS+ ray-tracing code \citep{becker2013}. 
In this application of \texttt{CALCLENS}, we use the pure spherical harmonic transform version with \verb+Nside=8192+. 
Appendix~\ref{app:simtests} presents tests of the underlying simulations in comparison to simple expectations from 
fitting functions to the matter power spectrum. We find that the simple expectations from matter power spectrum fitting 
functions agree with the simulation to within sample variance, but that some resolution issues remain on small scales. 
Note, however, that these small scales are excluded from the companion cosmological analysis \citep{deswlcosmo} and that despite the 
resolution issues, we find excellent agreement between the covariances computed from the mock catalogs and the halo model, 
as discussed below. Thus for purposes of computing covariance matrices, the mock catalogs we have constructed are 
sufficient. Future work may require higher-resolution shear fields for covariance estimation.

Finally, we generate the shape noise and other properties in the mock by randomly drawing from the observations 
separately for each tomographic bin. Importantly, we draw the intrinsic shape of each mock shear source 
separately from its other properties, like signal-to-noise, size, etc. Properties which have intrinsic spatial dependence 
in the survey (e.g. seeing, airmass, etc.) are drawn from the nearest real galaxy to each mock galaxy. See Section~\ref{sec:nctests} for more details.
These procedures randomise the shear field in the data and ensure that the mock catalogs have no correlations between the systematic parameters 
and the shear field. 

\section{Measurements of Cosmic Shear Two-point Statistics}\label{sec:2ptcov}
In this work, we focus on cosmic shear measurements made with two-point statistics, which are detailed in the 
following sections. A companion paper \citep{deswlcosmo} presents the associated cosmological parameter 
constraints using these measurements, which use the real-space two-point correlation functions as the fiducial two-point 
estimator. We summarize results from alternate estimators in Section~\ref{sec:altebstats} and Appendix~\ref{app:altebstats}. 
Note that although the choice of which two-point statistic to use is somewhat arbitrary, the companion cosmological analysis 
of this data \citep{deswlcosmo} demonstrates that the exact choice of two-point statistic does not change the cosmological parameter 
constraints from this data in a statistically significant way. 

\subsection{Real-space Two-point Function Estimators}\label{sec:2ptest}
We follow \citet{miller2013} and estimate the two-point functions with
\begin{eqnarray}
\xi_{\pm}&=&X_{+}\pm X_{\times}\\
X_{+/\times}&=&\frac{\sum_{i,j} w_{i}w_{j}(e-c)_{i,+/\times}(e-c)_{j,+/\times}}{\sum_{i,j} w_{i}w_{j}s_{i}s_{j}}\nonumber
\end{eqnarray}
where $i,j$ index the galaxies in the two sets we are correlating. 
Here $e_{+/\times}$ are the estimated shears from the lensing analysis projected into the $+$ (tangential) and $\times$ (cross) 
components rotated into the reference frame connecting each each pair of galaxies $\{i,j\}$ in the sum. The $w_{i}$ are weights 
applied to each galaxy (typically inverse variance weighting; see Sec. \ref{sec:data} for each lensing code). The $s_{i}$ are 
multiplicative noise bias and/or lensing sensitivity corrections that are applied to the shears. We follow \citet{miller2013} and 
apply these corrections to the entire population of shears as opposed to applying them to each shear individually. We compared several different 
methods for incorporating the sensitivities into the two-point function estimator and find that they differ by at most $\sim2\%$. The $c_{i}$ 
are the additive bias corrections used for \ims\ and are identically zero for \ngm\ per the definition of the \verb+lensfit+ method \citep{miller2007}. 
Finally, we use \verb+TreeCorr+\footnote{\url{https://github.com/rmjarvis/TreeCorr}} \citep{jarvis2004} to compute the shear correlation functions.

\begin{figure*}
\begin{center}
\includegraphics[width=\columnwidth]{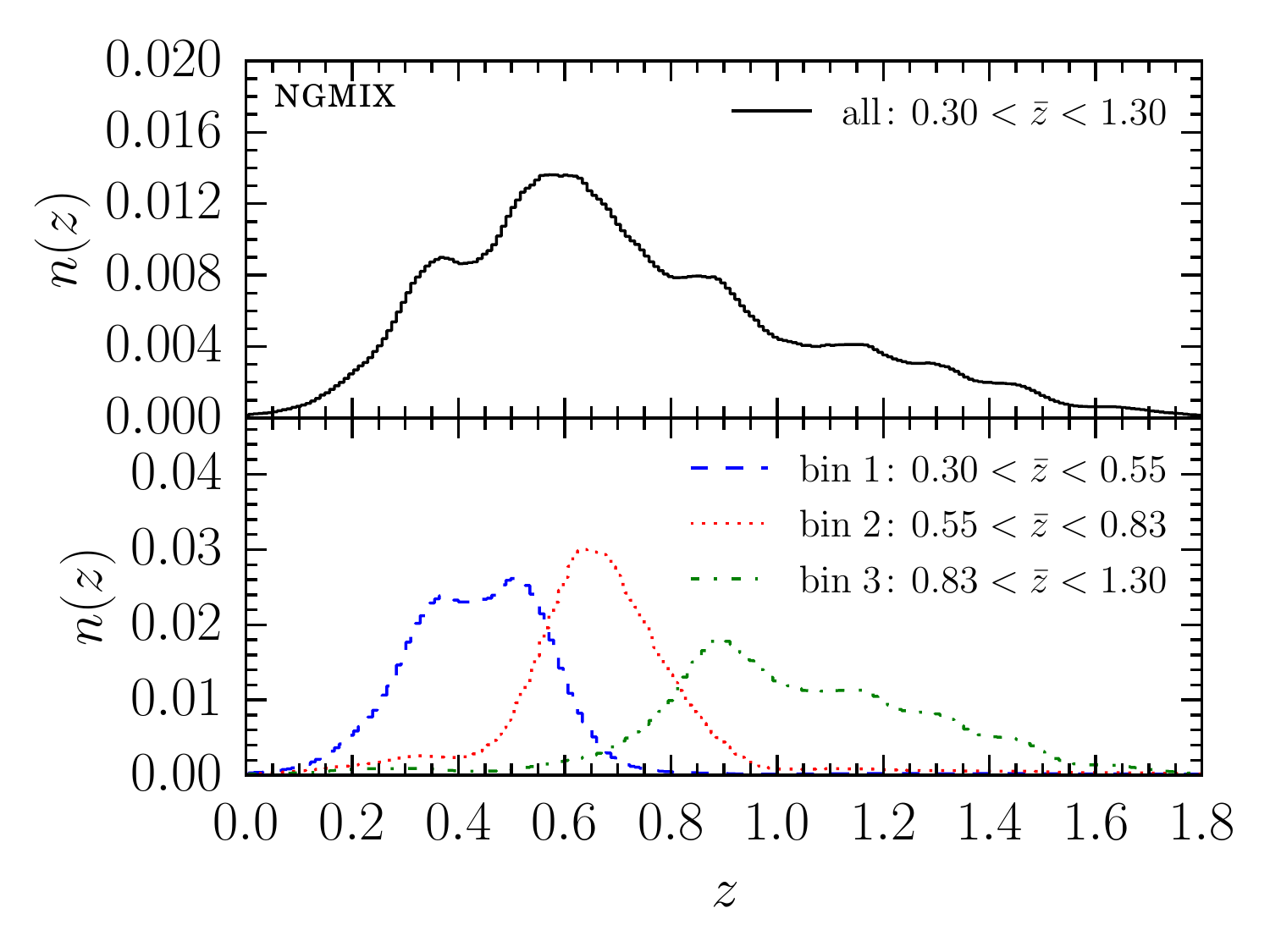}
\includegraphics[width=\columnwidth]{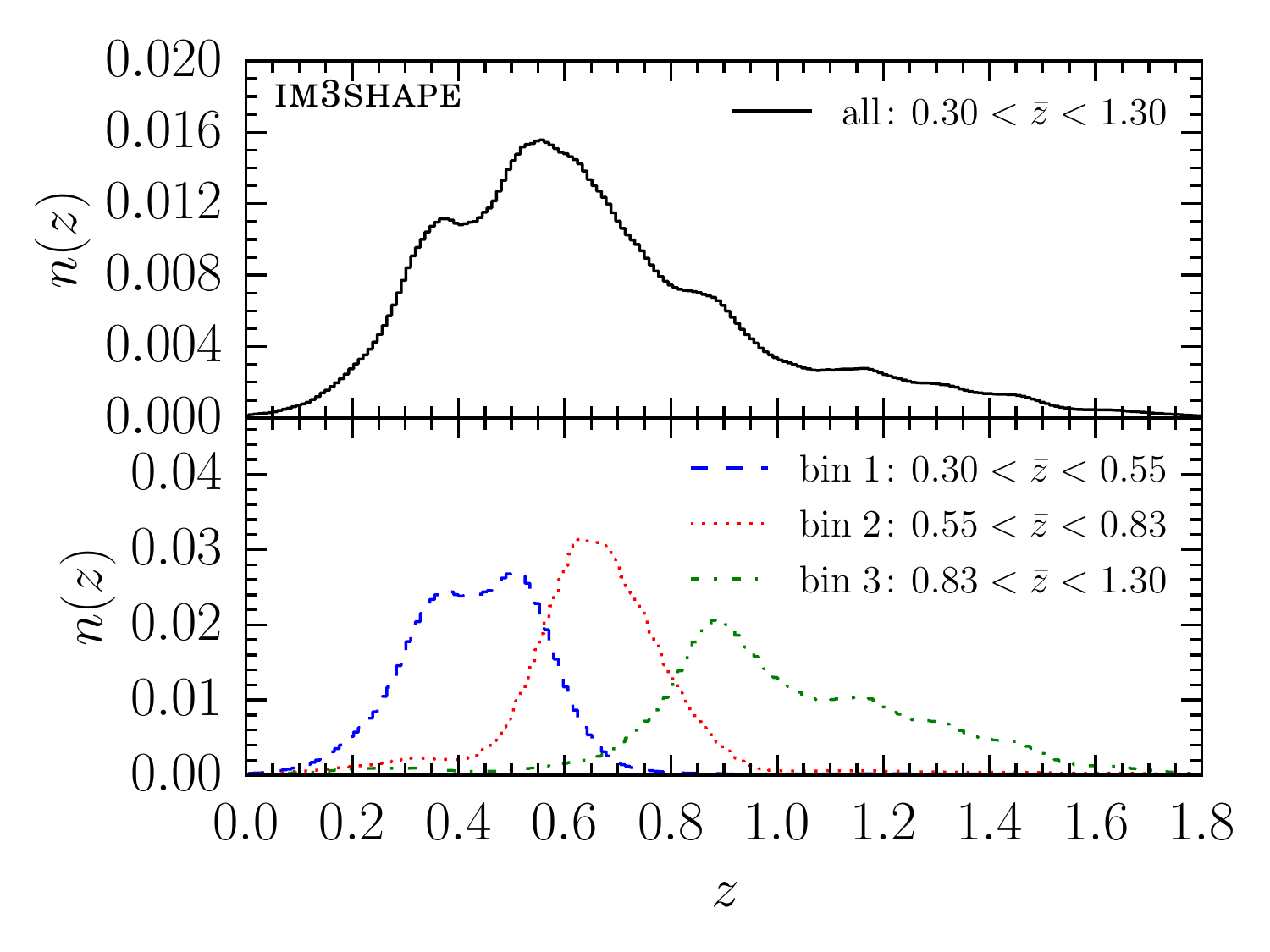}
\end{center}
\caption[]{The estimated redshift distributions from \texttt{SkyNet} for the \ngm\ shape catalog (left) and the \ims\ shape catalog (right). 
The full $n(z)$ for objects with mean redshifts in the redshift range $0.3<\bar{z}<1.3$ (top) and the $n(z)$ for three tomographic bins (bottom) 
are shown. The redshift distributions are estimated by summing and rescaling the photometric redshift probability distributions for each galaxy in the 
tomographic bin using the weights applied to the shear catalog.
\label{fig:geomphotoz}}
\end{figure*}

\subsection{Real-space Correlation Functions}\label{sec:goodstuff}
The real-space correlation functions without tomography are shown in Figure~\ref{fig:xinotomo}. We show \ngm\ 
on the left and \ims\ on the right, with $\xi_{+}$ in the top rows and $\xi_{-}$ in the bottom rows. Negative 
measurements are shown as upper limits. The redshift distribution of sources for the non-tomographic analysis 
is shown in the top panel of Figure~\ref{fig:geomphotoz} for the \texttt{SkyNet} code. It appears in Figure~\ref{fig:xinotomo} 
that the $\xi_{+}$ correlation function may approach a constant value at large scales. Interestingly, 
\citet{jarvis2015} find that the mean shear across the survey for \ngm\ and \ims\ is $\approx7-10\times10^{-4}$. This level of 
mean shear would produce a constant floor in the shear correlation functions of $\approx5-10\times10^{-7}$. For the DES SV 
survey, the root-mean-square mean shear just due to shape noise and cosmic variance is $\approx4\times10^{-4}$. Thus it is not clear 
if this feature is an indication of systematic effects or a few sigma fluctuation in the mean shear due to a real physical effect. However, in 
the cosmological analysis of this data, all $\xi_{+}$ data points above $60$ arcminutes were cut to avoid systematics in the PSF models 
\citep{jarvis2015,deswlcosmo}. Thus we do not explore this issue further in this work. 

We generate estimates of the $1\sigma$ 
uncertainties for each measurement by computing the covariance of the two-point functions over the simulation mock 
catalogs described in Section~\ref{sec:mocks}. These mock catalogs are built separately for each shear catalog in order to 
match the non-tomographic redshift distribution of the sources. The correction factor described in \citet{hartlap2007} 
is then applied to produce an unbiased estimate (see Section~\ref{sec:hmodcomp} for a further discussion of the statistical properties of the 
covariance matrix estimate from the mock catalogs). The significance of the resulting measurement is then calculated 
from this covariance as
\begin{equation}
S/N=\frac{\bxi_{\rm data} \mathbf{C}^{-1} \bxi_{\rm model}}{\sqrt{\bxi_{\rm model} \mathbf{C}^{-1} \bxi_{\rm model}}},
\end{equation}
where $\mathbf{C}^{-1}$ is the inverse covariance matrix estimated from the mock catalogs, $\bxi_{\rm data}$ is the vector 
of real-space shear two-point function measurements from the data, and $\bxi_{\rm model}$ is the vector 
of real-space shear two-point function measurements predicted from the cosmological model given above in Section~\ref{sec:mocks}. This quantity corresponds to 
the signal-to-noise of a least-squares estimate of a scaling parameter comparing our measurements to the theoretical model. This signal-to-noise 
measure will be an underestimate if the model employed is not well matched to the data. However, given the good match of our 
fiducial model to the data as shown in Figures~\ref{fig:xinotomo} and \ref{fig:xitomo}, the degree to which the signal-to-noise is underestimated is small in this case. 
We use the \texttt{COSMOSIS} package\footnote{\tt{https://bitbucket.org/joezuntz/cosmosis}} by \citet{cosmosis} to compute the shear correlation functions with 
the \citet{takahashi2012} non-linear power spectrum fitting function. See the companion 
paper \citep{deswlcosmo} presenting cosmological constraints from these measurements for additional details on 
the model correlation function $\bxi_{\rm model}$ computation. 
The covariance matrix has been validated through comparisons to both a detailed halo model prediction 
and jackknife estimates in single mock patches versus the survey data, which are discussed in detail in the Section~\ref{sec:covmatrix}. 
We find non-tomographic cosmic shear detections at 6.5$\sigma$ and 4.7$\sigma$ significance for \ngm\ and \ims\ respectively. 

Figure~\ref{fig:xitomo} shows the full three-bin tomographic shear correlation function measurements for \ngm\ on the left and 
\ims\ on the right. The redshift distributions of the three tomographic bins for the \texttt{SkyNet} code are given in the lower 
panels of Figure~\ref{fig:geomphotoz}. In order to compute the covariance matrix of these measurements, we use the same 
procedure in the mock catalogs as for the non-tomographic case, except that we use the tomographic redshift distributions to 
assign the mock galaxies to different tomographic bins. We additionally draw the shape noise in the mock from only the galaxies 
in the data in the same tomographic bin. We find overall tomographic cosmic shear detections of 9.7$\sigma$ and 7.0$\sigma$ 
for \ngm\ and \ims, respectively. Note that the \ngm\ catalog has more sources and extends to slightly higher redshift on average, 
yielding higher significance detections of cosmic shear. We have chosen three tomographic bins as a compromise between gaining 
signal-to-noise in the data and having too many data points in order to use the mocks to compute the covariance matrix of the data. 

In Figures~\ref{fig:xinotomo} and \ref{fig:xitomo}, the solid black line shows the expected amplitude and shape of the shear correlation functions 
in the cosmological model given above. This curve is not a fit, and is merely 
presented as a reference for comparison.  Due to the fact that the two catalogs have different redshift distributions, a direct comparison of the shear correlation 
functions between the two catalogs is not possible without further work matching the two catalogs and accounting for the shared shape noise, sample variance, 
and image noise between the two catalogs. This matched comparison is described further in Sec. \ref{sec:const}. 

\subsection{Alternative Two-point Statistics}\label{sec:altebstats}
In Appendix~\ref{app:altebstats}, we describe results from two alternative two-point statistics of the shear field. These include the 
methods of: (i) \citet{becker2014}, which use a weighting of the real-space correlation estimates to construct efficient estimates of the $C_{\ell}$ values 
and (ii) a second estimation of the spherical harmonic shear power spectrum using \texttt{PolSpice}\footnote{\url{http://www2.iap.fr/users/hivon/software/PolSpice/}} 
\citep{szapudi2001,chon2004}. Note that these estimators weight the data at different angular scales differently than the default two-point correlation functions so that 
we do not expect to get identical results in terms of the significance of the cosmic shear detection. We do find detections of cosmic shear that are consistent with the 
conventional real-space estimators we use by default, indicating no strong preference for any given estimator. Tests of B-mode statistics from these estimators are 
discussed in Sec. \ref{sec:bmodes}, where we again find consistency between different two-point function estimation methods. 

\section{Estimating and Validating the Covariance Matrix}\label{sec:covmatrix}
In this section, we present our covariance matrix and a set of validation tests. The fiducial covariance matrix for our measurements is estimated 
from the mock catalogs presented in Section~\ref{sec:mocks}. First we compare the covariance matrix from the mock catalogs to halo model computations. 
Second, we compare jackknife covariances in the data to the jackknife covariance computed from the mock catalogs. This procedure allows us 
to look for additional sources of noise and correlations in the data that are not present in the mock catalogs. 

\begin{figure*}
\begin{center}
\includegraphics[width=\columnwidth]{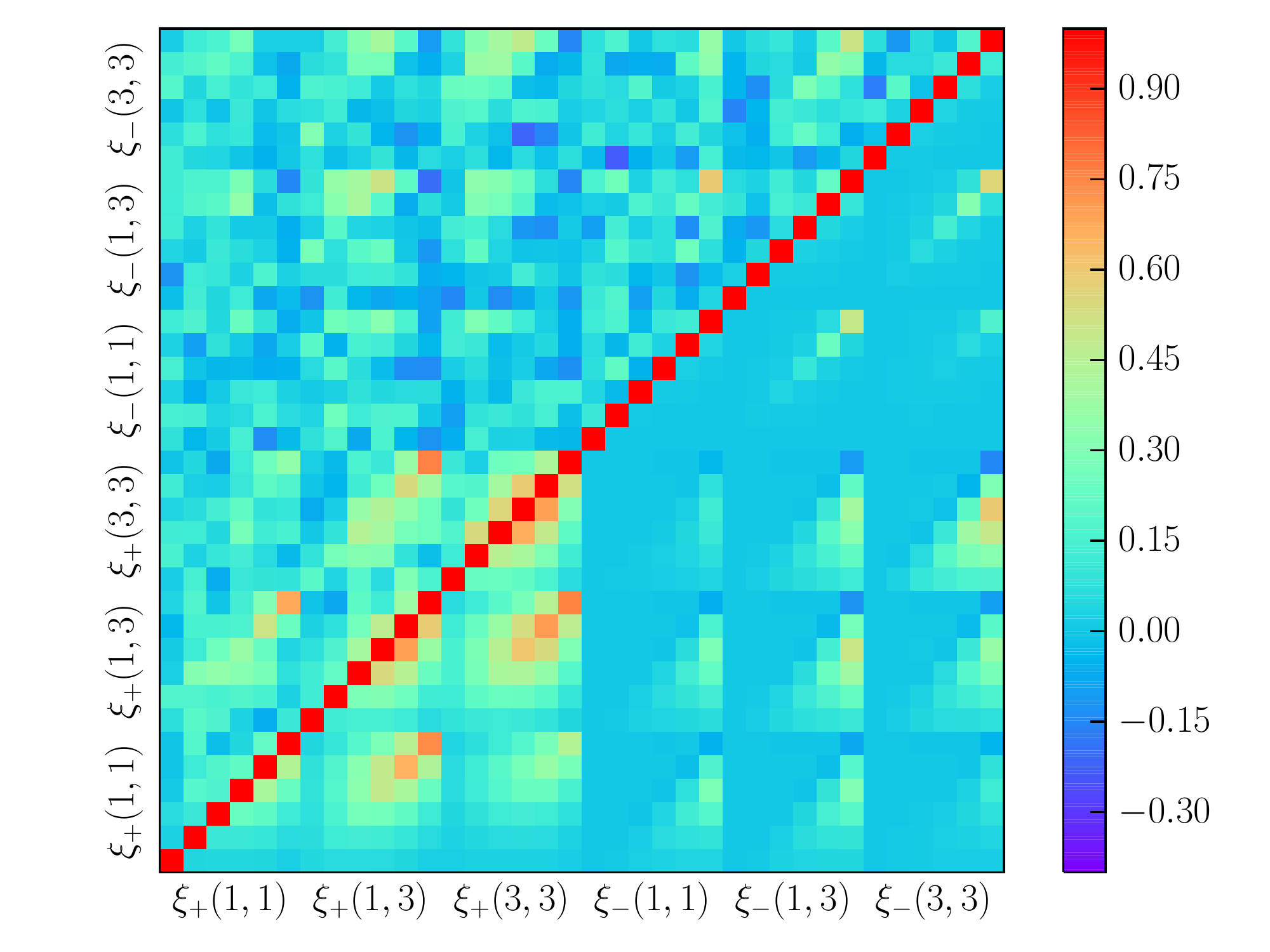}
\includegraphics[width=\columnwidth]{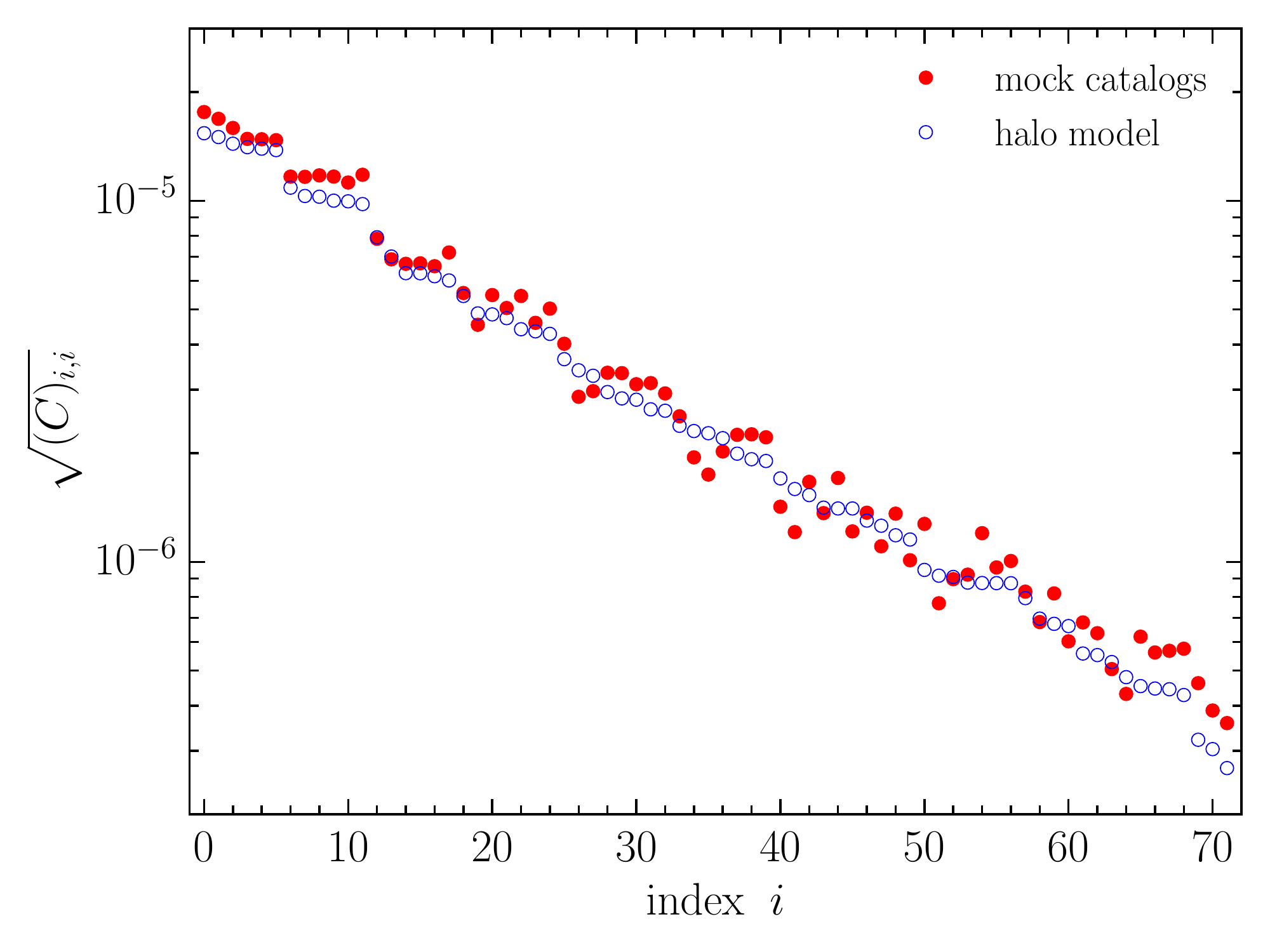}
\end{center}
\caption[]{Comparison of the tomographic shear correlation function correlation matrix estimated from the mock catalogs 
and calculated from the halo model. The left plot shows the correlation matrix from the 
mock catalogs (upper left) and halo model (lower right). We show only the components for the first and last tomographic bins, plus their cross correlations. On the right, we show the square root 
of the diagonal elements of both covariance matrices, sorted in reverse numerical order. The open symbols show the results from the halo model and the closed symbols show 
the results from the mock catalogs. \label{fig:covhmcompsimp}}
\end{figure*}

\begin{figure*}
\begin{center}
\includegraphics[width=\columnwidth]{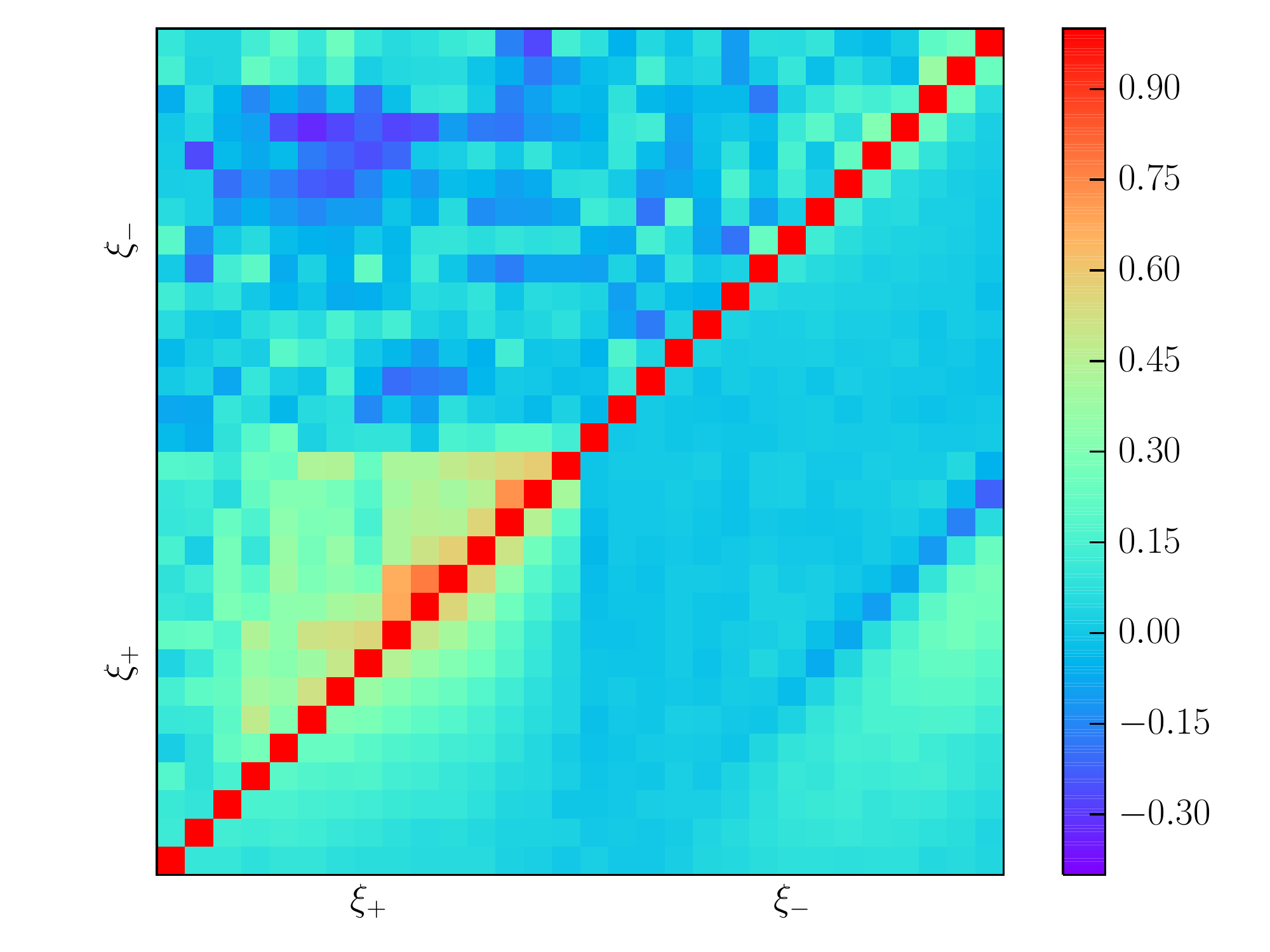}
\includegraphics[width=\columnwidth]{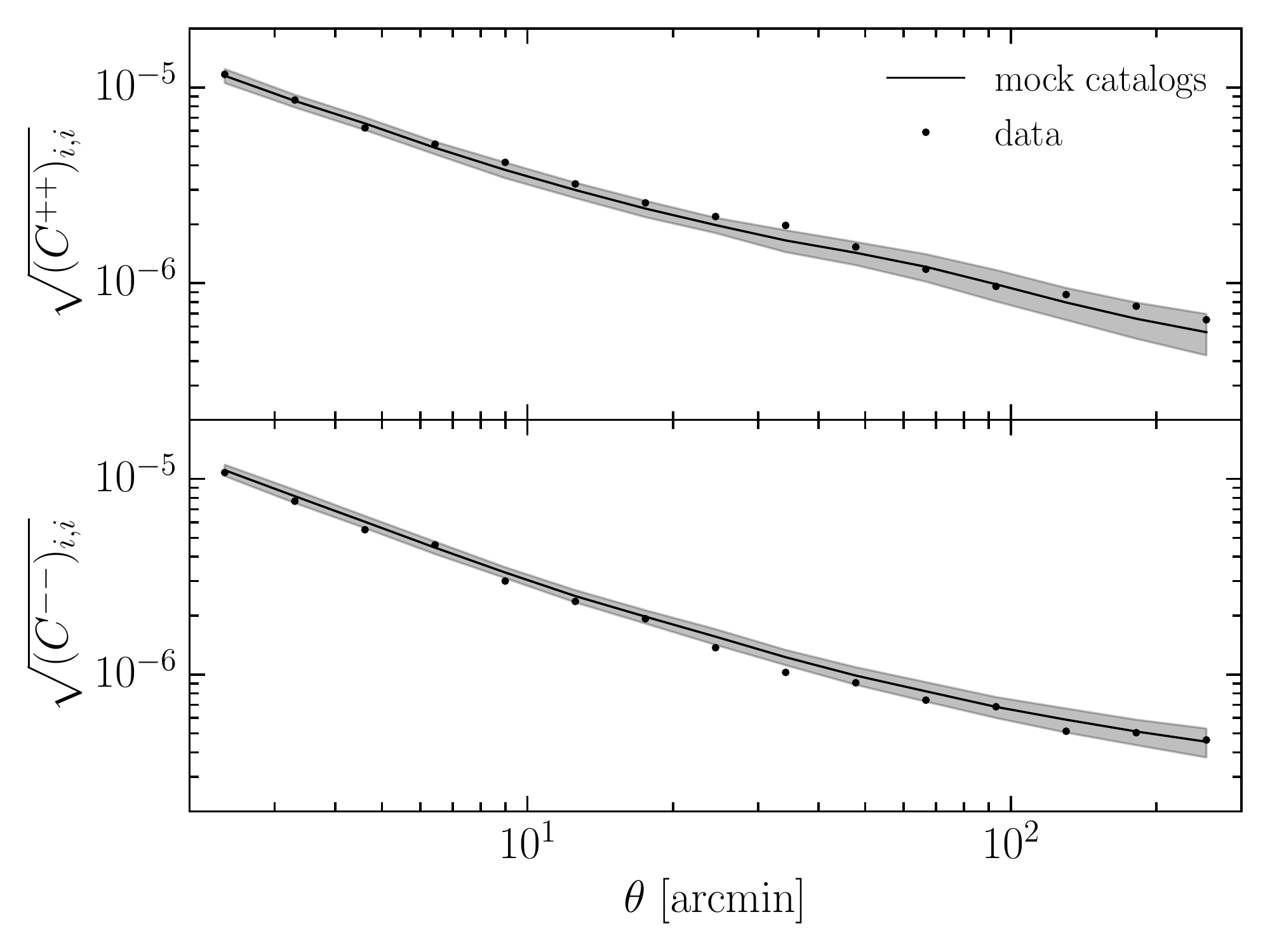}
\end{center}
\vspace{-2em}
\caption[]{Jackknife covariances in the mock catalogs and the data for the non-tomographic shear correlation functions. The left panel shows the correlation matrix using jackknifes in the 
data for \ngm\ (top left) and when averaging the jackknife covariances in the 126 mock catalogues (bottom right). 
The bottom left quadrant contains the $\xi_+$ correlations, the top right the $\xi_{-}$ correlations and the off diagonal components contain the cross-correlations. For each 
submatrix of the full correlation matrix, the angular scale increases from 2 arcminutes to 300 arcminutes. On the right, we show the diagonal 
elements of the jackknife covariance matrix in the data for \ngm\ (points) and when averaged over 126 mock catalogues (line). 
The grey band shows the standard deviation of diagonal elements over the 126 mock catalogs. \label{fig:jacktest}}
\end{figure*}

\subsection{Simulation and Halo Model Comparison}\label{sec:hmodcomp}
We compare the covariance matrix computed from the simulations to that obtained from a halo model in Figure~\ref{fig:covhmcompsimp}. 
The simulation-based covariance matrix is computed by populating the mock catalogs with shear sources as described above in 
Section~\ref{sec:mocks}, and then computing the covariance of the measurements performed on the full ensemble of mock catalogs. 
The halo model covariance was computed with the \verb+CosmoLike+ covariance module (see \citet{eifler2014} and 
\citet{krause2015} for details). Further details of our halo model 
computation and the full tomographic covariance matrix are given in Appendix~\ref{app:hmod}. Briefly, we include the Gaussian, non-Gaussian and halo sample variance terms 
\citep[e.g.,][]{takada2013} and compute the halo model covariance at the same cosmology and with the same redshift distribution as was used in the mock catalogs.  

We compare the general structure of the mock (upper triangle) and halo model (lower triangle) covariance in the left panel of 
Figure~\ref{fig:covhmcompsimp}, which shows part of the correlation matrix. Here we have shown a subset of the full set of tomographic bin 
combinations. The full correlation matrix is shown in Appendix~\ref{app:hmod}. The right panel compares the amplitude of the two covariances 
by plotting the variance. Overall, we find good agreement in both structure and amplitude. 

We quantitatively test the agreement using a Fisher matrix computation. We compute the expected error on the degenerate parameter 
combination $\sigma_{8}(\Omega_{m}/0.3)^{0.5}$, where $\sigma_{8}$ is the RMS 
amplitude of the linear matter power spectrum at redshift zero in a top hat window of 8 $h^{-1}$Mpc and $\Omega_{m}$ is the matter density 
in units of the critical density at $z=0$. This combination of parameters is typically the best constrained by low-redshift cosmic shear data 
sets like the DES SV data. The exact degeneracy is computed in the companion cosmological constraints paper to this work \citep{deswlcosmo}.
We use the standard Fisher matrix formalism for cosmic shear \citep[see, e.g.,][]{albrecht2009} and the same cosmological model as described above. 
We vary only the spectral index $n_{s}$, $\sigma_{8}$ and $\Omega_{m}$ in the Fisher matrix. 

We find that the error bars on $\sigma_{8}(\Omega_{m}/0.3)^{0.5}$ from the halo model and mock covariances agree to approximately $10\%$ without 
tomography, with the halo model yielding larger parameter uncertainties. When repeating the same exercise with tomography, we find a larger, 
$\approx35\%$ disagreement in the error bars, with the mocks yielding larger errors. 
However, we expect fluctuations in the uncertainties in parameters computed with the simulations due to the finite number of realizations 
used for the covariance computation. \citet{dodelson2013} estimate that this effect, in the Gaussian limit, increases the variance in the parameter estimates by a factor of 
\begin{displaymath}
\alpha = 1+\frac{(N_{d}-N_{p})(N_{s}-N_{d}-2)}{(N_{s}-N_{b}-1)(N_{s}-N_{b}-4)}
\end{displaymath}
where $N_{d}$ is the number of data points, $N_{s}$ is the number of simulations and $N_{p}$ is the number of parameters.
This factor is $\approx1+N_{d}/N_{s}$ in the limit that $N_{s}\gg N_{d}\gg N_{p}$. Thus we expect a fractional uncertainty in the 
parameter uncertainties of $\approx\sqrt{\alpha-1}$. In our case with tomography, $N_{d}=72$, $N_{s}=126$ and $N_{p}=1$. With these numbers, 
we get that the fractional uncertainty in the parameter uncertainty is $\approx118\%$. Thus the disagreement of $\approx35\%$ we find with the halo model with tomography 
is not statistically significant. Without tomography, we find a fractional uncertainty in the uncertainty of $\approx56\%$, again indicating consistency. 

Importantly, these numbers are the fractional uncertainty in the uncertainty. For parameter estimates, the fractional increase in the uncertainty on the parameter, equal to 
$\sqrt{\alpha}$, is the relevant quantity. For tomography, this fractional increase is $\approx55\%$ and without tomography it is $\approx15\%$. Furthermore, 
we have assumed that the tomographic analysis uses all 72 data points. As described in \citet{deswlcosmo}, only 36 of the 72 data points are used for tomography, 
bringing the fractional increase in the error due to the finite number of realizations down to only $\approx18\%$. Similar cuts are made for the non-tomographic analysis, 
using only 16 of the 30 data points. This number of data points results in a fractional increase of the parameter uncertainties of only $\approx7\%$ for the 
non-tomographic analysis. 

\subsection{Jackknife Comparisons to Data}\label{sec:cov}
While our mock catalogs include both sample variance and shape noise contributions, any spatially varying systematic effects, like errors in the shear calibration, should be included 
in the covariance matrix of the shear correlation functions as well. To search for these potential effects, we use the jackknife covariance matrix of the shear correlation functions as a 
statistic to be compared between the data and the mock catalogs. Any additional sources of noise in the data, which are captured by the spatial scale of our jackknife regions, will 
show up as a difference between the jackknife covariance as computed in the data versus the mock catalogs.

We estimate the jackknife covariances from the data and our mock catalogs as follows. We divide both the mock catalog and data into $100$ spatial sub-regions, 
employing the k-means algorithm.\footnote{Implemented for python by Erin Sheldon, \texttt{www.github.com/esheldon/kmeans\_radec}.} These regions are then 
used to perform jackknife resampling. For the details of jackknife covariance estimation for cosmic shear correlation functions, we refer the reader to a (technical) 
companion paper where these choices are examined in further detail (\citet{friedrich2016}, see also \citet{2009MNRAS.396...19N} for an application to 
galaxy clustering). We use the standard jackknife scheme, where all of the shear sources in an entire subregion are removed for each 
jackknife resampling, which is called the \emph{galaxy-jackknife} in \citet{friedrich2016}.

Note that we are not comparing jackknife covariances with the true covariances, but rather simply the co-variance in the shear correlation function across the DES SV 
survey to the same statistic computed with the mock catalogs. Thus the absolute correctness of the jackknife covariance matrix is not an issue for our test, since it is 
just a statistic that is sensitive to the effects for which we wish to search. The performance of empirical covariance measures for cosmic shear surveys is explored 
in \citet{friedrich2016}.

The comparison of our jackknife procedure between the mocks and the data is shown in Figure~\ref{fig:jacktest}. Here we plot the correlation matrix of the averaged jackknife 
covariance from the 126 mock \ngm\ catalogs (left panel, on the bottom right) and the same computation in the DES SV data (on the top left). 
The right panel compares the diagonal elements of the jackknife covariance for $\xi_{+}$ and $\xi_{-}$ when averaged 
over 126 mock catalogs and when computed from the data for \ngm. Using the Fisher matrix procedure described above, we find that the  
error on $\sigma_{8}(\Omega_{m}/0.3)^{0.5}$ from the data jackknife covariance matrix agrees with the mean of the ensemble of errors on this parameter from the mock 
jackknife covariances to within one standard deviation of the error over the ensemble. Thus we conclude that there are no statistically significant sources of 
additional variance in the data compared to the mock catalogs.

\begin{figure*}
\begin{center}
\includegraphics[width=\columnwidth]{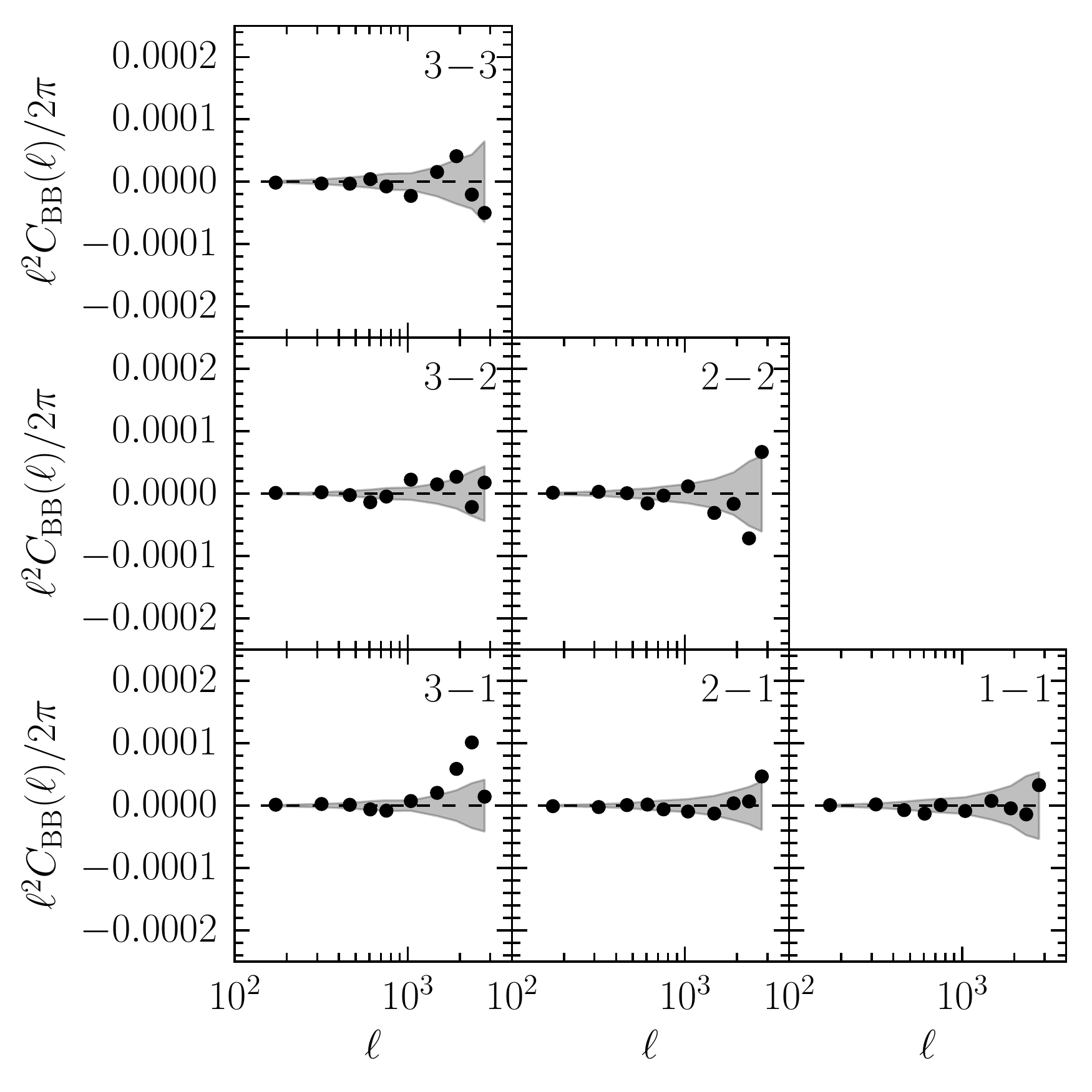}
\includegraphics[width=\columnwidth]{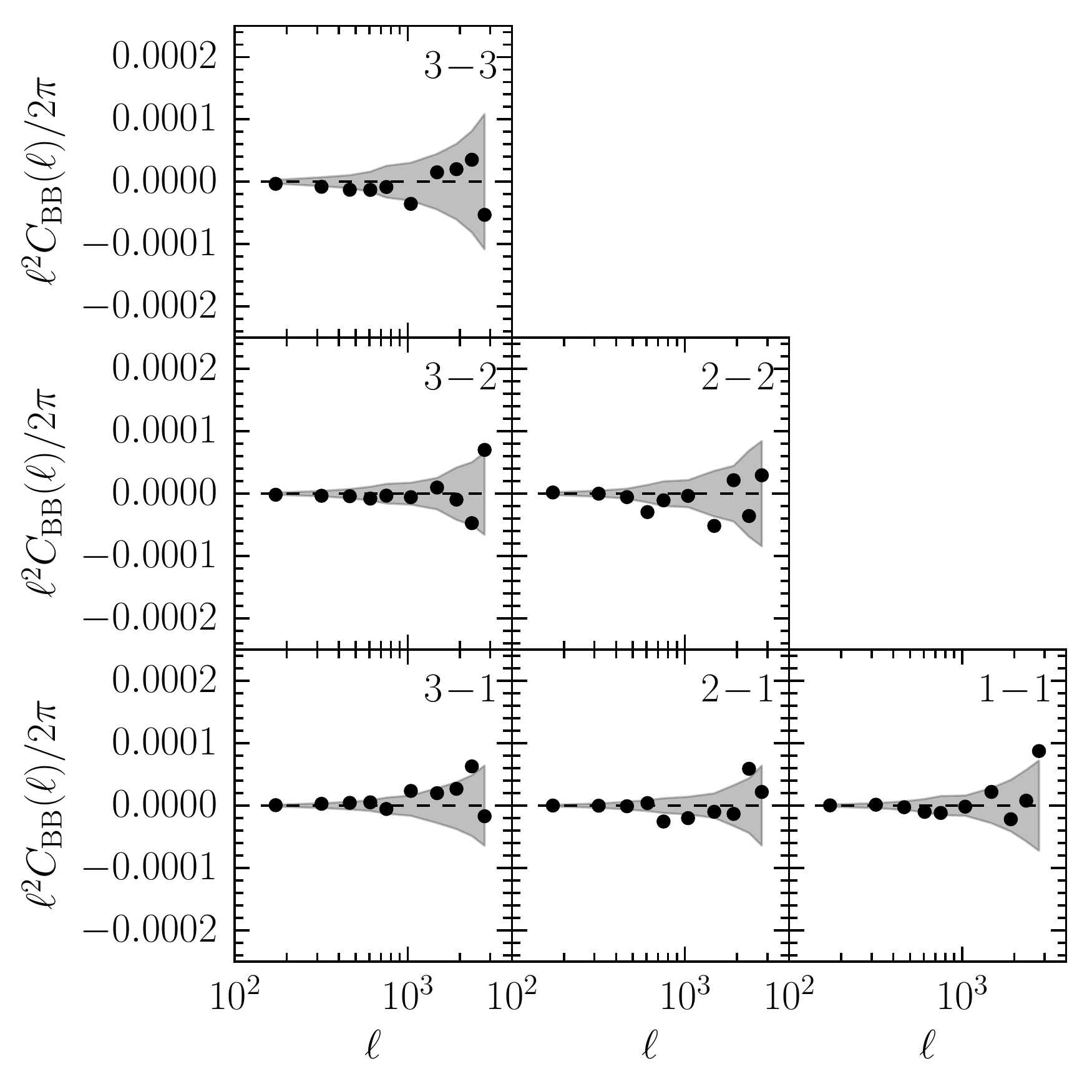}
\end{center}
\caption[]{Tomographic B-modes in DES SV data for \ngm\ (left) and \ims\ (right).
The error bars are calculated from the simulation realizations using the shape noise appropriate 
for each catalog. The tomographic bins correspond to those shown in Fig. \ref{fig:geomphotoz} and are labeled from 1 to 3, 
increasing with redshift. Thus, panel '3-2' shows the cross-correlation between the highest and middle redshift bins. The total $\chi^{2}/{\rm d.o.f.}$, 
accounting for the correlations between the points in each panel, for \ngm\ is 62.5/60 and for \ims\ is 41.2/60. \label{fig:bmodes}}
\end{figure*}

\section{Tests for Residual Systematic Errors in the Cosmic Shear Signal}\label{sec:systematics}
Systematic errors in shear measurements can be from a wide array of sources ranging from telescope optics and observing conditions to details in 
the modeling, measurement, and calibration of shapes. The development of tests to identify potential systematic errors is critical to verifying accurate 
measurement of cosmic shear. Toward this end, we devise a set of tests that should produce a null result when applied to true gravitational shear. 
The measurement of a significant non-zero result is then an indication of unresolved systematic errors in the shear catalog that could bias measurements.

The DES SV shear catalogs have passed a rigorous set of both traditional and novel null tests that lay the groundwork for validating the precise measurements 
that will be made with ongoing DES measurements during the main survey observing period. These tests are performed both at the catalog level and during 
the process of validating specific measurements based on the shear catalogs. We describe the methodology and results of both traditional and new null tests 
for sources of potential systematic errors in both the non-tomographic and tomographic measured cosmic shear signal in the next two sections.

Catalog-level tests were performed by \citet[][cf. their Section 8]{jarvis2015} and included tests for additive systematic errors related to spatial position, 
the PSF, and galaxy properties. These tests included the cross-correlation of the galaxies and the PSF. No significant additive systematic errors 
were found, and they put upper limits on the potential additive systematic contribution 
to $\xi_{+}$ in their Section 8.7. In addition, the overall multiplicative bias of the shear estimates was tested with simulations in \citet{jarvis2015}. \citet{jarvis2015} 
concludes that both catalogs are consistent with having small overall multiplicative bias, but due to uncertainties in their ability to constrain this value, they suggest 
marginalizing over a prior on the multiplicative bias with a standard deviation of 0.05 (see Equation 8-12 of \citet{jarvis2015}). This multiplicative systematic is treated in the 
cosmological analysis of this data \citep{deswlcosmo}, where it contributes to an increase in the uncertainties on the final cosmological parameters constrained with this data.

\subsection{B-mode Measurements}\label{sec:bmodes}
The cosmic shear field can be characterized by E- and B-modes which differ in parity. At first-order in the gravitational potential in General Relativity, cosmic shear 
produces a pure E-mode field \citep[see, e.g.,][]{bartelmann2010}. However, contaminating signals, like that from the telescope point-spread function, tend to 
contain both E- and B-modes. Thus one of the first suggested tests of cosmic shear detections was verifying that the B-mode signal is 
consistent with zero \citep{kaiser1992}.\footnote{Small levels of B-modes are produced at second order in the gravitational potential, but these are small enough not to spoil the null test 
\citep[see, e.g.,][]{hilbert2009,krause2010}.} Many methods have been suggested for B-mode estimation 
\citep[e.g.,][]{schneider1998,seljak1998,hu2001b,2002ApJ...568...20C,schneider2007,schneider2010}. 
Here we use the estimators from \citet{becker2014}, which estimate band-powers using linear combinations of the shear 
two-point functions that optimally separate E- and B-modes \citep{becker2013}. These estimators are
\begin{eqnarray}
E = \frac{1}{2}\left[\sum f_{+i}\xi_{+i}+\sum f_{-i}\xi_{-i}\right] \label{eq:bmodee}\\
B = \frac{1}{2}\left[\sum f_{+i}\xi_{+i}-\sum f_{-i}\xi_{-i}\right] \label{eq:bmodeb},
\end{eqnarray}
where the sums run over the angular bins of the shear two-point functions. The weight vectors $f_{+/-}$ are chosen to 
simultaneously minimize E- to B-mode mixing while also producing compact band-power estimates in Fourier-space. 
See Appendix~\ref{app:altebstats} for more details.

In Figure~\ref{fig:bmodes}, we show a measurement of the tomographic B-mode signal using the \citet{becker2014} band-powers. 
We find no statistically significant B-mode contamination, with a total $\chi^{2}/{\rm d.o.f.}$ for \ngm\ of 62.5/60 and for \ims\ of 41.2/60. The error bars in 
this case are computed using the mock catalogs above. In Appendix~\ref{app:altebstats}, we verify this conclusion by computing a complementary 
measurement of the non-tomographic B-mode signal using an alternate estimation of the spherical harmonic shear power spectrum. We find the B-modes 
from this alternate technique are consistent with zero with a $\chi^{2}/{\rm d.o.f.}=4.5/7$ for \ngm\ and 6.3/7 for \ims. Finally, note that 
\citet{becker2014} band-power measurements of the non-tomographic B-mode signal are presented in \citet{jarvis2015} using the methods and mock catalogs 
of this work. The non-tomographic B-mode measurements were again found to be consistent with zero, with $\chi^{2}/{\rm d.o.f.}=22.3/20$ 
for \ngm\ and $16.1/20$ for \ims.

\subsection{Consistency Between the Shear Pipelines}\label{sec:const}
We further test for consistency between the shear catalogs split into tomographic bins by selecting only sources which pass the selection cuts for both codes. For this subset of sources, 
we then compare the shear auto- and cross-correlation functions for each bin. Due to the fact that the two catalogs have the same sample variance, have similar shape noise 
and have correlated shear measurement errors, the error bars on the difference between the two correlation functions is much smaller than that on the correlation functions themselves. 
We account for this effect by constructing mock catalogs where a given mock galaxy is assigned its shape noise for each shear measurement code, \ngm\ or \ims, from the 
same real galaxy. 

\begin{figure}
\begin{center}
\includegraphics[width=\columnwidth]{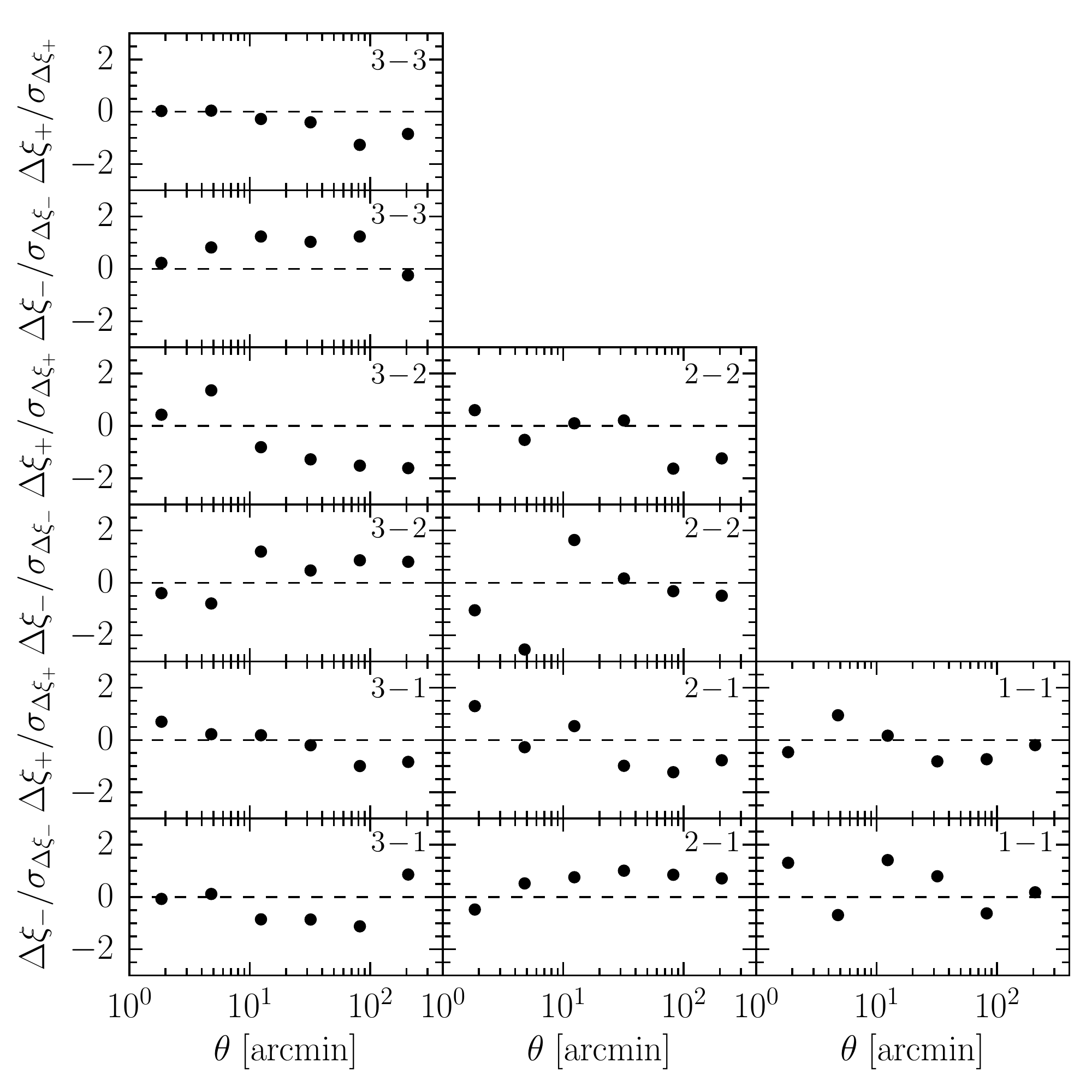}
\end{center}
\caption[]{Difference over error in the tomographic correlation functions for matched shear catalogues from \ngm\ and \ims. We show \ims\ minus \ngm. 
The total $\chi^{2}/{\rm d.o.f.}$ accounting for all correlations is 46.8/72.\label{fig:matchedxi}}
\end{figure}

This comparison is shown in Figure~\ref{fig:matchedxi} for the shear correlation function for \ims\ minus \ngm.\footnote{We have completed this test for 
the ratio of the shear correlation functions and without tomography, finding similar results.} We find that the shear correlation functions from the codes are statistically 
consistent over the full range of scales from 2 to 300 arcminutes, giving a $\chi^{2}/{\rm d.o.f.}=46.8/72$. Finally, note that this test is similar to the differenced shear correlation function test 
presented in Section 8.6 of \citet{jarvis2015}. For their test, they examine the shear correlation function of the 
the difference in the \ngm\ and \ims\ shear estimates using the matched catalogs. They find that below $\approx$3 arcminutes, the catalogs do not meet the requirements for 
additive systematic errors, set by the expected precision of the cosmological constraints. The test presented in this work is generally less sensitive, but complementary, to the 
differenced shear correlation function. 

\begin{figure*}
\begin{center}
\includegraphics[width=\columnwidth]{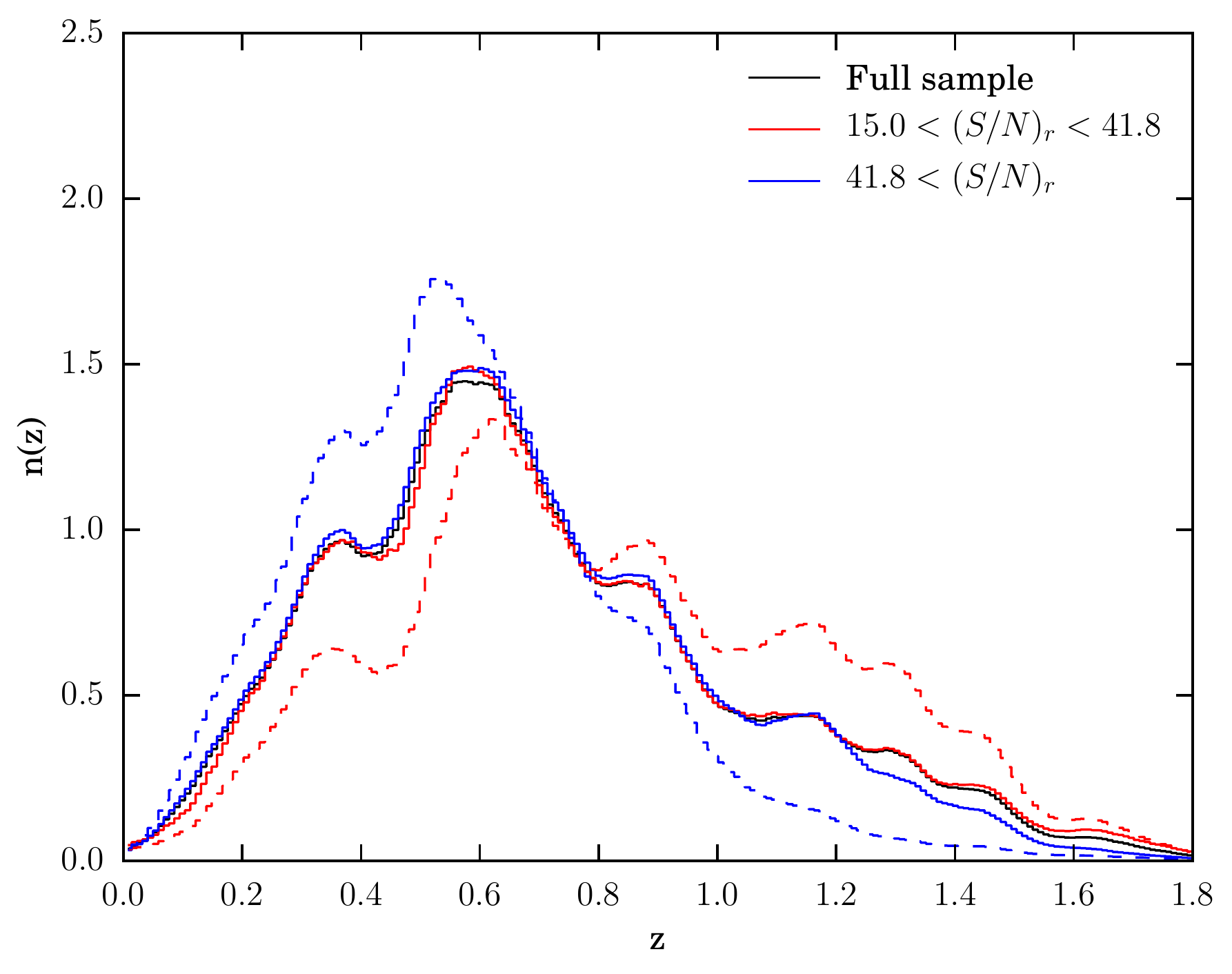}
\includegraphics[width=\columnwidth]{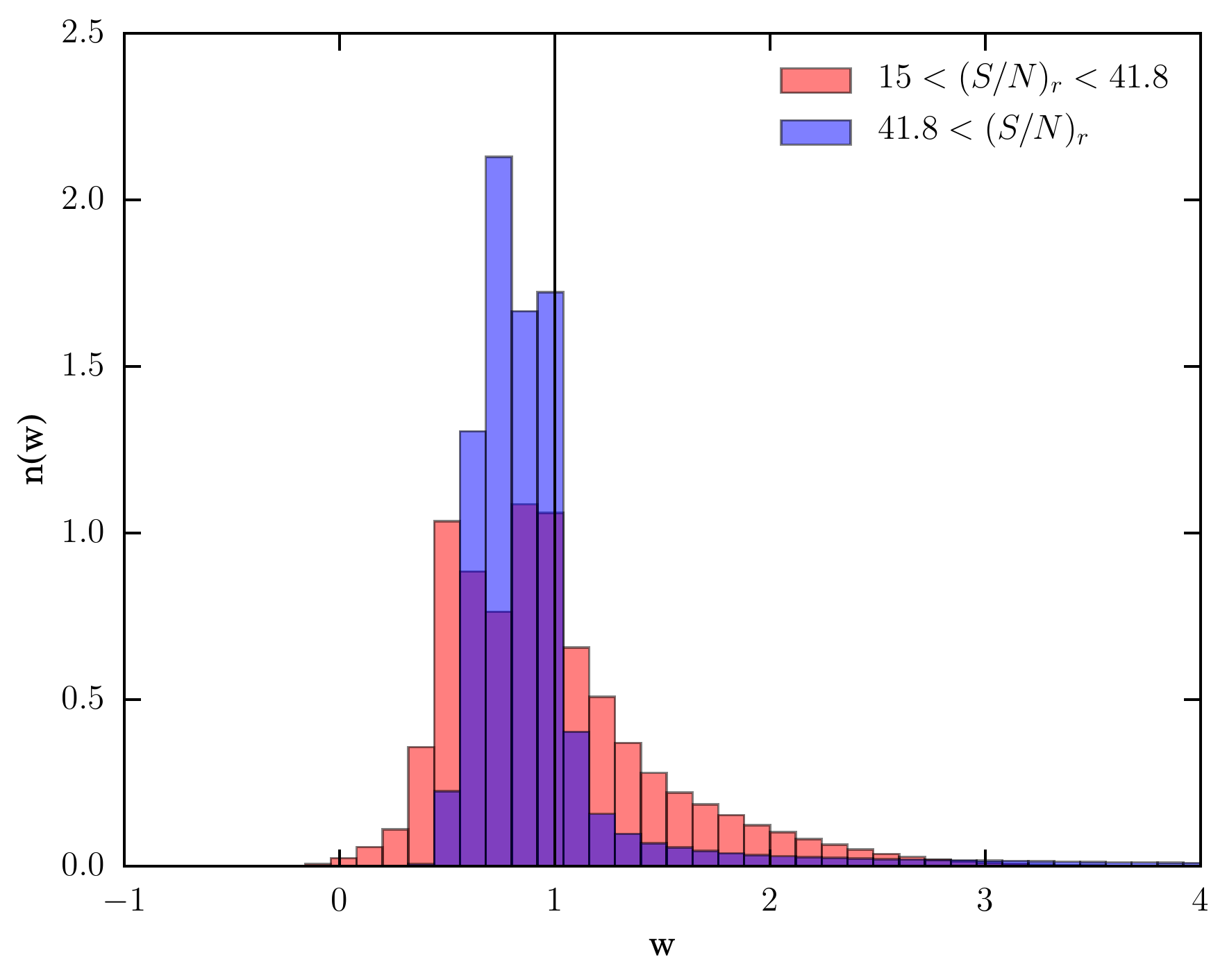}
\end{center}
\caption[]{An example of the redshift re-weighting procedure used when comparing the correlation function between galaxies split 
into bins of galaxy or survey properties. Left: The \textsc{SkyNet} redshift distribution for each half of the \ngm\ data, split into upper 
(blue) and lower (red) bins of signal-to-noise ratio $(S/N)_{r}$ before (dashed) and after (solid) re-weighting, compared to the full sample 
$n(z)$ (black solid curve). Right: The distribution of weights applied to each galaxy to produce the solid $n(z)$ lines, generated as described in Sec. \ref{sec:nctests}. \label{fig:nullexp}}
\end{figure*}

\begin{figure*}
\begin{center}
\includegraphics[width=.9\textwidth]{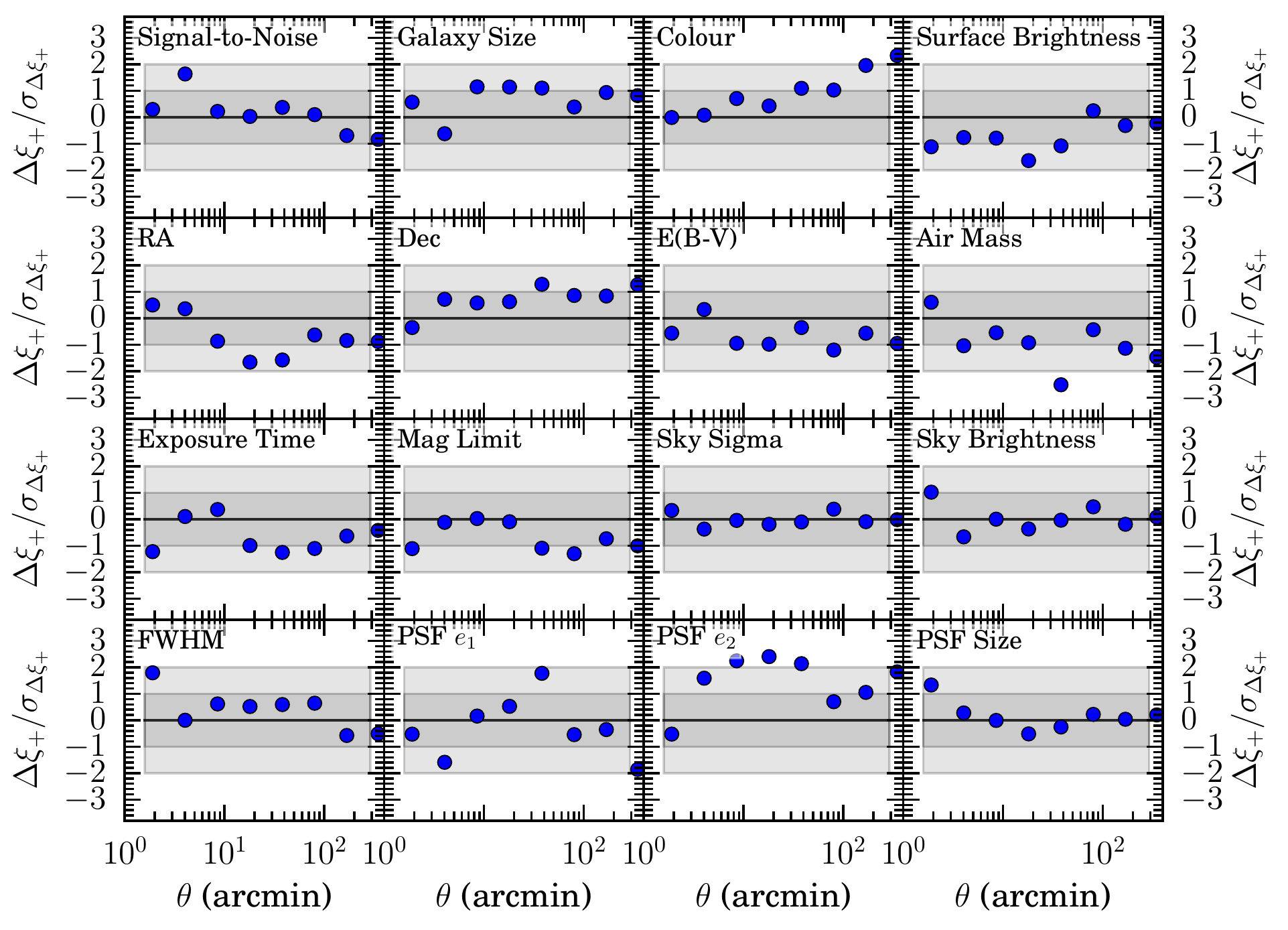}
\end{center}
\caption[]{Null tests for the \ngm\ two point correlation function based on a variety of catalog and survey properties as described in Table 1. 
Each panel for a given property shows the difference between the $\xi_{+}$ relative to its error for the galaxies in the upper and lower 
halves of the sample split into bins by the magnitude of the 
quantity. The two halves of the sample have been reweighted to have the same redshift distribution. The error on the difference is computed 
directly via the mock catalogs. Grey bands are shown 
representing the $1\sigma$ and $2\sigma$ variance at each value of $\theta$. Adjacent points in angle are correlated. \label{fig:ngnulls}}
\end{figure*}

\subsection{Two-Point Null Tests}\label{sec:nctests}
Even with a carefully chosen set of null tests at the catalog level, it is still possible that systematic errors, which can be due to 
complex interplays between different aspects of data and analysis, may influence the cosmic shear measurement. 
To test for any uncorrected systematic errors remaining in the measured cosmic shear signal, we attempt to 
measure the variation in $\xi_+$ as a function of survey and galaxy properties that may be correlated with sources systematic errors. 
For each survey or galaxy property, the shear data is split in half, and the correlation functions of each half are compared.
We use a reweighting method to ensure that the redshift distribution of each half is the same in order to remove 
any cosmological dependence from this null test. If the photo-$z$s and shear measurements are correct, then the shear correlation 
functions of the two halves should be consistent to within the noise of the shear measurements and the redshift reweighting. 
If they are not, this would indicate either uncorrected systematics, selection effects from the split, or non-shear differences in 
the two halves such as intrinsic alignments.

Due to the fact that each half is drawn from the same area in many 
cases, the standard error bars computed for the shear correlation functions are not correct for this test. We instead use the mock catalogs described 
above to compute the error on the difference between the two halves relative to the full sample, accounting for shared sample variance, as 
described below. It is important to note that this is a simultaneous test of both the photometric redshifts and the shear calibrations. This feature 
is in fact desirable because both of these quantities can contribute to biases in the shear correlation functions. We have used both the survey 
property maps described by \citet{leistedt2015} and also properties directly produced by the shape measurement codes. The 16 various 
systematic parameters are described in Table~\ref{table:nulltestsres}. Finally, \citet{jarvis2015} found that making cuts on signal-to-noise 
and size could lead to a selection bias in the population of shear values due to preferentially selecting 
galaxies that look more or less like the PSF.  We attempt to minimize this problem by using the ``round'' measures of signal-to-noise, size and surface brightness.

\subsubsection{Methodology}
The galaxies in each half-sample must be reweighted so that the total $n(z)$, computed from summing the individual $p(z)$ for each galaxy according to its 
weight, matches between the two half-samples. Matching the redshift distributions of the two halves removes any cosmological dependence in each null test.
For the data, the extra weights are computed using Ridge Regression (or Tikhonov regularization) \citep{scikit-learn}. We use the Ridge Regression algorithm to solve 
for an additional weight for each galaxy, which when used with the shear measurement weights described in Section~\ref{sec:data} to compute the $n(z)$, 
produces a matching redshift distribution between the two half-samples. The Ridge Regression algorithm solves the linear least-squares problem with an additional 
regularization parameter $\alpha$, minimizing
 \begin{equation}
||\mathbf{R}\mathbf{v}-\mathbf{t}|| + ||\alpha(\mathbf{v-I})||
 \end{equation}
 where $||...||$ denotes the least-squares norm, $\mathbf{R}$ is the matrix of galaxy $p(z)$'s each weighted by the lensing weights given in Section~\ref{sec:data},
 \begin{equation}
 \mathbf{R}=\left[\begin{array}{lllll}
w_{1}p_{11} & w_{2}p_{12} & w_{3}p_{13} & ... & w_{n}p_{1n}\\
w_{1}p_{21} & w_{2}p_{22} & w_{3}p_{23} & ... & w_{n}p_{2n}\\
w_{1}p_{31} & w_{2}p_{32} & w_{3}p_{33} & ... & w_{n}p_{3n}\\
... & ... & ... & ... & ...\\
w_{1}p_{m1} & w_{2}p_{m2} & w_{3}p_{m3} & ... & w_{n}p_{mn}
\end{array}\right]
\end{equation}
for $n$ galaxies and $m$ photo-$z$ bins with lensing weights $w_{i}$ and galaxy $p(z)$'s $p_{ji}$, $\mathbf{t}$ is the target photo-$z$ distribution, $\mathbf{v}$ is the vector of 
new weights for which we are solving and $\mathbf{I}$ is the identity vector. The parameter $\alpha$ governs the flexibility of the weight selection -- the smaller the 
value, the better matched the reweighed $n(z)$ are --- and is adjusted to prevent a significant contribution of negative or large weight values, which may impact the 
validity of the null tests.  We find that $\alpha=5\times10^{-11}$ produces an optimal match between the two half-samples while keeping the weights $\mathbf{v}$ 
sufficiently regular for our photo-$z$s and lensing weights. This value may not generalize to other lensing weights or photo-$z$s. We match the redshift distribution 
of each half-sample to that of the full sample (i.e. $\mathbf{t}$ is the redshift distribution of the full sample). This procedure is more stable than matching one half to 
another since smaller weights are needed for each half. The application of the Ridge Regression algorithm then produces a new weight $\mathbf{v}$, which is 
combined multiplicatively with the lensing weight in the calculation of the correlation functions. The resulting reweighting for galaxies split into bins of low and high 
galaxy detection signal-to-noise for \ngm\ is shown in Fig. \ref{fig:nullexp}. The left panel shows the $n(z)$ for each half before (dashed) and after (solid) reweighting 
compared to the full sample. The corresponding weight histograms are shown in the right panel. 

We use the 126 DES SV-shaped mock catalogs described above to compute the variance and significance of the differences 
between the shear correlation functions in each half-sample. In the mock catalogs, we select a subset of galaxies in narrow redshift 
slices to match the $n(z)$ distribution for the full galaxy catalog. 
Random shape noise is generated from the shear catalog and applied to the mock catalogs, and the property with which we split the galaxy 
sample in half is then mapped onto the galaxies in each 
mock via a nearest neighbour algorithm in angular position, and redshift. This preserves the same spatial patterns as exist in the data, 
but the shears have been randomised so that there is no 
correlation with this property. We then apply the same procedure to each mock as applied to the data to directly compute the error 
bars on the difference via Monte Carlo, with the exception of 
using the true mock point redshift values to reweight the $n(z)$ histograms of each half instead of a $p(z)$ estimate for each galaxy. We expect this difference will only underestimate the 
variance. Any statistically significant deviations then indicate that the there may be a residual systematic error in the shear catalogs related to the quantity split upon, which has affected the 
measured two-point correlation function. 

\begin{figure*}
\begin{center}
\includegraphics[width=.9\textwidth]{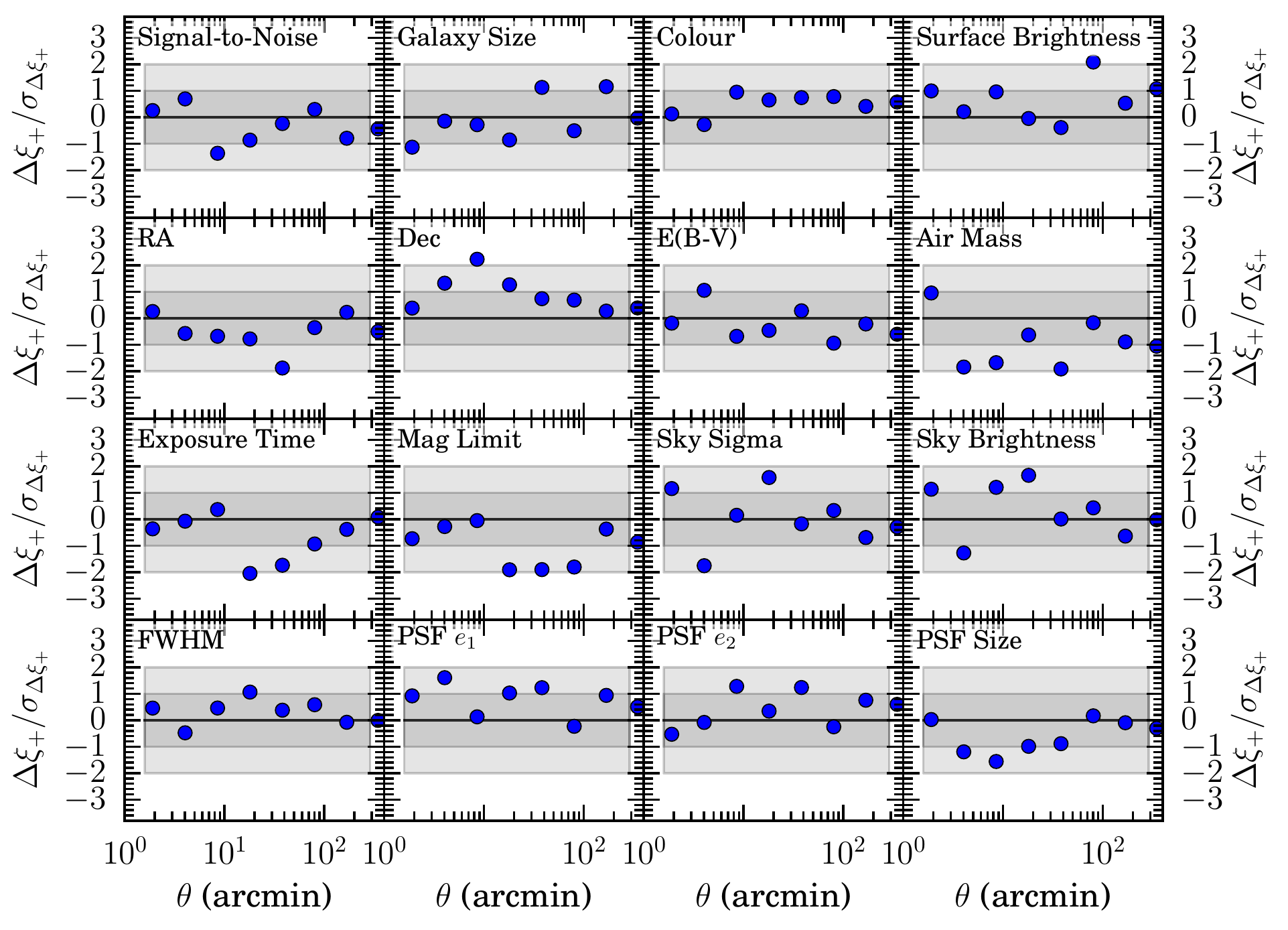}
\end{center}
\caption[]{Null tests for the \ims\ two point correlation function based on a variety of catalog and survey properties as described in Table 1. See Fig. \ref{fig:ngnulls} for details. \label{fig:i3nulls}}
\end{figure*}

\begin{table*}
\centering
\begin{tabular}{lccc}
\hline\hline
Property & $\chi^2$ $[\mathrm{d.o.f.}=8]$ & $\Delta \xi_{+}/\sigma(\xi_{+})$ & Description  \\
& \ngm\ (\ims) & \ngm\ (\ims)\\
\hline
Signal-to-Noise        & 4.9  ( 5.2 ) &  0.05  ( 0.49 ) & Signal-to-noise of galaxy detection \\
Galaxy Size            & 5.3  ( 10.7 ) &  -0.3  ( 0.15 ) & Galaxy size (deconvolved with PSF)  \\
Galaxy Colour          & 7.3  ( 2.2 ) &  -0.31  ( -0.32 ) & $g-z$ colour \\
Surface Brightness     &7.8  ( 8.7 ) &  0.33  ( -0.32 )& Galaxy surface brightness\\
RA                     & 7.0  ( 8.8 ) &  0.24  ( 0.28 ) & Galaxy right ascension \\
Dec                    & 4.0  ( 6.2 ) &  -0.24  ( -0.57 )& Galaxy declination\\
E(B-V)                 & 5.1  ( 6.2 ) &  0.23  ( 0.06 ) & Mean extinction \\
Air Mass               & 20.7  ( 13.8 ) &  0.31  ( 0.46 ) & Mean $r$-band air mass \\
Exposure Time          & 4.7  ( 6.8 ) &  0.18  ( 0.3 ) & Mean total $r$-band exposure time \\
Mag. Limit             & 4.4  ( 7.4 ) &  0.18  ( 0.45 ) & Mean $r$-band limiting magnitude \\
Sky Sigma              & 1.7  ( 13.0 ) &  -0.02  ( -0.08 ) & Mean $r$-band RMS sky brightness \\
Sky Brightness         & 5.0  ( 14.3 ) &  -0.05  ( -0.27 )& Mean $r$-band sky brightness \\
FWHM                   & 6.4  ( 3.3 ) &  -0.23  ( -0.13 ) & Mean $r$-band PSF FWHM\\
PSF $e_1$              & 16.8  ( 13.5 ) &  0.12  ( -0.37 ) & Galaxy PSF $e_1$\\
PSF $e_2$              & 17.1  ( 7.5 ) &  -0.58  ( -0.22 ) & Galaxy PSF $e_2$\\
PSF Size               & 2.6  ( 5.6 ) &  -0.1  ( 0.42 ) & Galaxy PSF size \\
\hline\hline
\end{tabular}
\caption{Summary of null tests for \ngm\ and \ims. Results are given as \ngm\ (\ims). The $\chi^2$ values are given for the differences between the two-point correlation function calculated from galaxies that fall within one of two bins in each catalog or survey property. Also shown is the magnitude of the difference relative to the 1$\sigma$ error of the measurement of $\xi_+$ on the full sample. \label{table:nulltestsres}}
\end{table*}

\subsubsection{Results}
The split null tests on $\xi_+$ are presented in Figure~\ref{fig:ngnulls} for \ngm\ and Figure~\ref{fig:i3nulls} for \ims. This is repeated in 
Appendix~\ref{app:nullxim} for $\xi_-$. For each quantity (panel), the difference in $\xi_+$ is shown at each value of $\theta$ relative to 
the $1 \sigma$ error in the difference from the mock catalogs. Grey bands corresponding to $1\sigma$ and $2\sigma$ errors are shown for 
comparison. The corresponding statistical significance of the null tests for \ims\ and \ngm\ are given in Table~\ref{table:nulltestsres}. We 
find that for both \ngm\ and \ims\ the null tests pass with deviations smaller than 2$\sigma$ ($\chi^{2}/{\rm d.o.f.}=17.8/8$) for all tests 
except for \ngm\ airmass. Note that because \ngm\ has a higher source density, it is generally more sensitive to residual systematic errors in 
these tests. While this detection is still weak, it warrants evaluating whether this difference in the galaxy population halves 
will have a significant bias on the correlation function. To test this, we also show in Table~\ref{table:nulltestsres} the difference 
$\Delta \xi_{+}=\xi_{+({\rm upper})}-\xi_{+({\rm lower})}$ relative to the 1$\sigma$ error on the full sample measurement. For \ngm\ airmass, this difference 
is approximately one-third of the statistical error on the measurements and consistent with the level of bias in several other quantities. Of slightly 
lesser significance are splits in the magnitude of \ngm\ PSF $e_1$ and $e_2$, for which PSF $e_2$ has the largest difference in $\xi_+$ between 
upper and lower halves --- though still small compared to the statistical error. 

There is some subtlety in interpreting the significance of these null tests. First, due to physical effects not accounted for in the simulations, some tests could 
yield non-zero results but not indicate systematic errors in the data analysis itself. For example, if the level of intrinsic alignments differs between galaxies 
split by colour, then these null tests could fail and yet the shear measurements themselves could be free of systematic errors. Second, these tests could 
also, in principle, flag differences between the shear calibrations of galaxies of different types, which although interesting, may not ultimately impact cosmological 
constraints from the full sample, which could be unbiased on average. Third, as stated above, it is not clear from these tests alone if any deviations are due 
to the shear measurements or the photometric redshifts. Finally, note that the $\chi^{2}$ values from these tests are not independent, due to correlations in the underlying 
quantities used to construct the tests (e.g., the survey depth is correlated with the seeing). We have performed a large number of null tests, so to the extent that the 
$\chi^{2}$ values between many of the tests should be independent, we do expect some apparent deviations purely from statistical fluctuations. However, we have not attempted 
to combine the tests in order to quote an overall significance.

\section{Conclusions}\label{sec:conclusions}
In this work, we present cosmic shear two-point measurements from Dark Energy Survey Science Verification data. 
We find an overall detection significance of $9.7\sigma$ for our higher source density catalog, \ngm. We additionally 
present multiple advances in band-power estimation, covariance estimation, simulations versus theory, and null tests for 
shear two-point correlations. Through this work we demonstrate that our measurements are robust and free of statistically 
significant systematic errors. 

We demonstrate that the covariance matrices derived from the DES SV mock simulations presented in this work are  
consistent with the halo model, including the halo sample variance terms. We also compare the variance in the mock catalogs to 
the variance in the DES SV data by comparing jackknife covariances computed in the data and mock catalogs. The structure of the 
covariance matrices is very similar and we detect no statistically significant sources of additional variance in the data.

We find that the B-mode signals in the data are consistent with zero and that the two shear estimation codes agree well. We additionally present a set of 
simultaneous null tests of the photo-$z$s and shear measurements, performed by splitting the shear sample in half according to some 
parameter and comparing the shear correlation functions of the halves. We find that these tests pass with no statistically significant indications 
of biases. We expect null tests similar to those developed here to have increased utility in future cosmic shear analyses, where the statistical 
power is larger and the requirements for controlling systematic errors and shear selection effects are more stringent. The DES itself will have nearly 
$\approx36\times$ more data and will measure cosmic shear at significantly higher signal-to-noise, so that these tests will be very useful. 

Future cosmic shear two-point function measurements in the Dark Energy Survey face a variety of challenges. First, while we have a sufficient 
number of simulations for the SV data, simulating the increased area of the full DES will present a significant computational challenge. This 
challenge will need to be met by a combination of large simulation campaigns, information compression schemes applied directly to 
the data, and combinations of theoretical models for the covariances with simulations in order to reduce the noise in the covariance matrix 
elements.  Second, in order to use simulations to evaluate the statistical significance of null tests on future DES data, like those presented in 
this work, we will need to increase the fidelity of the treatment of both the galaxies and the shear signals.  Third, we must better address the formal 
aspects of the construction of the two-point function statistic estimators in order to make higher precision measurements. 
Finally, while this work has focused exclusively on broad-bin tomography of the two-point function measurements of cosmic shear, future exploration 
of higher order correlation functions and finer tomographic binning will be needed to extract the full amount of cosmological information from cosmic 
shear data.  Fortunately, none of these issues are fundamentally intractable and we expect that the new techniques presented in this work will be 
of great assistance in making future cosmic shear measurements with DES data.

\section*{Acknowledgements}
We are grateful for the extraordinary contributions of our CTIO colleagues and the DECam 
Construction, Commissioning and Science Verification teams in achieving the excellent 
instrument and telescope conditions that have made this work possible. The success of this 
project also relies critically on the expertise and dedication of the DES Data Management group.

MRB is grateful for the support of the University of Chicago Research Computing Center, and especially 
Doug Rudd, for the time used to carry out the N-body simulations carried out in this work. MRB would also like to thank 
Stewart Marshall for his ongoing assistance in using SLAC computing resources. This work used the Extreme Science and 
Engineering Discovery Environment (XSEDE), which is supported by National Science Foundation grant number ACI-1053575.
JAZ, MAT, SLB acknowledge support from the European Research Council in the form of a Starting Grant with number 240672.
MRB and RHW received partial support from NSF-AST-1211838 and from a DOE SciDAC grant.
OF and DG were supported by SFB-Transregio 33 'The Dark Universe' by the Deutsche Forschungsgemeinaft (DFG) 
and the DFG cluster of excellence 'Origin and Structure of the Universe'
AA, AR, AN are supported in part by grants 20021\_14944 and 20021\_1439606 from the Swiss National Foundation. 
Jarvis has been supported on this project by NSF grants AST-0812790 and AST-1138729. 
Jarvis, Bernstein, and Jain are partially supported by DoE grant DE-SC0007901. 
ML is partially supported by FAPESP and CNPq.
This work made extensive use of the NASA Astrophysics Data System and \texttt{arXiv.org} preprint server.

Funding for the DES Projects has been provided by the U.S. Department of Energy, the U.S. National Science 
Foundation, the Ministry of Science and Education of Spain, the Science and Technology Facilities Council of 
the United Kingdom, the Higher Education Funding Council for England, the National Center for Supercomputing 
Applications at the University of Illinois at Urbana-Champaign, the Kavli Institute of Cosmological Physics 
at the University of Chicago, the Center for Cosmology and Astro-Particle Physics at the Ohio State University,
the Mitchell Institute for Fundamental Physics and Astronomy at Texas A\&M University, Financiadora de 
Estudos e Projetos, Funda{\c c}{\~a}o Carlos Chagas Filho de Amparo {\`a} Pesquisa do Estado do Rio de 
Janeiro, Conselho Nacional de Desenvolvimento Cient{\'i}fico e Tecnol{\'o}gico and the Minist{\'e}rio da 
Ci{\^e}ncia e Tecnologia, the Deutsche Forschungsgemeinschaft and the Collaborating Institutions in the 
Dark Energy Survey. 

The DES data management system is supported by the National Science Foundation under Grant Number 
AST-1138766. The DES participants from Spanish institutions are partially supported by MINECO under 
grants AYA2012-39559, ESP2013-48274, FPA2013-47986, and Centro de Excelencia Severo Ochoa 
SEV-2012-0234, some of which include ERDF funds from the European Union.

The Collaborating Institutions are Argonne National Laboratory, the University of California at Santa Cruz, 
the University of Cambridge, Centro de Investigaciones Energeticas, Medioambientales y Tecnologicas-Madrid, 
the University of Chicago, University College London, the DES-Brazil Consortium, the Eidgen{\"o}ssische 
Technische Hochschule (ETH) Z{\"u}rich, Fermi National Accelerator Laboratory,
the University of Edinburgh, 
the University of Illinois at Urbana-Champaign, the Institut de Ci\`encies de l'Espai (IEEC/CSIC), 
the Institut de F\'{\i}sica d'Altes Energies, Lawrence Berkeley National Laboratory, the Ludwig-Maximilians 
Universit{\"a}t and the associated Excellence Cluster Universe, the University of Michigan, the National Optical 
Astronomy Observatory, the University of Nottingham, The Ohio State University, the University of Pennsylvania, 
the University of Portsmouth, SLAC National Accelerator Laboratory, Stanford University, the University of 
Sussex, and Texas A\&M University.

This paper is Fermilab publication \texttt{FERMILAB-PUB-15-303-AE} and DES publication \texttt{DES-2015-0061}.
This paper has gone through internal review by the DES collaboration.

\appendix

\section{Alternative E- and B-mode Statistics}\label{app:altebstats}
In this appendix we consider alternative statistics of the shear field, verifying that our conclusions above, especially that the B-modes are consistent with zero, do not depend on the choice of 
statistic. These alternative statistics include the band-powers of \citet{becker2014} and power spectra band-powers estimated with 
\texttt{PolSpice}\footnote{\url{http://www2.iap.fr/users/hivon/software/PolSpice/}} \citep{szapudi2001,chon2004}.

\subsection{Band-powers}
The band-powers of \citet{becker2014} use the methods of \citet{becker2013} to estimate Fourier-space band-powers directly from linear combinations of the real-space two-point functions. The final band-power estimates can be computed from the underlying E-mode power spectrum as
\be
E = \int \frac{d\ln\ell\,\ell^{2}}{2\pi}C_{EE}(\ell)W_{+}(\ell)
\ee
where $W_{+}(\ell)$ is the band-power window function computed from the coefficients $\{f_{+i},f_{-i}\}$ in Eqs. \ref{eq:bmodee} \& \ref{eq:bmodeb}. See \citet{becker2014} for more details. 
The optimal computation of the band-powers requires computing the effective radial bin window functions of the shear correlation function points. 
Instead in this work we just use the geometric approximation to the bin window functions to compute the amplitudes $\{f_{+i},f_{-i}\}$. 
This procedure means that the band-powers do not separate E- and B-modes as well as they could in principle. However, 
when comparing to a fiducial cosmological model below, we do compute the band-power window function using estimates of 
radial bin window functions from the data. These window functions are computed via interpolating the weighted counts in each radial bin of 
the estimated shear two-point function. We have compared the results of this procedure for computing the window functions 
to estimates of the window functions from counts in finer bins. We find unsurprisingly that the bin window functions are quite smooth 
and thus the interpolation is accurate enough for our purpose. 

\begin{figure}
\includegraphics[width=\columnwidth]{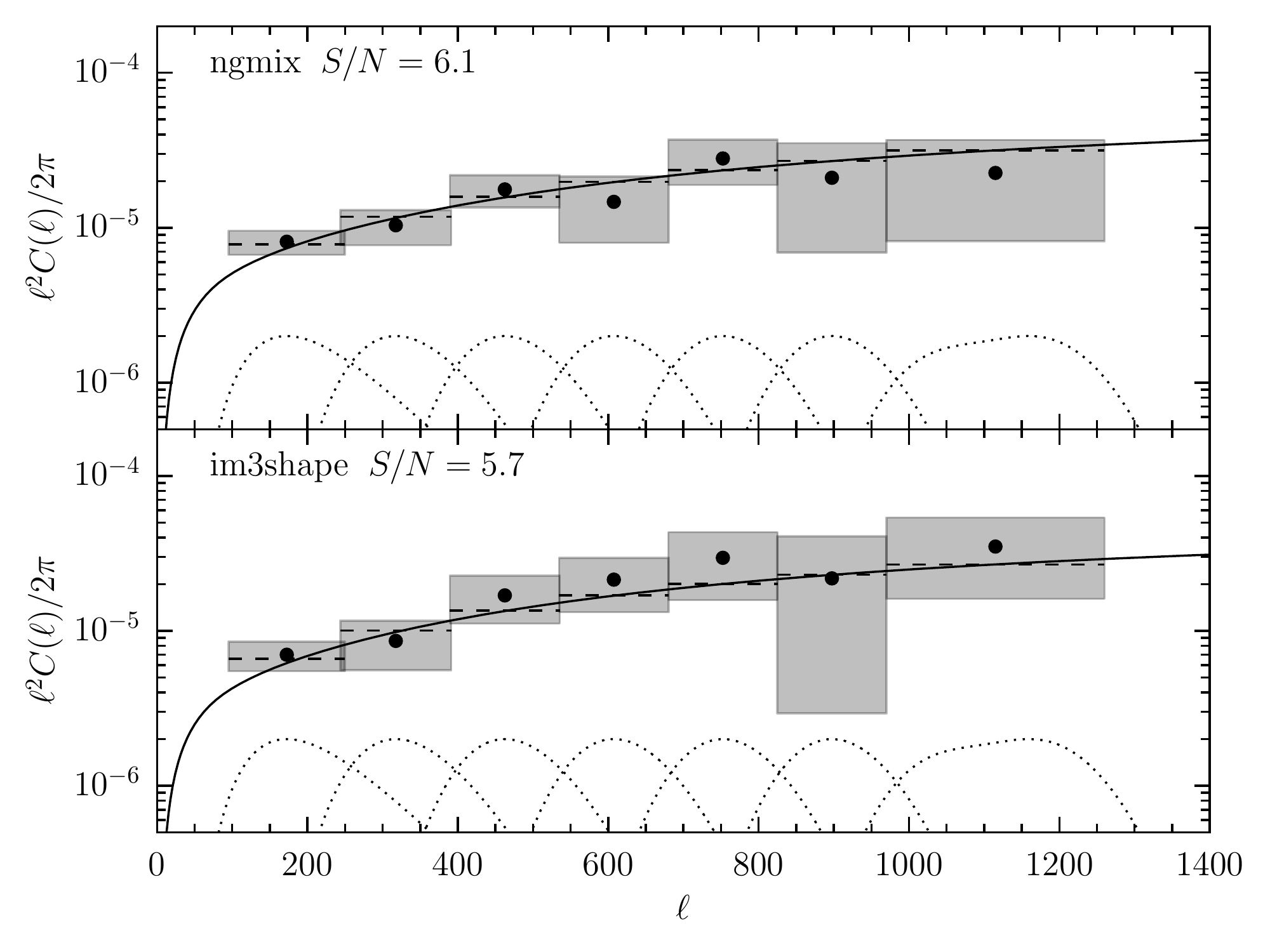}
\caption[]{Band-powers in DES SV data for \ngm\ (top) and \ims\ (bottom).
The error bars indicated by the grey bands are calculated from the simulation realizations using the shape noise appropriate 
for each catalog. The dotted lines show the band-power window functions $W_{+}(\ell)$ scaled so that their peak values are $2\times10^{-6}$. 
The solid line is the prediction for the shear power spectrum for the flat, $\Lambda$CDM model given above. The dashed line 
shows the integral of the band-power window functions over the shear power spectrum. 
\label{fig:bandpows}}
\end{figure}

\begin{figure}
\includegraphics[width=\columnwidth]{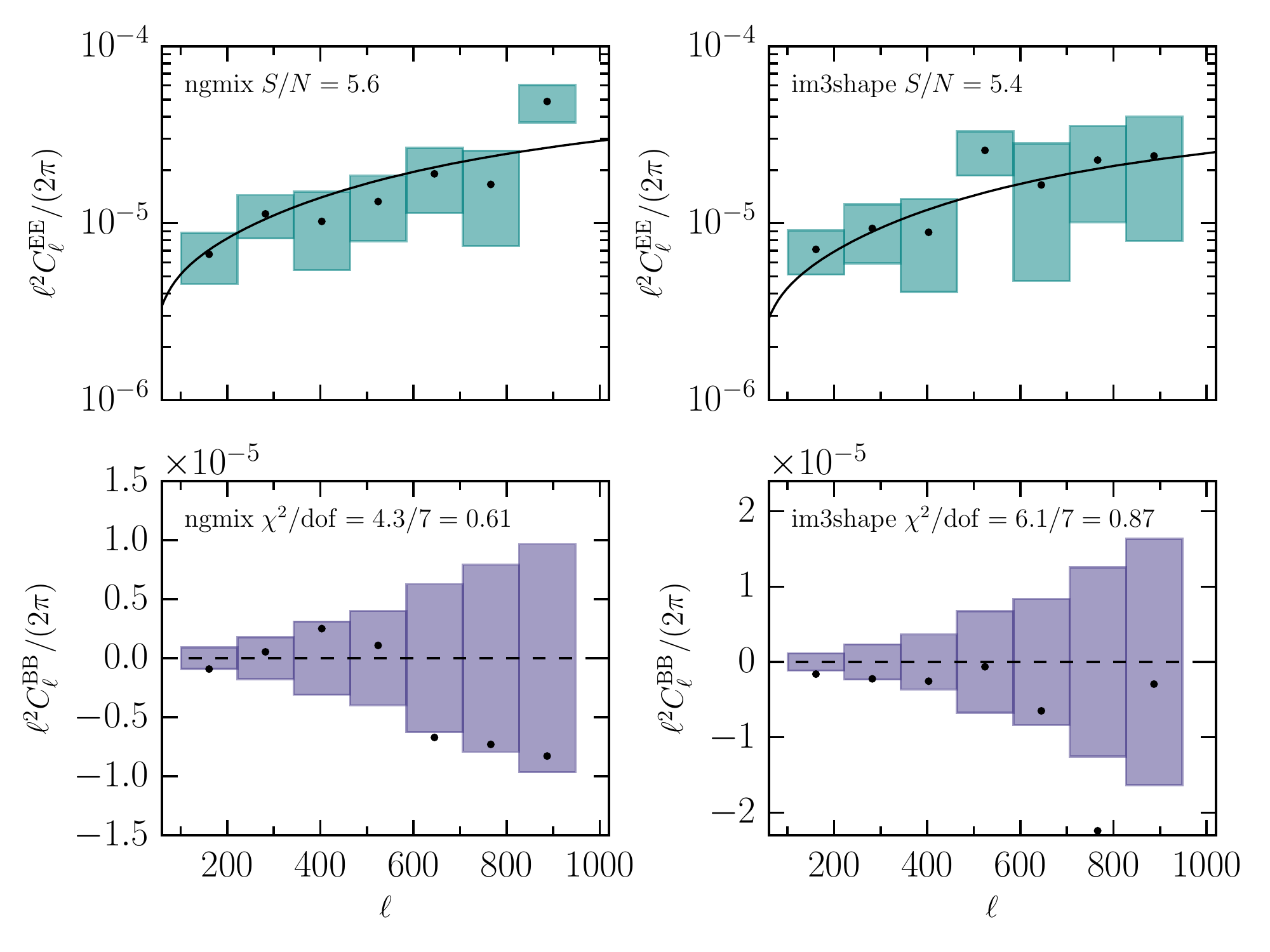}
\caption[]{Spherical harmonic shear power spectrum estimated using \texttt{PolSpice}. The left and right panels correspond to the \ngm\ and \ims\ catalogs, respectively. The top and bottom 
panels show the E- and B-modes, respectively. The measurement uncertainties are estimated using the mock catalogs. The black solid lines show the predictions for the flat, $\Lambda$CDM 
model given above. Note that the theoretical prediction has been convolved with the \texttt{PolSpice} kernels, which relate the true to measured power spectra. The $S/N$ values for the 
E-modes are computed as outlined in Section~\ref{sec:2ptest} and the $\chi^{2}$ values for the B-modes indicate consistency with zero. The reported values take into account correlations 
between the band-powers. \label{fig:pcl}}
\end{figure}

\subsection{Spherical Harmonic Power Spectrum} 
The cosmic shear power spectrum can also be estimated in spherical harmonic space, which has the advantage of being faster 
and less memory intensive than working in real-space. In view of upcoming wide field galaxy lensing surveys, e.g. the full five year DES dataset, 
we therefore investigate the applicability of standard spherical harmonic space methods to weak lensing. For this purpose, we use the 
\texttt{PolSpice} \citep{szapudi2001,chon2004} code together with the \texttt{HEALPix} \citep{gorski2005} package, which has been applied to, amongst 
other things, CMB polarization data \citep[e.g.,][]{chiang2010}. \texttt{PolSpice} is based on the fast correlation function approach 
described in \cite{szapudi2001} and \cite{chon2004}. The method is designed to exploit the advantages of both real and spherical harmonic space: 
to limit computation time and resources, the data are analyzed in spherical harmonic space. In order to facilitate demasking, the power spectrum is 
transformed to real-space in an intermediate step. In real-space, the survey mask can simply be corrected for, since the masked correlation function 
is the product of the unmasked correlation function and the correlation function of the mask. More precisely, the algorithm first calculates pseudo-$C_{\ell}$'s 
from pixelized and masked galaxy ellipticity maps which are then transformed to the correlation function. The real-space correlation function is then divided 
by the correlation function of the mask to correct for finite survey effects and inverted to obtain the full-sky power spectrum, removing E- to B-mode leakage 
in the mean. Incomplete sky coverage implies that the inversion can only be performed on angular scales for which the correlation function can be estimated 
thus introducing Fourier ringing in the inversion process, which can be reduced by apodizing the correlation function. Both the apodization and finite integration 
range introduce kernels which relate the power spectra measured by \texttt{PolSpice} to the underlying true power spectra. These kernels can be computed 
for a given apodization scheme and integration range and can therefore be corrected for when comparing measurement to theory (for details see \citet{chon2004}). 

For our analysis, we pixelize the galaxy ellipticities onto a \texttt{HEALPix} pixelization of the sphere with a resolution of \verb+Nside=1024+, where each pixel 
covers a solid angle of $\approx$11.8 arcmin$^{2}$. In order to obtain a robust estimate of the shear field, we need to correct for multiplicative bias in the 
measured ellipticities. Since the correction factors described in Sections~\ref{sec:ngmix} and \ref{sec:im3shape} are noisy estimates of the true corrections, 
we determine the mean sensitivity or multiplicative bias correction for our galaxy sample and apply this mean correction to the pixelized maps. As the power 
spectrum is estimated from maps constructed from the discrete values of the galaxy ellipticities, we apply a conservative masking scheme to maximize galaxy 
number density. We therefore adopt the DES SV LSS mask used for galaxy clustering measurements \citep{crocce2016}. This mask is identical 
to the DES SV mask used for weak lensing except that it restricts analyses to the largest contiguous region overlapping the SPT-E field by selecting the area with 
$60 < \mr{ra}\ [\mr{deg}] < 95$ and $-60 < \mr{dec}\ [\mr{deg}] < -40$. It further considers only regions with survey limiting magnitude in the i-band $> 22.5$ 
(i.e. all regions considered to provide at least 10$\sigma$ measurements for objects at i-band $= 22.5$; \citep{crocce2016}). For the power spectrum 
measurement, we limit all integrations to scales smaller than $\theta_\mr{max} = 15$ degrees and we apodize the correlation function with a Gaussian window of 
$\theta_\mr{FWHM} =10$ degrees. Finally, we correct the measured power spectra for the \texttt{HEALPix} pixel window function and compress them into 7 band-powers with \texttt{PolSpice} band-power kernels.

The noise power spectrum needs to be computed from simulations. In order to produce noise only maps from the data, we remove correlations in the ellipticity maps 
by rotating each galaxy ellipticity by a random angle. We then estimate the noise power spectrum as the mean of the power spectra of 100 such random realizations. 
This procedure yields shape noise estimates consistent with $C_{\ell, \mr{SN}} = \frac{\sigma^{2}_{\epsilon,\mr{pix}}}{n_{\mr{pix}}}$ where $\sigma^{2}_{\epsilon,\mr{pix}}$ 
is the variance of either component of the mean ellipticity per pixel and $n_{\mr{pix}}$ is the angular number density of \texttt{HEALPix} pixels. Comparing the measured shape 
noise to the galaxy-based Gaussian shape noise estimate $C_{\ell, \mr{SN}} = \frac{\sigma^{2}_{\epsilon, \mathrm{gal}}}{n_{\mathrm{gal}}}$, where $\sigma^{2}_{\epsilon, \mathrm{gal}}$ 
is the variance of either component of the galaxy ellipticities and $n_{\mathrm{gal}}$ denotes the galaxy number density, we find that the latter underestimates the 
measured shape noise. This suggests that the galaxy ellipticity distribution is non-Gaussian and the Gaussian approximation can therefore only be applied after averaging 
the galaxy ellipticities over pixels. We test the pipeline using Gaussian field realizations and the mock catalogs.

\subsection{Results}
Figure~\ref{fig:bandpows} shows the non-tomographic band-powers using the methods of \citet{becker2014}, their window 
functions as the dotted lines, and their error bars computed with the mock catalogs as the grey bands. We find a detection significance 
6.1$\sigma$ and 5.7$\sigma$ for \ngm\ and \ims, respectively. These detection significances are similar to the real-space two-point functions. Finally, 
the solid line shows the expected shear power spectrum amplitude assuming the flat, $\Lambda$CDM model given above. The dashed line 
shows for each band-power the integral of the band-power window function over the shear power spectrum. 

Figure~\ref{fig:pcl} shows the results for the \texttt{PolSpice} statistics. We find a detection of cosmic shear of 5.6$\sigma$ and 5.4$\sigma$ 
for \ngm\ and \ims\ respectively for the \texttt{PolSpice} statistics. Note that the \texttt{PolSpice} statistics do not use as many high-$\ell$ modes 
as the real-space band-powers or the real-space correlation functions, so that one expects a lower detection significance. 
We also find that the B-modes are statistically consistent with zero for the \texttt{PolSpice} statistics.

Finally, note that these two estimators process the data in different ways (e.g., averaging the data in pixels versus computing real-space correlation functions), 
have different sensitivities to shot noise, and have different Fourier-space window functions. We thus do not expect them to give precisely the same results in 
Fourier-space for the shear power spectrum. However, we do expect that when treated self-consistently they should give statistically 
consistent results for cosmological parameters, as demonstrated in the accompanying cosmological analysis of this data \citep{deswlcosmo}.
~\\

\section{Validation of the Mock Catalogs}\label{app:simtests}

\begin{figure*}
\begin{center}
\includegraphics[width=\textwidth]{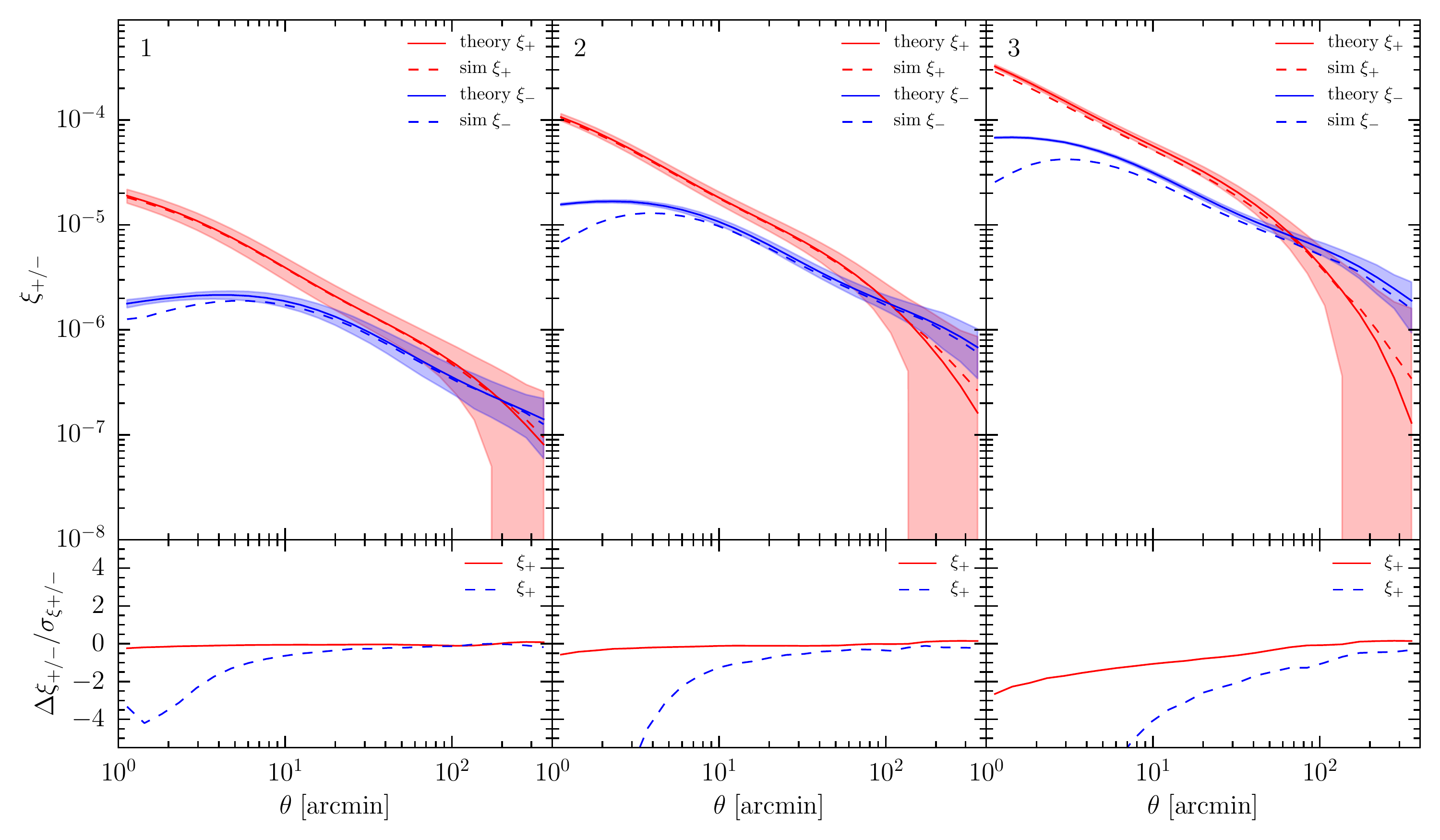}
\end{center}
\caption[]{The shear correlation functions in the mock catalogs compared to the expected values from \citet{takahashi2012} for all three tomographic bins (labeled in the top 
left corner from left to right). In the top panel, the solid lines show the theoretical expectation, the bands show the $1\sigma$ sample variance estimate and the dashed line 
shows the mean from the mock catalogs. $\xi_{+}$ is in red and $\xi_{-}$ is in blue. In the bottom panels, we show the fractional deviation of the mean signal in the mock 
catalogs from the expected values from \citet{takahashi2012} in units of the sample variance. $\xi_{+}$ data below $\approx2-4$ arcminutes and $\xi-$ data below 
$\approx25-55$ arcminutes is not used for the final cosmological analysis in \citet{deswlcosmo} due to the expected baryonic effects in the matter power spectrum. 
\label{fig:mocktest}}
\end{figure*}

In this section we present a validation test for the mock catalogs. We compare the shear correlation functions measured in the mock catalogs in tomographic bins 
with the theoretical expectation from the \citet{takahashi2012} fitting function for the matter power spectrum. The result of this test is shown in Figure~\ref{fig:mocktest}. 
We find that at high redshift the small-scale shear correlation functions are suppressed relative to the theoretical expectation. Note however that this numerical effect 
is below the scales where the two-point functions are being used for cosmological parameter estimation ($\approx2-4$ arcminutes for $\xi_{+}$ and $\approx25-55$ 
arcminutes for $\xi_{-}$; see Table 2 of \citet{deswlcosmo}). Additionally, we only estimate the covariance of the two-point functions from the mock catalogs, not the mean 
signal. Within the noise of our mock covariance matrix, the overall parameter uncertainties are consistent when using the halo model versus the simulation covariance 
(see Sec. \ref{sec:covmatrix} for a quantitative comparison). This fact may indicate that the covariance is less sensitive to these numerical effects than the mean signal. 
Future work may require higher-resolution shear fields for covariance estimation. 

\section{Detailed Covariance Matrix Validation}\label{app:hmod}
In this section, we present further details of the validation of the covariance matrices, including our tomographic halo 
model computations and the comparison to the simulations. The halo model covariance was computed with the 
\verb+CosmoLike+ covariance module (see \citet{eifler2014} and \citet{krause2015} for details).

\begin{figure*}
\begin{center}
\includegraphics[width=\textwidth]{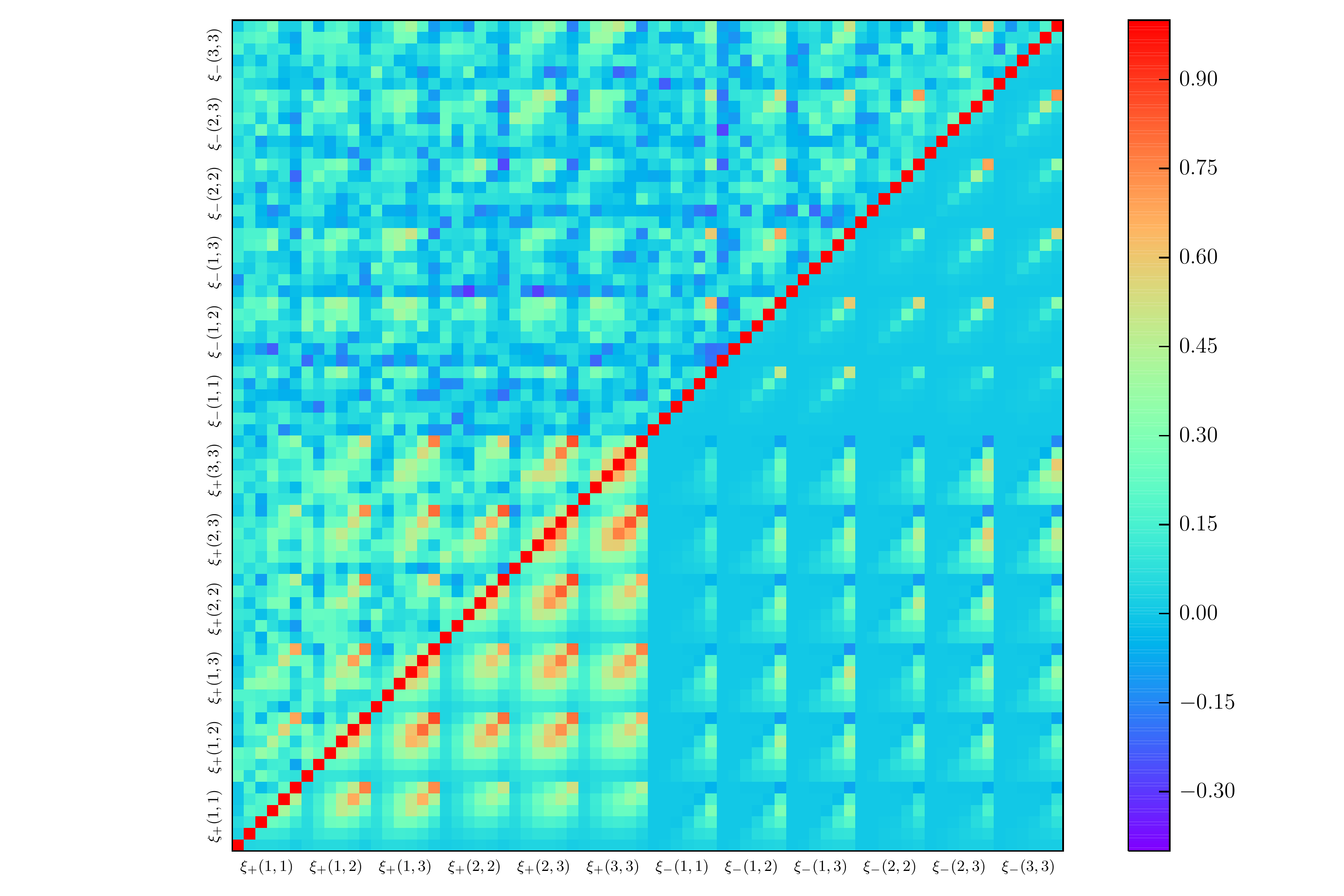}
\end{center}
\caption[]{Comparison of the shear correlation function correlation matrix estimated from mock catalogs and calculated
 from the halo model. Figure~\ref{fig:covhmcompsimp} shows a subset, those for tomographic bin combinations 
 (1,1), (1,3) and (3,3), of the covariance matrix elements shown in this figure. The correlation matrix from mock catalogs is on the upper-left 
and that from the halo model is on the lower-right.\label{fig:covhmcomp}}
\end{figure*}

In the halo model, the covariance of tomographic shear power spectra $C_\kappa^{ij}(l)$ is given by \citep{ch01cov,huj04,sht09}
\begin{widetext}
\bea
\label{eq:covC}
\lefteqn{\mr{Cov}\left( C_\kappa^{ij} (l_1), C_\kappa^{kl} (l_2) \right)=} &&\nonumber \\
\nonumber &&\frac{2 \pi\delta_{ l_1 l_2}}{ \Omega_{\mathrm{s}} l_1 \Delta l_1}  
\left[\left(C_\kappa^{ik}(l_1)+ \delta_{ik} \frac{\sigma_\eps^2}{2n^{i}}\right) \left(C_\kappa^{jl}(l_2)+ \delta_{jl} \frac{\sigma_\eps^2}{2n^{j}}\right) 
+\left(C_\kappa^{il}(l_1)+ \delta_{il} \frac{\sigma_\eps^2}{2n^{i}}\right) \left(C_\kappa^{jk}(l_2)+ \delta_{jk} \frac{\sigma_\eps^2}{2n^{j}}\right) \right]\\
&& +\int_{|\mathbf l|\in l_1}\frac{d^2\mathbf l}{A(l_1)}\int_{|\mathbf l'|\in l_2}\frac{d^2\mathbf l'}{A(l_2)} 
\left[\frac{1}{\Omega_{\mr s}}T_{\kappa,0}^{ijkl}(\mathbf l,-\mathbf l,\mathbf l',-\mathbf l') + T_{\kappa,\rm{HSV}}^{ijkl}(\mathbf l,-\mathbf l,\mathbf l',-\mathbf l') \right]\,,
\eea
\end{widetext}
with $n^i$ the number of source galaxies in tomography bin $i$, $\sigma_\epsilon$ the ellipticity dispersion, 
$A(l_i) = \int_{|\mathbf l|\in l_i}d^2\mathbf l \approx 2 \pi l_i\Delta l_i$ the integration area associated with a 
power spectrum bin centered at $l_i$ and width $\Delta l_i$, and $T_{\kappa,0}^{ijkl}$ and $T_{\kappa,\rm{HSV}}^{ijkl}$ 
the convergence trispectrum of source redshift bins $i,j,k$ and $l$ in the absence of finite volume effects and the halo sample variance contribution 
to the trispectrum \citep{sht09,takada2013}. Our halo model implementation for these terms is described in \citet{ekd14}. 

Note that Equation~\ref{eq:covC} ignores the so called finite-area effect (cf. \citet{sato2011} or \citet{friedrich2016}), linear beat-coupling terms 
\citep[e.g.,][]{takada2009} and linear dilation terms \citet[e.g.,][]{li2014}. For a survey of the size of DES-SV the finite-area effect is expected to be negligible. 
Furthermore, ignoring this effect is at most conservative since it will slightly overestimate the statistical uncertainties. The beat-coupling terms are negligible 
compared to the halo sample variance terms (and even the non-Gaussian terms, see e.g., \citet{takada2009}). Further, the linear dilation terms reduce the 
effect of the beat-coupling terms and are negligible \citep{li2014}. Finally, we have ignored the effects of masking (except for the total area of the survey in the 
halo sample variance terms). We have found with Gaussian simulations that the effects of the details of the mask, besides the overall survey area, are negligible 
when computing cosmological constraints.

The covariance of angular shear correlation functions is then given by
\begin{widetext}
\be
\mr{Cov}\left( \xi_\pm^{ij} (\theta_1), \xi_\pm^{kl} (\theta_2) \right) = \int \frac{d l}{2\pi} l J_{0/4}\left(l \theta_1\right)\int 
\frac{d l'}{2\pi} l' J_{0/4}\left(l' \theta_2\right)\, \mr{Cov}\left( C_\kappa^{ij} (l_1), C_\kappa^{kl} (l_2) \right)\,
\ee
\end{widetext}
where we use the results of \citet{JSE08} to simplify the calculation of the Gaussian part of the covariance.

Figure~\ref{fig:covhmcomp} shows the full tomographic correlation matrix, comparing 
the halo model on the lower-right and the mock catalogs on the upper-left. The overall structure of the covariance matrices is similar in both computations, 
but the mock catalogs exhibit more noise in the off-diagonal components.

\section{Additional Two-Point Null Tests of $\xi_-$}\label{app:nullxim}

\begin{figure*}
\begin{center}
\includegraphics[width=.9\textwidth]{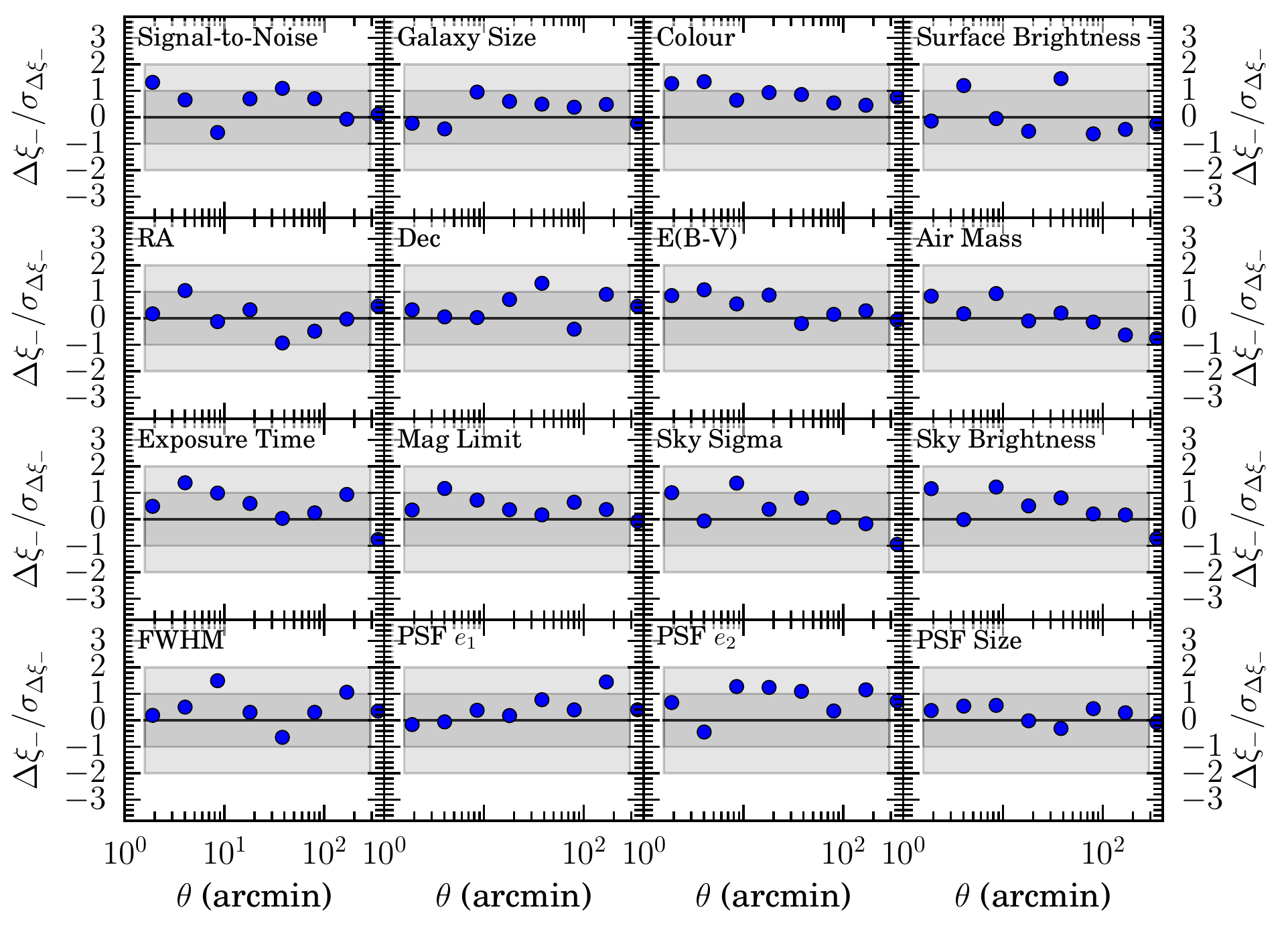}
\end{center}
\caption[]{Null tests for the \ngm\ two point correlation function based on a variety of catalog and survey properties as described in Table 1. See Fig. \ref{fig:ngnulls} for details.  
\label{fig:ngnullsb}}
\end{figure*}

\begin{figure*}
\begin{center}
\includegraphics[width=.9\textwidth]{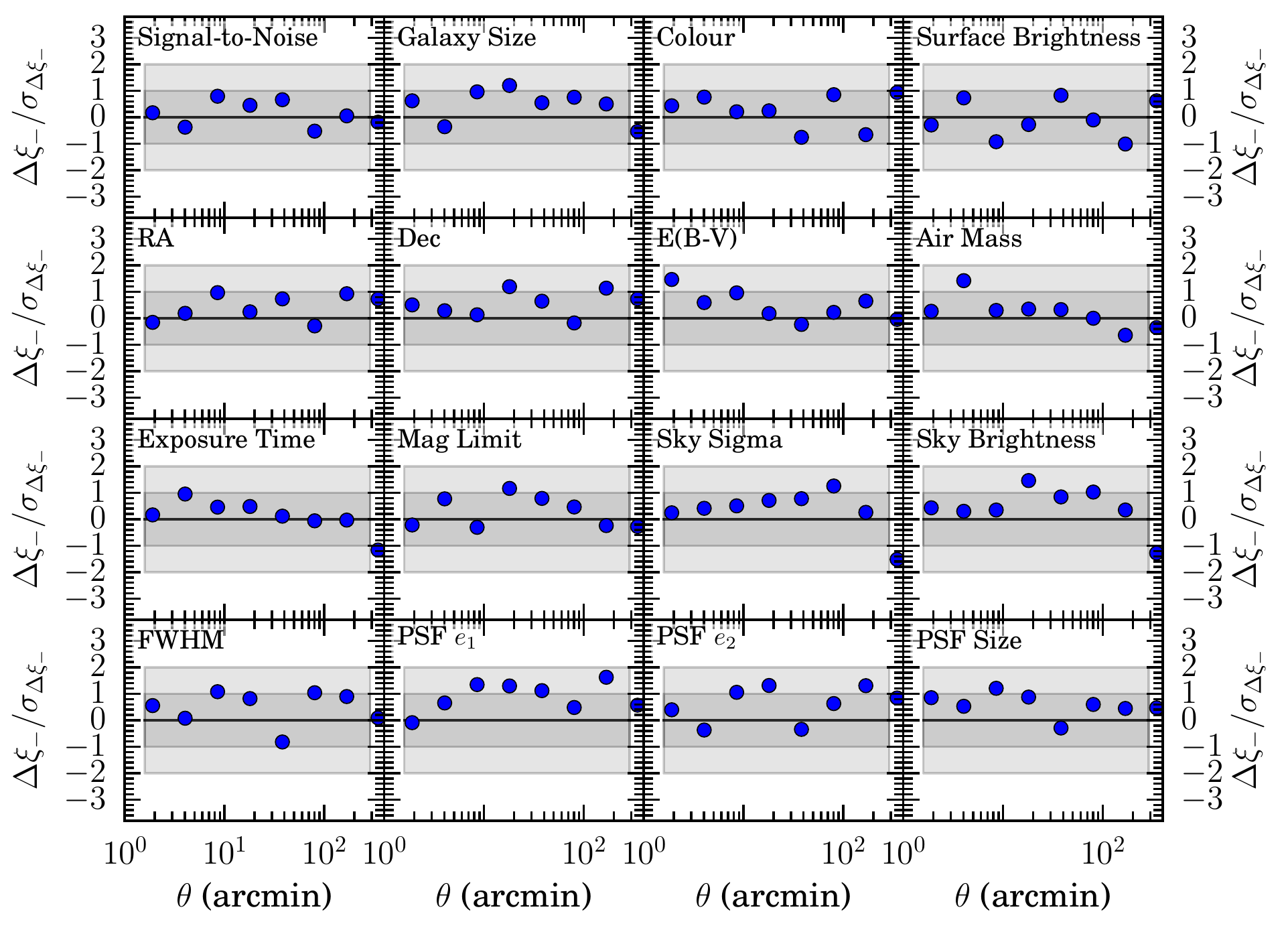}
\end{center}
\caption[]{Null tests for the \ims\ two point correlation function based on a variety of catalog and survey properties as described in Table 1. 
See Fig. \ref{fig:ngnulls} for details. \label{fig:i3nullsb}}
\end{figure*}

We have repeated an identical analysis for $\xi_-$ to that described in Sec. \ref{sec:nctests} for $\xi_+$. We show 
the results of the tests for \ims\ in Fig. \ref{fig:ngnullsb} and for \ngm\ in Fig. \ref{fig:i3nullsb}. Qualitatively, comparing 
to Figs. \ref{fig:ngnulls} \& \ref{fig:i3nulls}, there is an indication that some of the larger deviations in the figures for 
$\xi_+$ may be due to additive systematic errors. For example, there is an offset in the difference of $\xi_+$ based on values 
of airmass at the 2$\sigma$ level that disappears for $\xi_-$. The corresponding $\chi^2$ and difference values are given 
in Table \ref{table:nulltestsres2}. There are no significant indications of systematic errors in these null tests for $\xi_-$, though 
this may simply be due to the poorer constraining power of $\xi_-$.

\begin{table*}
\centering
\begin{tabular}{lccc}
\hline\hline
Property & $\chi^2$ $[\mathrm{d.o.f.}=8]$ & $\Delta \xi_{-}/\sigma(\xi_{-})$ & Description  \\
& \ngm\ (\ims) & \ngm\ (\ims)\\
\hline
Signal-to-Noise        &5.8  ( 1.8 ) &  -0.07  ( 0.03 )& Signal-to-noise of galaxy detection \\
Galaxy Size            & 2.5  ( 5.0 ) &  -0.23  ( -0.35 ) & Galaxy size (deconvolved with PSF)  \\
Galaxy Colour          & 7.1  ( 3.8 ) &  -0.3  ( 0.04 ) & $g-z$ colour\\
Surface Brightness     & 4.4  ( 5.2 ) &  -0.04  ( -0.06 ) & Galaxy surface brightness\\
RA                     & 2.9  ( 3.0 ) &  0.06  ( -0.22 ) & Galaxy right ascension \\
Dec                    & 4.9  ( 3.5 ) &  -0.35  ( -0.37 ) & Galaxy declination\\
E(B-V)                 & 2.8  ( 4.9 ) &  -0.22  ( -0.02 )& Mean extinction \\
Air Mass               & 2.7  ( 3.4 ) &  -0.01  ( -0.08 ) & Mean $r$-band air mass \\
Exposure Time          & 4.5  ( 2.5 ) &  -0.35  ( 0.0 )& Mean total $r$-band exposure time \\
Mag. Limit             & 2.2  ( 3.3 ) &  -0.29  ( -0.43 )& Mean $r$-band limiting magnitude \\
Sky Sigma              & 3.8  ( 5.6 ) &  -0.21  ( -0.3 ) & Mean $r$-band RMS sky brightness \\
Sky Brightness         & 4.0  ( 6.2 ) &  -0.27  ( -0.42 )& Mean $r$-band sky brightness \\
FWHM                   & 4.1  ( 4.5 ) &  -0.2  ( -0.08 ) & Mean $r$-band PSF FWHM\\
PSF $e_1$              & 2.7  ( 7.9 ) &  -0.37  ( -0.55 ) & Galaxy PSF $e_1$\\
PSF $e_2$              & 6.8  ( 5.8 ) &  -0.5  ( -0.33 ) & Galaxy PSF $e_2$\\
PSF Size               & 1.2  ( 3.8 ) &  -0.08  ( -0.1 ) & Galaxy PSF size \\
\hline\hline
\end{tabular}
\caption{Summary of null tests for \ngm\ and \ims. Results for \ngm\ and \ims\ are given as \ngm\ (\ims). The $\chi^2$ values are given for the differences between the two-point correlation 
function calculated from galaxies that fall within one of two bins in each catalog or survey property. Also shown is the magnitude of the difference relative to the 1$\sigma$ error of the 
measurement of $\xi_-$ on the full sample. \label{table:nulltestsres2}}
\end{table*}

\bibliography{refs}

\begin{thebibliography}{78}%
\makeatletter
\providecommand \@ifxundefined [1]{%
 \@ifx{#1\undefined}
}%
\providecommand \@ifnum [1]{%
 \ifnum #1\expandafter \@firstoftwo
 \else \expandafter \@secondoftwo
 \fi
}%
\providecommand \@ifx [1]{%
 \ifx #1\expandafter \@firstoftwo
 \else \expandafter \@secondoftwo
 \fi
}%
\providecommand \natexlab [1]{#1}%
\providecommand \enquote  [1]{``#1''}%
\providecommand \bibnamefont  [1]{#1}%
\providecommand \bibfnamefont [1]{#1}%
\providecommand \citenamefont [1]{#1}%
\providecommand \href@noop [0]{\@secondoftwo}%
\providecommand \href [0]{\begingroup \@sanitize@url \@href}%
\providecommand \@href[1]{\@@startlink{#1}\@@href}%
\providecommand \@@href[1]{\endgroup#1\@@endlink}%
\providecommand \@sanitize@url [0]{\catcode `\\12\catcode `\$12\catcode
  `\&12\catcode `\#12\catcode `\^12\catcode `\_12\catcode `\%12\relax}%
\providecommand \@@startlink[1]{}%
\providecommand \@@endlink[0]{}%
\providecommand \url  [0]{\begingroup\@sanitize@url \@url }%
\providecommand \@url [1]{\endgroup\@href {#1}{\urlprefix }}%
\providecommand \urlprefix  [0]{URL }%
\providecommand \Eprint [0]{\href }%
\providecommand \doibase [0]{http://dx.doi.org/}%
\providecommand \selectlanguage [0]{\@gobble}%
\providecommand \bibinfo  [0]{\@secondoftwo}%
\providecommand \bibfield  [0]{\@secondoftwo}%
\providecommand \translation [1]{[#1]}%
\providecommand \BibitemOpen [0]{}%
\providecommand \bibitemStop [0]{}%
\providecommand \bibitemNoStop [0]{.\EOS\space}%
\providecommand \EOS [0]{\spacefactor3000\relax}%
\providecommand \BibitemShut  [1]{\csname bibitem#1\endcsname}%
\let\auto@bib@innerbib\@empty
\bibitem [{\citenamefont {{The Dark Energy Survey Collaboration}}\ \emph
  {et~al.}(2015)\citenamefont {{The Dark Energy Survey Collaboration}},
  \citenamefont {{Abbott}}, \citenamefont {{Abdalla}}, \citenamefont {{Allam}},
  \citenamefont {{Amara}}, \citenamefont {{Annis}}, \citenamefont
  {{Armstrong}}, \citenamefont {{Bacon}}, \citenamefont {{Banerji}},
  \citenamefont {{Bauer}}, \citenamefont {{Baxter}}, \citenamefont {{Becker}},
  \citenamefont {{Benoit-L{\'e}vy}}, \citenamefont {{Bernstein}}, \citenamefont
  {{Bernstein}}, \citenamefont {{Bertin}}, \citenamefont {{Blazek}},
  \citenamefont {{Bonnett}}, \citenamefont {{Bridle}}, \citenamefont
  {{Brooks}}, \citenamefont {{Bruderer}}, \citenamefont {{Buckley-Geer}},
  \citenamefont {{Burke}}, \citenamefont {{Busha}}, \citenamefont {{Capozzi}},
  \citenamefont {{Carnero Rosell}}, \citenamefont {{Carrasco Kind}},
  \citenamefont {{Carretero}}, \citenamefont {{Castander}}, \citenamefont
  {{Chang}}, \citenamefont {{Clampitt}}, \citenamefont {{Crocce}},
  \citenamefont {{Cunha}}, \citenamefont {{D'Andrea}}, \citenamefont {{da
  Costa}}, \citenamefont {{Das}}, \citenamefont {{DePoy}}, \citenamefont
  {{Desai}}, \citenamefont {{Diehl}}, \citenamefont {{Dietrich}}, \citenamefont
  {{Dodelson}}, \citenamefont {{Doel}}, \citenamefont {{Drlica-Wagner}},
  \citenamefont {{Efstathiou}}, \citenamefont {{Eifler}}, \citenamefont
  {{Erickson}}, \citenamefont {{Estrada}}, \citenamefont {{Evrard}},
  \citenamefont {{Fausti Neto}}, \citenamefont {{Fernandez}}, \citenamefont
  {{Finley}}, \citenamefont {{Flaugher}}, \citenamefont {{Fosalba}},
  \citenamefont {{Friedrich}}, \citenamefont {{Frieman}}, \citenamefont
  {{Gangkofner}}, \citenamefont {{Garcia-Bellido}}, \citenamefont
  {{Gaztanaga}}, \citenamefont {{Gerdes}}, \citenamefont {{Gruen}},
  \citenamefont {{Gruendl}}, \citenamefont {{Gutierrez}}, \citenamefont
  {{Hartley}}, \citenamefont {{Hirsch}}, \citenamefont {{Honscheid}},
  \citenamefont {{Huff}}, \citenamefont {{Jain}}, \citenamefont {{James}},
  \citenamefont {{Jarvis}}, \citenamefont {{Kacprzak}}, \citenamefont {{Kent}},
  \citenamefont {{Kirk}}, \citenamefont {{Krause}}, \citenamefont {{Kravtsov}},
  \citenamefont {{Kuehn}}, \citenamefont {{Kuropatkin}}, \citenamefont
  {{Kwan}}, \citenamefont {{Lahav}}, \citenamefont {{Leistedt}}, \citenamefont
  {{Li}}, \citenamefont {{Lima}}, \citenamefont {{Lin}}, \citenamefont
  {{MacCrann}}, \citenamefont {{March}}, \citenamefont {{Marshall}},
  \citenamefont {{Martini}}, \citenamefont {{McMahon}}, \citenamefont
  {{Melchior}}, \citenamefont {{Miller}}, \citenamefont {{Miquel}},
  \citenamefont {{Mohr}}, \citenamefont {{Neilsen}}, \citenamefont {{Nichol}},
  \citenamefont {{Nicola}}, \citenamefont {{Nord}}, \citenamefont {{Ogando}},
  \citenamefont {{Palmese}}, \citenamefont {{Peiris}}, \citenamefont
  {{Plazas}}, \citenamefont {{Refregier}}, \citenamefont {{Roe}}, \citenamefont
  {{Romer}}, \citenamefont {{Roodman}}, \citenamefont {{Rowe}}, \citenamefont
  {{Rykoff}}, \citenamefont {{Sabiu}}, \citenamefont {{Sadeh}}, \citenamefont
  {{Sako}}, \citenamefont {{Samuroff}}, \citenamefont {{S{\'a}nchez}},
  \citenamefont {{Sanchez}}, \citenamefont {{Seo}}, \citenamefont
  {{Sevilla-Noarbe}}, \citenamefont {{Sheldon}}, \citenamefont {{Smith}},
  \citenamefont {{Soares-Santos}}, \citenamefont {{Sobreira}}, \citenamefont
  {{Suchyta}}, \citenamefont {{Swanson}}, \citenamefont {{Tarle}},
  \citenamefont {{Thaler}}, \citenamefont {{Thomas}}, \citenamefont {{Troxel}},
  \citenamefont {{Vikram}}, \citenamefont {{Walker}}, \citenamefont
  {{Wechsler}}, \citenamefont {{Weller}}, \citenamefont {{Zhang}},\ and\
  \citenamefont {{Zuntz}}}]{deswlcosmo}%
  \BibitemOpen
  \bibfield  {author} {\bibinfo {author} {\bibnamefont {{The Dark Energy Survey
  Collaboration}}}, \bibinfo {author} {\bibfnamefont {T.}~\bibnamefont
  {{Abbott}}}, \bibinfo {author} {\bibfnamefont {F.~B.}\ \bibnamefont
  {{Abdalla}}}, \bibinfo {author} {\bibfnamefont {S.}~\bibnamefont {{Allam}}},
  \bibinfo {author} {\bibfnamefont {A.}~\bibnamefont {{Amara}}}, \bibinfo
  {author} {\bibfnamefont {J.}~\bibnamefont {{Annis}}}, \bibinfo {author}
  {\bibfnamefont {R.}~\bibnamefont {{Armstrong}}}, \bibinfo {author}
  {\bibfnamefont {D.}~\bibnamefont {{Bacon}}}, \bibinfo {author} {\bibfnamefont
  {M.}~\bibnamefont {{Banerji}}}, \bibinfo {author} {\bibfnamefont {A.~H.}\
  \bibnamefont {{Bauer}}}, \bibinfo {author} {\bibfnamefont {E.}~\bibnamefont
  {{Baxter}}}, \bibinfo {author} {\bibfnamefont {M.~R.}\ \bibnamefont
  {{Becker}}}, \bibinfo {author} {\bibfnamefont {A.}~\bibnamefont
  {{Benoit-L{\'e}vy}}}, \bibinfo {author} {\bibfnamefont {R.~A.}\ \bibnamefont
  {{Bernstein}}}, \bibinfo {author} {\bibfnamefont {G.~M.}\ \bibnamefont
  {{Bernstein}}}, \bibinfo {author} {\bibfnamefont {E.}~\bibnamefont
  {{Bertin}}}, \bibinfo {author} {\bibfnamefont {J.}~\bibnamefont {{Blazek}}},
  \bibinfo {author} {\bibfnamefont {C.}~\bibnamefont {{Bonnett}}}, \bibinfo
  {author} {\bibfnamefont {S.~L.}\ \bibnamefont {{Bridle}}}, \bibinfo {author}
  {\bibfnamefont {D.}~\bibnamefont {{Brooks}}}, \bibinfo {author}
  {\bibfnamefont {C.}~\bibnamefont {{Bruderer}}}, \bibinfo {author}
  {\bibfnamefont {E.}~\bibnamefont {{Buckley-Geer}}}, \bibinfo {author}
  {\bibfnamefont {D.~L.}\ \bibnamefont {{Burke}}}, \bibinfo {author}
  {\bibfnamefont {M.~T.}\ \bibnamefont {{Busha}}}, \bibinfo {author}
  {\bibfnamefont {D.}~\bibnamefont {{Capozzi}}}, \bibinfo {author}
  {\bibfnamefont {A.}~\bibnamefont {{Carnero Rosell}}}, \bibinfo {author}
  {\bibfnamefont {M.}~\bibnamefont {{Carrasco Kind}}}, \bibinfo {author}
  {\bibfnamefont {J.}~\bibnamefont {{Carretero}}}, \bibinfo {author}
  {\bibfnamefont {F.~J.}\ \bibnamefont {{Castander}}}, \bibinfo {author}
  {\bibfnamefont {C.}~\bibnamefont {{Chang}}}, \bibinfo {author} {\bibfnamefont
  {J.}~\bibnamefont {{Clampitt}}}, \bibinfo {author} {\bibfnamefont
  {M.}~\bibnamefont {{Crocce}}}, \bibinfo {author} {\bibfnamefont {C.~E.}\
  \bibnamefont {{Cunha}}}, \bibinfo {author} {\bibfnamefont {C.~B.}\
  \bibnamefont {{D'Andrea}}}, \bibinfo {author} {\bibfnamefont {L.~N.}\
  \bibnamefont {{da Costa}}}, \bibinfo {author} {\bibfnamefont
  {R.}~\bibnamefont {{Das}}}, \bibinfo {author} {\bibfnamefont {D.~L.}\
  \bibnamefont {{DePoy}}}, \bibinfo {author} {\bibfnamefont {S.}~\bibnamefont
  {{Desai}}}, \bibinfo {author} {\bibfnamefont {H.~T.}\ \bibnamefont
  {{Diehl}}}, \bibinfo {author} {\bibfnamefont {J.~P.}\ \bibnamefont
  {{Dietrich}}}, \bibinfo {author} {\bibfnamefont {S.}~\bibnamefont
  {{Dodelson}}}, \bibinfo {author} {\bibfnamefont {P.}~\bibnamefont {{Doel}}},
  \bibinfo {author} {\bibfnamefont {A.}~\bibnamefont {{Drlica-Wagner}}},
  \bibinfo {author} {\bibfnamefont {G.}~\bibnamefont {{Efstathiou}}}, \bibinfo
  {author} {\bibfnamefont {T.~F.}\ \bibnamefont {{Eifler}}}, \bibinfo {author}
  {\bibfnamefont {B.}~\bibnamefont {{Erickson}}}, \bibinfo {author}
  {\bibfnamefont {J.}~\bibnamefont {{Estrada}}}, \bibinfo {author}
  {\bibfnamefont {A.~E.}\ \bibnamefont {{Evrard}}}, \bibinfo {author}
  {\bibfnamefont {A.}~\bibnamefont {{Fausti Neto}}}, \bibinfo {author}
  {\bibfnamefont {E.}~\bibnamefont {{Fernandez}}}, \bibinfo {author}
  {\bibfnamefont {D.~A.}\ \bibnamefont {{Finley}}}, \bibinfo {author}
  {\bibfnamefont {B.}~\bibnamefont {{Flaugher}}}, \bibinfo {author}
  {\bibfnamefont {P.}~\bibnamefont {{Fosalba}}}, \bibinfo {author}
  {\bibfnamefont {O.}~\bibnamefont {{Friedrich}}}, \bibinfo {author}
  {\bibfnamefont {J.}~\bibnamefont {{Frieman}}}, \bibinfo {author}
  {\bibfnamefont {C.}~\bibnamefont {{Gangkofner}}}, \bibinfo {author}
  {\bibfnamefont {J.}~\bibnamefont {{Garcia-Bellido}}}, \bibinfo {author}
  {\bibfnamefont {E.}~\bibnamefont {{Gaztanaga}}}, \bibinfo {author}
  {\bibfnamefont {D.~W.}\ \bibnamefont {{Gerdes}}}, \bibinfo {author}
  {\bibfnamefont {D.}~\bibnamefont {{Gruen}}}, \bibinfo {author} {\bibfnamefont
  {R.~A.}\ \bibnamefont {{Gruendl}}}, \bibinfo {author} {\bibfnamefont
  {G.}~\bibnamefont {{Gutierrez}}}, \bibinfo {author} {\bibfnamefont
  {W.}~\bibnamefont {{Hartley}}}, \bibinfo {author} {\bibfnamefont
  {M.}~\bibnamefont {{Hirsch}}}, \bibinfo {author} {\bibfnamefont
  {K.}~\bibnamefont {{Honscheid}}}, \bibinfo {author} {\bibfnamefont {E.~M.}\
  \bibnamefont {{Huff}}}, \bibinfo {author} {\bibfnamefont {B.}~\bibnamefont
  {{Jain}}}, \bibinfo {author} {\bibfnamefont {D.~J.}\ \bibnamefont {{James}}},
  \bibinfo {author} {\bibfnamefont {M.}~\bibnamefont {{Jarvis}}}, \bibinfo
  {author} {\bibfnamefont {T.}~\bibnamefont {{Kacprzak}}}, \bibinfo {author}
  {\bibfnamefont {S.}~\bibnamefont {{Kent}}}, \bibinfo {author} {\bibfnamefont
  {D.}~\bibnamefont {{Kirk}}}, \bibinfo {author} {\bibfnamefont
  {E.}~\bibnamefont {{Krause}}}, \bibinfo {author} {\bibfnamefont
  {A.}~\bibnamefont {{Kravtsov}}}, \bibinfo {author} {\bibfnamefont
  {K.}~\bibnamefont {{Kuehn}}}, \bibinfo {author} {\bibfnamefont
  {N.}~\bibnamefont {{Kuropatkin}}}, \bibinfo {author} {\bibfnamefont
  {J.}~\bibnamefont {{Kwan}}}, \bibinfo {author} {\bibfnamefont
  {O.}~\bibnamefont {{Lahav}}}, \bibinfo {author} {\bibfnamefont
  {B.}~\bibnamefont {{Leistedt}}}, \bibinfo {author} {\bibfnamefont {T.~S.}\
  \bibnamefont {{Li}}}, \bibinfo {author} {\bibfnamefont {M.}~\bibnamefont
  {{Lima}}}, \bibinfo {author} {\bibfnamefont {H.}~\bibnamefont {{Lin}}},
  \bibinfo {author} {\bibfnamefont {N.}~\bibnamefont {{MacCrann}}}, \bibinfo
  {author} {\bibfnamefont {M.}~\bibnamefont {{March}}}, \bibinfo {author}
  {\bibfnamefont {J.~L.}\ \bibnamefont {{Marshall}}}, \bibinfo {author}
  {\bibfnamefont {P.}~\bibnamefont {{Martini}}}, \bibinfo {author}
  {\bibfnamefont {R.~G.}\ \bibnamefont {{McMahon}}}, \bibinfo {author}
  {\bibfnamefont {P.}~\bibnamefont {{Melchior}}}, \bibinfo {author}
  {\bibfnamefont {C.~J.}\ \bibnamefont {{Miller}}}, \bibinfo {author}
  {\bibfnamefont {R.}~\bibnamefont {{Miquel}}}, \bibinfo {author}
  {\bibfnamefont {J.~J.}\ \bibnamefont {{Mohr}}}, \bibinfo {author}
  {\bibfnamefont {E.}~\bibnamefont {{Neilsen}}}, \bibinfo {author}
  {\bibfnamefont {R.~C.}\ \bibnamefont {{Nichol}}}, \bibinfo {author}
  {\bibfnamefont {A.}~\bibnamefont {{Nicola}}}, \bibinfo {author}
  {\bibfnamefont {B.}~\bibnamefont {{Nord}}}, \bibinfo {author} {\bibfnamefont
  {R.}~\bibnamefont {{Ogando}}}, \bibinfo {author} {\bibfnamefont
  {A.}~\bibnamefont {{Palmese}}}, \bibinfo {author} {\bibfnamefont {H.~V.}\
  \bibnamefont {{Peiris}}}, \bibinfo {author} {\bibfnamefont {A.~A.}\
  \bibnamefont {{Plazas}}}, \bibinfo {author} {\bibfnamefont {A.}~\bibnamefont
  {{Refregier}}}, \bibinfo {author} {\bibfnamefont {N.}~\bibnamefont {{Roe}}},
  \bibinfo {author} {\bibfnamefont {A.~K.}\ \bibnamefont {{Romer}}}, \bibinfo
  {author} {\bibfnamefont {A.}~\bibnamefont {{Roodman}}}, \bibinfo {author}
  {\bibfnamefont {B.}~\bibnamefont {{Rowe}}}, \bibinfo {author} {\bibfnamefont
  {E.~S.}\ \bibnamefont {{Rykoff}}}, \bibinfo {author} {\bibfnamefont
  {C.}~\bibnamefont {{Sabiu}}}, \bibinfo {author} {\bibfnamefont
  {I.}~\bibnamefont {{Sadeh}}}, \bibinfo {author} {\bibfnamefont
  {M.}~\bibnamefont {{Sako}}}, \bibinfo {author} {\bibfnamefont
  {S.}~\bibnamefont {{Samuroff}}}, \bibinfo {author} {\bibfnamefont
  {C.}~\bibnamefont {{S{\'a}nchez}}}, \bibinfo {author} {\bibfnamefont
  {E.}~\bibnamefont {{Sanchez}}}, \bibinfo {author} {\bibfnamefont
  {H.}~\bibnamefont {{Seo}}}, \bibinfo {author} {\bibfnamefont
  {I.}~\bibnamefont {{Sevilla-Noarbe}}}, \bibinfo {author} {\bibfnamefont
  {E.}~\bibnamefont {{Sheldon}}}, \bibinfo {author} {\bibfnamefont {R.~C.}\
  \bibnamefont {{Smith}}}, \bibinfo {author} {\bibfnamefont {M.}~\bibnamefont
  {{Soares-Santos}}}, \bibinfo {author} {\bibfnamefont {F.}~\bibnamefont
  {{Sobreira}}}, \bibinfo {author} {\bibfnamefont {E.}~\bibnamefont
  {{Suchyta}}}, \bibinfo {author} {\bibfnamefont {M.~E.~C.}\ \bibnamefont
  {{Swanson}}}, \bibinfo {author} {\bibfnamefont {G.}~\bibnamefont {{Tarle}}},
  \bibinfo {author} {\bibfnamefont {J.}~\bibnamefont {{Thaler}}}, \bibinfo
  {author} {\bibfnamefont {D.}~\bibnamefont {{Thomas}}}, \bibinfo {author}
  {\bibfnamefont {M.~A.}\ \bibnamefont {{Troxel}}}, \bibinfo {author}
  {\bibfnamefont {V.}~\bibnamefont {{Vikram}}}, \bibinfo {author}
  {\bibfnamefont {A.~R.}\ \bibnamefont {{Walker}}}, \bibinfo {author}
  {\bibfnamefont {R.~H.}\ \bibnamefont {{Wechsler}}}, \bibinfo {author}
  {\bibfnamefont {J.}~\bibnamefont {{Weller}}}, \bibinfo {author}
  {\bibfnamefont {Y.}~\bibnamefont {{Zhang}}}, \ and\ \bibinfo {author}
  {\bibfnamefont {J.}~\bibnamefont {{Zuntz}}},\ }\href@noop {} {\bibfield
  {journal} {\bibinfo  {journal} {arXiv:astro-ph/1507.05552}\ } (\bibinfo
  {year} {2015})},\ \Eprint {http://arxiv.org/abs/1507.05552}
  {arXiv:1507.05552} \BibitemShut {NoStop}%
\bibitem [{\citenamefont {{Albrecht}}\ \emph {et~al.}(2006)\citenamefont
  {{Albrecht}}, \citenamefont {{Bernstein}}, \citenamefont {{Cahn}},
  \citenamefont {{Freedman}}, \citenamefont {{Hewitt}}, \citenamefont {{Hu}},
  \citenamefont {{Huth}}, \citenamefont {{Kamionkowski}}, \citenamefont
  {{Kolb}}, \citenamefont {{Knox}}, \citenamefont {{Mather}}, \citenamefont
  {{Staggs}},\ and\ \citenamefont {{Suntzeff}}}]{detf}%
  \BibitemOpen
  \bibfield  {author} {\bibinfo {author} {\bibfnamefont {A.}~\bibnamefont
  {{Albrecht}}}, \bibinfo {author} {\bibfnamefont {G.}~\bibnamefont
  {{Bernstein}}}, \bibinfo {author} {\bibfnamefont {R.}~\bibnamefont {{Cahn}}},
  \bibinfo {author} {\bibfnamefont {W.~L.}\ \bibnamefont {{Freedman}}},
  \bibinfo {author} {\bibfnamefont {J.}~\bibnamefont {{Hewitt}}}, \bibinfo
  {author} {\bibfnamefont {W.}~\bibnamefont {{Hu}}}, \bibinfo {author}
  {\bibfnamefont {J.}~\bibnamefont {{Huth}}}, \bibinfo {author} {\bibfnamefont
  {M.}~\bibnamefont {{Kamionkowski}}}, \bibinfo {author} {\bibfnamefont
  {E.~W.}\ \bibnamefont {{Kolb}}}, \bibinfo {author} {\bibfnamefont
  {L.}~\bibnamefont {{Knox}}}, \bibinfo {author} {\bibfnamefont {J.~C.}\
  \bibnamefont {{Mather}}}, \bibinfo {author} {\bibfnamefont {S.}~\bibnamefont
  {{Staggs}}}, \ and\ \bibinfo {author} {\bibfnamefont {N.~B.}\ \bibnamefont
  {{Suntzeff}}},\ }\href@noop {} {\bibfield  {journal} {\bibinfo  {journal}
  {arXiv:astro-ph/0609591}\ } (\bibinfo {year} {2006})},\ \Eprint
  {http://arxiv.org/abs/astro-ph/0609591} {astro-ph/0609591} \BibitemShut
  {NoStop}%
\bibitem [{\citenamefont {{Peacock}}\ and\ \citenamefont
  {{Schneider}}(2006)}]{esoesa}%
  \BibitemOpen
  \bibfield  {author} {\bibinfo {author} {\bibfnamefont {J.}~\bibnamefont
  {{Peacock}}}\ and\ \bibinfo {author} {\bibfnamefont {P.}~\bibnamefont
  {{Schneider}}},\ }\href@noop {} {\bibfield  {journal} {\bibinfo  {journal}
  {The Messenger}\ }\textbf {\bibinfo {volume} {125}},\ \bibinfo {pages} {48}
  (\bibinfo {year} {2006})}\BibitemShut {NoStop}%
\bibitem [{\citenamefont {{Weinberg}}\ \emph {et~al.}(2013)\citenamefont
  {{Weinberg}}, \citenamefont {{Mortonson}}, \citenamefont {{Eisenstein}},
  \citenamefont {{Hirata}}, \citenamefont {{Riess}},\ and\ \citenamefont
  {{Rozo}}}]{weinberg2013}%
  \BibitemOpen
  \bibfield  {author} {\bibinfo {author} {\bibfnamefont {D.~H.}\ \bibnamefont
  {{Weinberg}}}, \bibinfo {author} {\bibfnamefont {M.~J.}\ \bibnamefont
  {{Mortonson}}}, \bibinfo {author} {\bibfnamefont {D.~J.}\ \bibnamefont
  {{Eisenstein}}}, \bibinfo {author} {\bibfnamefont {C.}~\bibnamefont
  {{Hirata}}}, \bibinfo {author} {\bibfnamefont {A.~G.}\ \bibnamefont
  {{Riess}}}, \ and\ \bibinfo {author} {\bibfnamefont {E.}~\bibnamefont
  {{Rozo}}},\ }\href {\doibase 10.1016/j.physrep.2013.05.001} {\bibfield
  {journal} {\bibinfo  {journal} {\physrep}\ }\textbf {\bibinfo {volume}
  {530}},\ \bibinfo {pages} {87} (\bibinfo {year} {2013})},\ \Eprint
  {http://arxiv.org/abs/1201.2434} {arXiv:1201.2434} \BibitemShut {NoStop}%
\bibitem [{\citenamefont {{Kilbinger}}(2015)}]{kilbinger2014}%
  \BibitemOpen
  \bibfield  {author} {\bibinfo {author} {\bibfnamefont {M.}~\bibnamefont
  {{Kilbinger}}},\ }\href {\doibase 10.1088/0034-4885/78/8/086901} {\bibfield
  {journal} {\bibinfo  {journal} {Reports on Progress in Physics}\ }\textbf
  {\bibinfo {volume} {78}},\ \bibinfo {eid} {086901} (\bibinfo {year}
  {2015})},\ \Eprint {http://arxiv.org/abs/1411.0115} {arXiv:1411.0115}
  \BibitemShut {NoStop}%
\bibitem [{\citenamefont {{Bacon}}\ \emph {et~al.}(2000)\citenamefont
  {{Bacon}}, \citenamefont {{Refregier}},\ and\ \citenamefont
  {{Ellis}}}]{Bacon:2000yp}%
  \BibitemOpen
  \bibfield  {author} {\bibinfo {author} {\bibfnamefont {D.~J.}\ \bibnamefont
  {{Bacon}}}, \bibinfo {author} {\bibfnamefont {A.~R.}\ \bibnamefont
  {{Refregier}}}, \ and\ \bibinfo {author} {\bibfnamefont {R.~S.}\ \bibnamefont
  {{Ellis}}},\ }\href@noop {} {\bibfield  {journal} {\bibinfo  {journal}
  {\mnras}\ }\textbf {\bibinfo {volume} {318}},\ \bibinfo {pages} {625}
  (\bibinfo {year} {2000})},\ \Eprint
  {http://arxiv.org/abs/arXiv:astro-ph/0003008} {arXiv:astro-ph/0003008}
  \BibitemShut {NoStop}%
\bibitem [{\citenamefont {{Kaiser}}\ \emph {et~al.}(2000)\citenamefont
  {{Kaiser}}, \citenamefont {{Wilson}},\ and\ \citenamefont
  {{Luppino}}}]{Kaiser:2000if}%
  \BibitemOpen
  \bibfield  {author} {\bibinfo {author} {\bibfnamefont {N.}~\bibnamefont
  {{Kaiser}}}, \bibinfo {author} {\bibfnamefont {G.}~\bibnamefont {{Wilson}}},
  \ and\ \bibinfo {author} {\bibfnamefont {G.~A.}\ \bibnamefont {{Luppino}}},\
  }\href@noop {} {\bibfield  {journal} {\bibinfo  {journal}
  {arXiv:astro-ph/0003338}\ } (\bibinfo {year} {2000})},\ \Eprint
  {http://arxiv.org/abs/astro-ph/0003338} {astro-ph/0003338} \BibitemShut
  {NoStop}%
\bibitem [{\citenamefont {Wittman}\ \emph {et~al.}(2000)\citenamefont
  {Wittman}, \citenamefont {Tyson}, \citenamefont {Kirkman}, \citenamefont
  {Dell'Antonio},\ and\ \citenamefont {Bernstein}}]{Wittman:2000tc}%
  \BibitemOpen
  \bibfield  {author} {\bibinfo {author} {\bibfnamefont {D.~M.}\ \bibnamefont
  {Wittman}}, \bibinfo {author} {\bibfnamefont {J.~A.}\ \bibnamefont {Tyson}},
  \bibinfo {author} {\bibfnamefont {D.}~\bibnamefont {Kirkman}}, \bibinfo
  {author} {\bibfnamefont {I.}~\bibnamefont {Dell'Antonio}}, \ and\ \bibinfo
  {author} {\bibfnamefont {G.}~\bibnamefont {Bernstein}},\ }\href@noop {}
  {\bibfield  {journal} {\bibinfo  {journal} {Nature}\ }\textbf {\bibinfo
  {volume} {405}},\ \bibinfo {pages} {143} (\bibinfo {year} {2000})},\ \Eprint
  {http://arxiv.org/abs/astro-ph/0003014} {astro-ph/0003014} \BibitemShut
  {NoStop}%
\bibitem [{\citenamefont {van Waerbeke}\ \emph {et~al.}(2000)\citenamefont {van
  Waerbeke} \emph {et~al.}}]{van_Waerbeke:2000rm}%
  \BibitemOpen
  \bibfield  {author} {\bibinfo {author} {\bibfnamefont {L.}~\bibnamefont {van
  Waerbeke}} \emph {et~al.},\ }\href@noop {} {\bibfield  {journal} {\bibinfo
  {journal} {Astron. Astrophys.}\ }\textbf {\bibinfo {volume} {358}},\ \bibinfo
  {pages} {30} (\bibinfo {year} {2000})},\ \Eprint
  {http://arxiv.org/abs/astro-ph/0002500} {astro-ph/0002500} \BibitemShut
  {NoStop}%
\bibitem [{\citenamefont {{Jee}}\ \emph {et~al.}(2013)\citenamefont {{Jee}},
  \citenamefont {{Tyson}}, \citenamefont {{Schneider}}, \citenamefont
  {{Wittman}}, \citenamefont {{Schmidt}},\ and\ \citenamefont
  {{Hilbert}}}]{jee2013}%
  \BibitemOpen
  \bibfield  {author} {\bibinfo {author} {\bibfnamefont {M.~J.}\ \bibnamefont
  {{Jee}}}, \bibinfo {author} {\bibfnamefont {J.~A.}\ \bibnamefont {{Tyson}}},
  \bibinfo {author} {\bibfnamefont {M.~D.}\ \bibnamefont {{Schneider}}},
  \bibinfo {author} {\bibfnamefont {D.}~\bibnamefont {{Wittman}}}, \bibinfo
  {author} {\bibfnamefont {S.}~\bibnamefont {{Schmidt}}}, \ and\ \bibinfo
  {author} {\bibfnamefont {S.}~\bibnamefont {{Hilbert}}},\ }\href {\doibase
  10.1088/0004-637X/765/1/74} {\bibfield  {journal} {\bibinfo  {journal}
  {\apj}\ }\textbf {\bibinfo {volume} {765}},\ \bibinfo {eid} {74} (\bibinfo
  {year} {2013})},\ \Eprint {http://arxiv.org/abs/1210.2732} {arXiv:1210.2732
  [astro-ph.CO]} \BibitemShut {NoStop}%
\bibitem [{\citenamefont {{Lin}}\ \emph {et~al.}(2012)\citenamefont {{Lin}},
  \citenamefont {{Dodelson}}, \citenamefont {{Seo}}, \citenamefont
  {{Soares-Santos}}, \citenamefont {{Annis}}, \citenamefont {{Hao}},
  \citenamefont {{Johnston}}, \citenamefont {{Kubo}}, \citenamefont {{Reis}},\
  and\ \citenamefont {{Simet}}}]{lin2012}%
  \BibitemOpen
  \bibfield  {author} {\bibinfo {author} {\bibfnamefont {H.}~\bibnamefont
  {{Lin}}}, \bibinfo {author} {\bibfnamefont {S.}~\bibnamefont {{Dodelson}}},
  \bibinfo {author} {\bibfnamefont {H.-J.}\ \bibnamefont {{Seo}}}, \bibinfo
  {author} {\bibfnamefont {M.}~\bibnamefont {{Soares-Santos}}}, \bibinfo
  {author} {\bibfnamefont {J.}~\bibnamefont {{Annis}}}, \bibinfo {author}
  {\bibfnamefont {J.}~\bibnamefont {{Hao}}}, \bibinfo {author} {\bibfnamefont
  {D.}~\bibnamefont {{Johnston}}}, \bibinfo {author} {\bibfnamefont {J.~M.}\
  \bibnamefont {{Kubo}}}, \bibinfo {author} {\bibfnamefont {R.~R.~R.}\
  \bibnamefont {{Reis}}}, \ and\ \bibinfo {author} {\bibfnamefont
  {M.}~\bibnamefont {{Simet}}},\ }\href {\doibase 10.1088/0004-637X/761/1/15}
  {\bibfield  {journal} {\bibinfo  {journal} {\apj}\ }\textbf {\bibinfo
  {volume} {761}},\ \bibinfo {eid} {15} (\bibinfo {year} {2012})},\ \Eprint
  {http://arxiv.org/abs/1111.6622} {arXiv:1111.6622} \BibitemShut {NoStop}%
\bibitem [{\citenamefont {{Huff}}\ \emph {et~al.}(2014)\citenamefont {{Huff}},
  \citenamefont {{Eifler}}, \citenamefont {{Hirata}}, \citenamefont
  {{Mandelbaum}}, \citenamefont {{Schlegel}},\ and\ \citenamefont
  {{Seljak}}}]{huff2014}%
  \BibitemOpen
  \bibfield  {author} {\bibinfo {author} {\bibfnamefont {E.~M.}\ \bibnamefont
  {{Huff}}}, \bibinfo {author} {\bibfnamefont {T.}~\bibnamefont {{Eifler}}},
  \bibinfo {author} {\bibfnamefont {C.~M.}\ \bibnamefont {{Hirata}}}, \bibinfo
  {author} {\bibfnamefont {R.}~\bibnamefont {{Mandelbaum}}}, \bibinfo {author}
  {\bibfnamefont {D.}~\bibnamefont {{Schlegel}}}, \ and\ \bibinfo {author}
  {\bibfnamefont {U.}~\bibnamefont {{Seljak}}},\ }\href {\doibase
  10.1093/mnras/stu145} {\bibfield  {journal} {\bibinfo  {journal} {\mnras}\
  }\textbf {\bibinfo {volume} {440}},\ \bibinfo {pages} {1322} (\bibinfo {year}
  {2014})}\BibitemShut {NoStop}%
\bibitem [{\citenamefont {{Kuijken}}\ \emph {et~al.}(2015)\citenamefont
  {{Kuijken}}, \citenamefont {{Heymans}}, \citenamefont {{Hildebrandt}},
  \citenamefont {{Nakajima}}, \citenamefont {{Erben}}, \citenamefont {{de
  Jong}}, \citenamefont {{Viola}}, \citenamefont {{Choi}}, \citenamefont
  {{Hoekstra}}, \citenamefont {{Miller}}, \citenamefont {{van Uitert}},
  \citenamefont {{Amon}}, \citenamefont {{Blake}}, \citenamefont {{Brouwer}},
  \citenamefont {{Buddendiek}}, \citenamefont {{Conti}}, \citenamefont
  {{Eriksen}}, \citenamefont {{Grado}}, \citenamefont {{Harnois-D{\'e}raps}},
  \citenamefont {{Helmich}}, \citenamefont {{Herbonnet}}, \citenamefont
  {{Irisarri}}, \citenamefont {{Kitching}}, \citenamefont {{Klaes}},
  \citenamefont {{La Barbera}}, \citenamefont {{Napolitano}}, \citenamefont
  {{Radovich}}, \citenamefont {{Schneider}}, \citenamefont {{Sif{\'o}n}},
  \citenamefont {{Sikkema}}, \citenamefont {{Simon}}, \citenamefont
  {{Tudorica}}, \citenamefont {{Valentijn}}, \citenamefont {{Verdoes Kleijn}},\
  and\ \citenamefont {{van Waerbeke}}}]{kuijken2015}%
  \BibitemOpen
  \bibfield  {author} {\bibinfo {author} {\bibfnamefont {K.}~\bibnamefont
  {{Kuijken}}}, \bibinfo {author} {\bibfnamefont {C.}~\bibnamefont
  {{Heymans}}}, \bibinfo {author} {\bibfnamefont {H.}~\bibnamefont
  {{Hildebrandt}}}, \bibinfo {author} {\bibfnamefont {R.}~\bibnamefont
  {{Nakajima}}}, \bibinfo {author} {\bibfnamefont {T.}~\bibnamefont {{Erben}}},
  \bibinfo {author} {\bibfnamefont {J.~T.~A.}\ \bibnamefont {{de Jong}}},
  \bibinfo {author} {\bibfnamefont {M.}~\bibnamefont {{Viola}}}, \bibinfo
  {author} {\bibfnamefont {A.}~\bibnamefont {{Choi}}}, \bibinfo {author}
  {\bibfnamefont {H.}~\bibnamefont {{Hoekstra}}}, \bibinfo {author}
  {\bibfnamefont {L.}~\bibnamefont {{Miller}}}, \bibinfo {author}
  {\bibfnamefont {E.}~\bibnamefont {{van Uitert}}}, \bibinfo {author}
  {\bibfnamefont {A.}~\bibnamefont {{Amon}}}, \bibinfo {author} {\bibfnamefont
  {C.}~\bibnamefont {{Blake}}}, \bibinfo {author} {\bibfnamefont
  {M.}~\bibnamefont {{Brouwer}}}, \bibinfo {author} {\bibfnamefont
  {A.}~\bibnamefont {{Buddendiek}}}, \bibinfo {author} {\bibfnamefont {I.~F.}\
  \bibnamefont {{Conti}}}, \bibinfo {author} {\bibfnamefont {M.}~\bibnamefont
  {{Eriksen}}}, \bibinfo {author} {\bibfnamefont {A.}~\bibnamefont {{Grado}}},
  \bibinfo {author} {\bibfnamefont {J.}~\bibnamefont {{Harnois-D{\'e}raps}}},
  \bibinfo {author} {\bibfnamefont {E.}~\bibnamefont {{Helmich}}}, \bibinfo
  {author} {\bibfnamefont {R.}~\bibnamefont {{Herbonnet}}}, \bibinfo {author}
  {\bibfnamefont {N.}~\bibnamefont {{Irisarri}}}, \bibinfo {author}
  {\bibfnamefont {T.}~\bibnamefont {{Kitching}}}, \bibinfo {author}
  {\bibfnamefont {D.}~\bibnamefont {{Klaes}}}, \bibinfo {author} {\bibfnamefont
  {F.}~\bibnamefont {{La Barbera}}}, \bibinfo {author} {\bibfnamefont
  {N.}~\bibnamefont {{Napolitano}}}, \bibinfo {author} {\bibfnamefont
  {M.}~\bibnamefont {{Radovich}}}, \bibinfo {author} {\bibfnamefont
  {P.}~\bibnamefont {{Schneider}}}, \bibinfo {author} {\bibfnamefont
  {C.}~\bibnamefont {{Sif{\'o}n}}}, \bibinfo {author} {\bibfnamefont
  {G.}~\bibnamefont {{Sikkema}}}, \bibinfo {author} {\bibfnamefont
  {P.}~\bibnamefont {{Simon}}}, \bibinfo {author} {\bibfnamefont
  {A.}~\bibnamefont {{Tudorica}}}, \bibinfo {author} {\bibfnamefont
  {E.}~\bibnamefont {{Valentijn}}}, \bibinfo {author} {\bibfnamefont
  {G.}~\bibnamefont {{Verdoes Kleijn}}}, \ and\ \bibinfo {author}
  {\bibfnamefont {L.}~\bibnamefont {{van Waerbeke}}},\ }\href {\doibase
  10.1093/mnras/stv2140} {\bibfield  {journal} {\bibinfo  {journal} {\mnras}\
  }\textbf {\bibinfo {volume} {454}},\ \bibinfo {pages} {3500} (\bibinfo {year}
  {2015})},\ \Eprint {http://arxiv.org/abs/1507.00738} {arXiv:1507.00738}
  \BibitemShut {NoStop}%
\bibitem [{\citenamefont {{Kilbinger}}\ \emph {et~al.}(2013)\citenamefont
  {{Kilbinger}}, \citenamefont {{Fu}}, \citenamefont {{Heymans}}, \citenamefont
  {{Simpson}}, \citenamefont {{Benjamin}}, \citenamefont {{Erben}},
  \citenamefont {{Harnois-D{\'e}raps}}, \citenamefont {{Hoekstra}},
  \citenamefont {{Hildebrandt}}, \citenamefont {{Kitching}}, \citenamefont
  {{Mellier}}, \citenamefont {{Miller}}, \citenamefont {{Van Waerbeke}},
  \citenamefont {{Benabed}}, \citenamefont {{Bonnett}}, \citenamefont
  {{Coupon}}, \citenamefont {{Hudson}}, \citenamefont {{Kuijken}},
  \citenamefont {{Rowe}}, \citenamefont {{Schrabback}}, \citenamefont
  {{Semboloni}}, \citenamefont {{Vafaei}},\ and\ \citenamefont
  {{Velander}}}]{kilbinger13}%
  \BibitemOpen
  \bibfield  {author} {\bibinfo {author} {\bibfnamefont {M.}~\bibnamefont
  {{Kilbinger}}}, \bibinfo {author} {\bibfnamefont {L.}~\bibnamefont {{Fu}}},
  \bibinfo {author} {\bibfnamefont {C.}~\bibnamefont {{Heymans}}}, \bibinfo
  {author} {\bibfnamefont {F.}~\bibnamefont {{Simpson}}}, \bibinfo {author}
  {\bibfnamefont {J.}~\bibnamefont {{Benjamin}}}, \bibinfo {author}
  {\bibfnamefont {T.}~\bibnamefont {{Erben}}}, \bibinfo {author} {\bibfnamefont
  {J.}~\bibnamefont {{Harnois-D{\'e}raps}}}, \bibinfo {author} {\bibfnamefont
  {H.}~\bibnamefont {{Hoekstra}}}, \bibinfo {author} {\bibfnamefont
  {H.}~\bibnamefont {{Hildebrandt}}}, \bibinfo {author} {\bibfnamefont {T.~D.}\
  \bibnamefont {{Kitching}}}, \bibinfo {author} {\bibfnamefont
  {Y.}~\bibnamefont {{Mellier}}}, \bibinfo {author} {\bibfnamefont
  {L.}~\bibnamefont {{Miller}}}, \bibinfo {author} {\bibfnamefont
  {L.}~\bibnamefont {{Van Waerbeke}}}, \bibinfo {author} {\bibfnamefont
  {K.}~\bibnamefont {{Benabed}}}, \bibinfo {author} {\bibfnamefont
  {C.}~\bibnamefont {{Bonnett}}}, \bibinfo {author} {\bibfnamefont
  {J.}~\bibnamefont {{Coupon}}}, \bibinfo {author} {\bibfnamefont {M.~J.}\
  \bibnamefont {{Hudson}}}, \bibinfo {author} {\bibfnamefont {K.}~\bibnamefont
  {{Kuijken}}}, \bibinfo {author} {\bibfnamefont {B.}~\bibnamefont {{Rowe}}},
  \bibinfo {author} {\bibfnamefont {T.}~\bibnamefont {{Schrabback}}}, \bibinfo
  {author} {\bibfnamefont {E.}~\bibnamefont {{Semboloni}}}, \bibinfo {author}
  {\bibfnamefont {S.}~\bibnamefont {{Vafaei}}}, \ and\ \bibinfo {author}
  {\bibfnamefont {M.}~\bibnamefont {{Velander}}},\ }\href {\doibase
  10.1093/mnras/stt041} {\bibfield  {journal} {\bibinfo  {journal} {\mnras}\
  }\textbf {\bibinfo {volume} {430}},\ \bibinfo {pages} {2200} (\bibinfo {year}
  {2013})},\ \Eprint {http://arxiv.org/abs/1212.3338} {arXiv:1212.3338
  [astro-ph.CO]} \BibitemShut {NoStop}%
\bibitem [{\citenamefont {{Heymans}}\ \emph {et~al.}(2013)\citenamefont
  {{Heymans}}, \citenamefont {{Grocutt}}, \citenamefont {{Heavens}},
  \citenamefont {{Kilbinger}}, \citenamefont {{Kitching}}, \citenamefont
  {{Simpson}}, \citenamefont {{Benjamin}}, \citenamefont {{Erben}},
  \citenamefont {{Hildebrandt}}, \citenamefont {{Hoekstra}}, \citenamefont
  {{Mellier}}, \citenamefont {{Miller}}, \citenamefont {{Van Waerbeke}},
  \citenamefont {{Brown}}, \citenamefont {{Coupon}}, \citenamefont {{Fu}},
  \citenamefont {{Harnois-D{\'e}raps}}, \citenamefont {{Hudson}}, \citenamefont
  {{Kuijken}}, \citenamefont {{Rowe}}, \citenamefont {{Schrabback}},
  \citenamefont {{Semboloni}}, \citenamefont {{Vafaei}},\ and\ \citenamefont
  {{Velander}}}]{heymans13}%
  \BibitemOpen
  \bibfield  {author} {\bibinfo {author} {\bibfnamefont {C.}~\bibnamefont
  {{Heymans}}}, \bibinfo {author} {\bibfnamefont {E.}~\bibnamefont
  {{Grocutt}}}, \bibinfo {author} {\bibfnamefont {A.}~\bibnamefont
  {{Heavens}}}, \bibinfo {author} {\bibfnamefont {M.}~\bibnamefont
  {{Kilbinger}}}, \bibinfo {author} {\bibfnamefont {T.~D.}\ \bibnamefont
  {{Kitching}}}, \bibinfo {author} {\bibfnamefont {F.}~\bibnamefont
  {{Simpson}}}, \bibinfo {author} {\bibfnamefont {J.}~\bibnamefont
  {{Benjamin}}}, \bibinfo {author} {\bibfnamefont {T.}~\bibnamefont {{Erben}}},
  \bibinfo {author} {\bibfnamefont {H.}~\bibnamefont {{Hildebrandt}}}, \bibinfo
  {author} {\bibfnamefont {H.}~\bibnamefont {{Hoekstra}}}, \bibinfo {author}
  {\bibfnamefont {Y.}~\bibnamefont {{Mellier}}}, \bibinfo {author}
  {\bibfnamefont {L.}~\bibnamefont {{Miller}}}, \bibinfo {author}
  {\bibfnamefont {L.}~\bibnamefont {{Van Waerbeke}}}, \bibinfo {author}
  {\bibfnamefont {M.~L.}\ \bibnamefont {{Brown}}}, \bibinfo {author}
  {\bibfnamefont {J.}~\bibnamefont {{Coupon}}}, \bibinfo {author}
  {\bibfnamefont {L.}~\bibnamefont {{Fu}}}, \bibinfo {author} {\bibfnamefont
  {J.}~\bibnamefont {{Harnois-D{\'e}raps}}}, \bibinfo {author} {\bibfnamefont
  {M.~J.}\ \bibnamefont {{Hudson}}}, \bibinfo {author} {\bibfnamefont
  {K.}~\bibnamefont {{Kuijken}}}, \bibinfo {author} {\bibfnamefont
  {B.}~\bibnamefont {{Rowe}}}, \bibinfo {author} {\bibfnamefont
  {T.}~\bibnamefont {{Schrabback}}}, \bibinfo {author} {\bibfnamefont
  {E.}~\bibnamefont {{Semboloni}}}, \bibinfo {author} {\bibfnamefont
  {S.}~\bibnamefont {{Vafaei}}}, \ and\ \bibinfo {author} {\bibfnamefont
  {M.}~\bibnamefont {{Velander}}},\ }\href {\doibase 10.1093/mnras/stt601}
  {\bibfield  {journal} {\bibinfo  {journal} {\mnras}\ }\textbf {\bibinfo
  {volume} {432}},\ \bibinfo {pages} {2433} (\bibinfo {year} {2013})},\ \Eprint
  {http://arxiv.org/abs/1303.1808} {arXiv:1303.1808 [astro-ph.CO]} \BibitemShut
  {NoStop}%
\bibitem [{\citenamefont {{Heitmann}}\ \emph {et~al.}(2014)\citenamefont
  {{Heitmann}}, \citenamefont {{Lawrence}}, \citenamefont {{Kwan}},
  \citenamefont {{Habib}},\ and\ \citenamefont {{Higdon}}}]{heitmann2014}%
  \BibitemOpen
  \bibfield  {author} {\bibinfo {author} {\bibfnamefont {K.}~\bibnamefont
  {{Heitmann}}}, \bibinfo {author} {\bibfnamefont {E.}~\bibnamefont
  {{Lawrence}}}, \bibinfo {author} {\bibfnamefont {J.}~\bibnamefont {{Kwan}}},
  \bibinfo {author} {\bibfnamefont {S.}~\bibnamefont {{Habib}}}, \ and\
  \bibinfo {author} {\bibfnamefont {D.}~\bibnamefont {{Higdon}}},\ }\href
  {\doibase 10.1088/0004-637X/780/1/111} {\bibfield  {journal} {\bibinfo
  {journal} {\apj}\ }\textbf {\bibinfo {volume} {780}},\ \bibinfo {eid} {111}
  (\bibinfo {year} {2014})},\ \Eprint {http://arxiv.org/abs/1304.7849}
  {arXiv:1304.7849 [astro-ph.CO]} \BibitemShut {NoStop}%
\bibitem [{\citenamefont {{Troxel}}\ and\ \citenamefont
  {{Ishak}}(2015)}]{Troxel20151}%
  \BibitemOpen
  \bibfield  {author} {\bibinfo {author} {\bibfnamefont {M.~A.}\ \bibnamefont
  {{Troxel}}}\ and\ \bibinfo {author} {\bibfnamefont {M.}~\bibnamefont
  {{Ishak}}},\ }\href {\doibase 10.1016/j.physrep.2014.11.001} {\bibfield
  {journal} {\bibinfo  {journal} {\physrep}\ }\textbf {\bibinfo {volume}
  {558}},\ \bibinfo {pages} {1} (\bibinfo {year} {2015})},\ \Eprint
  {http://arxiv.org/abs/1407.6990} {arXiv:1407.6990} \BibitemShut {NoStop}%
\bibitem [{\citenamefont {{Kirk}}\ \emph {et~al.}(2015)\citenamefont {{Kirk}},
  \citenamefont {{Brown}}, \citenamefont {{Hoekstra}}, \citenamefont
  {{Joachimi}}, \citenamefont {{Kitching}}, \citenamefont {{Mandelbaum}},
  \citenamefont {{Sif{\'o}n}}, \citenamefont {{Cacciato}}, \citenamefont
  {{Choi}}, \citenamefont {{Kiessling}}, \citenamefont {{Leonard}},
  \citenamefont {{Rassat}},\ and\ \citenamefont {{Sch{\"a}fer}}}]{kirk2015}%
  \BibitemOpen
  \bibfield  {author} {\bibinfo {author} {\bibfnamefont {D.}~\bibnamefont
  {{Kirk}}}, \bibinfo {author} {\bibfnamefont {M.~L.}\ \bibnamefont {{Brown}}},
  \bibinfo {author} {\bibfnamefont {H.}~\bibnamefont {{Hoekstra}}}, \bibinfo
  {author} {\bibfnamefont {B.}~\bibnamefont {{Joachimi}}}, \bibinfo {author}
  {\bibfnamefont {T.~D.}\ \bibnamefont {{Kitching}}}, \bibinfo {author}
  {\bibfnamefont {R.}~\bibnamefont {{Mandelbaum}}}, \bibinfo {author}
  {\bibfnamefont {C.}~\bibnamefont {{Sif{\'o}n}}}, \bibinfo {author}
  {\bibfnamefont {M.}~\bibnamefont {{Cacciato}}}, \bibinfo {author}
  {\bibfnamefont {A.}~\bibnamefont {{Choi}}}, \bibinfo {author} {\bibfnamefont
  {A.}~\bibnamefont {{Kiessling}}}, \bibinfo {author} {\bibfnamefont
  {A.}~\bibnamefont {{Leonard}}}, \bibinfo {author} {\bibfnamefont
  {A.}~\bibnamefont {{Rassat}}}, \ and\ \bibinfo {author} {\bibfnamefont
  {B.~M.}\ \bibnamefont {{Sch{\"a}fer}}},\ }\href {\doibase
  10.1007/s11214-015-0213-4} {\bibfield  {journal} {\bibinfo  {journal} {\ssr}\
  }\textbf {\bibinfo {volume} {193}},\ \bibinfo {pages} {139} (\bibinfo {year}
  {2015})},\ \Eprint {http://arxiv.org/abs/1504.05465} {arXiv:1504.05465}
  \BibitemShut {NoStop}%
\bibitem [{\citenamefont {{Jarvis}}\ \emph {et~al.}(2015)\citenamefont
  {{Jarvis}}, \citenamefont {{Sheldon}}, \citenamefont {{Zuntz}}, \citenamefont
  {{Kacprzak}}, \citenamefont {{Bridle}}, \citenamefont {{Amara}},
  \citenamefont {{Armstrong}}, \citenamefont {{Becker}}, \citenamefont
  {{Bernstein}}, \citenamefont {{Bonnett}}, \citenamefont {{Chang}},
  \citenamefont {{Das}}, \citenamefont {{Dietrich}}, \citenamefont
  {{Drlica-Wagner}}, \citenamefont {{Eifler}}, \citenamefont {{Gangkofner}},
  \citenamefont {{Gruen}}, \citenamefont {{Hirsch}}, \citenamefont {{Huff}},
  \citenamefont {{Jain}}, \citenamefont {{Kent}}, \citenamefont {{Kirk}},
  \citenamefont {{MacCrann}}, \citenamefont {{Melchior}}, \citenamefont
  {{Plazas}}, \citenamefont {{Refregier}}, \citenamefont {{Rowe}},
  \citenamefont {{Rykoff}}, \citenamefont {{Samuroff}}, \citenamefont
  {{S{\'a}nchez}}, \citenamefont {{Suchyta}}, \citenamefont {{Troxel}},
  \citenamefont {{Vikram}}, \citenamefont {{Abbott}}, \citenamefont
  {{Abdalla}}, \citenamefont {{Allam}}, \citenamefont {{Annis}}, \citenamefont
  {{Benoit-L{\'e}vy}}, \citenamefont {{Bertin}}, \citenamefont {{Brooks}},
  \citenamefont {{Buckley-Geer}}, \citenamefont {{Burke}}, \citenamefont
  {{Capozzi}}, \citenamefont {{Carnero Rosell}}, \citenamefont {{Carrasco
  Kind}}, \citenamefont {{Carretero}}, \citenamefont {{Castander}},
  \citenamefont {{Crocce}}, \citenamefont {{Cunha}}, \citenamefont
  {{D'Andrea}}, \citenamefont {{da Costa}}, \citenamefont {{DePoy}},
  \citenamefont {{Desai}}, \citenamefont {{Diehl}}, \citenamefont {{Doel}},
  \citenamefont {{Fausti Neto}}, \citenamefont {{Flaugher}}, \citenamefont
  {{Fosalba}}, \citenamefont {{Frieman}}, \citenamefont {{Gaztanaga}},
  \citenamefont {{Gerdes}}, \citenamefont {{Gruendl}}, \citenamefont
  {{Gutierrez}}, \citenamefont {{Honscheid}}, \citenamefont {{James}},
  \citenamefont {{Kuehn}}, \citenamefont {{Kuropatkin}}, \citenamefont
  {{Lahav}}, \citenamefont {{Li}}, \citenamefont {{Lima}}, \citenamefont
  {{March}}, \citenamefont {{Martini}}, \citenamefont {{Miquel}}, \citenamefont
  {{Mohr}}, \citenamefont {{Neilsen}}, \citenamefont {{Nord}}, \citenamefont
  {{Ogando}}, \citenamefont {{Reil}}, \citenamefont {{Romer}}, \citenamefont
  {{Roodman}}, \citenamefont {{Sako}}, \citenamefont {{Sanchez}}, \citenamefont
  {{Scarpine}}, \citenamefont {{Schubnell}}, \citenamefont {{Sevilla-Noarbe}},
  \citenamefont {{Smith}}, \citenamefont {{Soares-Santos}}, \citenamefont
  {{Sobreira}}, \citenamefont {{Swanson}}, \citenamefont {{Tarle}},
  \citenamefont {{Thaler}}, \citenamefont {{Thomas}}, \citenamefont
  {{Walker}},\ and\ \citenamefont {{Wechsler}}}]{jarvis2015}%
  \BibitemOpen
  \bibfield  {author} {\bibinfo {author} {\bibfnamefont {M.}~\bibnamefont
  {{Jarvis}}}, \bibinfo {author} {\bibfnamefont {E.}~\bibnamefont {{Sheldon}}},
  \bibinfo {author} {\bibfnamefont {J.}~\bibnamefont {{Zuntz}}}, \bibinfo
  {author} {\bibfnamefont {T.}~\bibnamefont {{Kacprzak}}}, \bibinfo {author}
  {\bibfnamefont {S.~L.}\ \bibnamefont {{Bridle}}}, \bibinfo {author}
  {\bibfnamefont {A.}~\bibnamefont {{Amara}}}, \bibinfo {author} {\bibfnamefont
  {R.}~\bibnamefont {{Armstrong}}}, \bibinfo {author} {\bibfnamefont {M.~R.}\
  \bibnamefont {{Becker}}}, \bibinfo {author} {\bibfnamefont {G.~M.}\
  \bibnamefont {{Bernstein}}}, \bibinfo {author} {\bibfnamefont
  {C.}~\bibnamefont {{Bonnett}}}, \bibinfo {author} {\bibfnamefont
  {C.}~\bibnamefont {{Chang}}}, \bibinfo {author} {\bibfnamefont
  {R.}~\bibnamefont {{Das}}}, \bibinfo {author} {\bibfnamefont {J.~P.}\
  \bibnamefont {{Dietrich}}}, \bibinfo {author} {\bibfnamefont
  {A.}~\bibnamefont {{Drlica-Wagner}}}, \bibinfo {author} {\bibfnamefont
  {T.~F.}\ \bibnamefont {{Eifler}}}, \bibinfo {author} {\bibfnamefont
  {C.}~\bibnamefont {{Gangkofner}}}, \bibinfo {author} {\bibfnamefont
  {D.}~\bibnamefont {{Gruen}}}, \bibinfo {author} {\bibfnamefont
  {M.}~\bibnamefont {{Hirsch}}}, \bibinfo {author} {\bibfnamefont {E.~M.}\
  \bibnamefont {{Huff}}}, \bibinfo {author} {\bibfnamefont {B.}~\bibnamefont
  {{Jain}}}, \bibinfo {author} {\bibfnamefont {S.}~\bibnamefont {{Kent}}},
  \bibinfo {author} {\bibfnamefont {D.}~\bibnamefont {{Kirk}}}, \bibinfo
  {author} {\bibfnamefont {N.}~\bibnamefont {{MacCrann}}}, \bibinfo {author}
  {\bibfnamefont {P.}~\bibnamefont {{Melchior}}}, \bibinfo {author}
  {\bibfnamefont {A.~A.}\ \bibnamefont {{Plazas}}}, \bibinfo {author}
  {\bibfnamefont {A.}~\bibnamefont {{Refregier}}}, \bibinfo {author}
  {\bibfnamefont {B.}~\bibnamefont {{Rowe}}}, \bibinfo {author} {\bibfnamefont
  {E.~S.}\ \bibnamefont {{Rykoff}}}, \bibinfo {author} {\bibfnamefont
  {S.}~\bibnamefont {{Samuroff}}}, \bibinfo {author} {\bibfnamefont
  {C.}~\bibnamefont {{S{\'a}nchez}}}, \bibinfo {author} {\bibfnamefont
  {E.}~\bibnamefont {{Suchyta}}}, \bibinfo {author} {\bibfnamefont {M.~A.}\
  \bibnamefont {{Troxel}}}, \bibinfo {author} {\bibfnamefont {V.}~\bibnamefont
  {{Vikram}}}, \bibinfo {author} {\bibfnamefont {T.}~\bibnamefont {{Abbott}}},
  \bibinfo {author} {\bibfnamefont {F.~B.}\ \bibnamefont {{Abdalla}}}, \bibinfo
  {author} {\bibfnamefont {S.}~\bibnamefont {{Allam}}}, \bibinfo {author}
  {\bibfnamefont {J.}~\bibnamefont {{Annis}}}, \bibinfo {author} {\bibfnamefont
  {A.}~\bibnamefont {{Benoit-L{\'e}vy}}}, \bibinfo {author} {\bibfnamefont
  {E.}~\bibnamefont {{Bertin}}}, \bibinfo {author} {\bibfnamefont
  {D.}~\bibnamefont {{Brooks}}}, \bibinfo {author} {\bibfnamefont
  {E.}~\bibnamefont {{Buckley-Geer}}}, \bibinfo {author} {\bibfnamefont
  {D.~L.}\ \bibnamefont {{Burke}}}, \bibinfo {author} {\bibfnamefont
  {D.}~\bibnamefont {{Capozzi}}}, \bibinfo {author} {\bibfnamefont
  {A.}~\bibnamefont {{Carnero Rosell}}}, \bibinfo {author} {\bibfnamefont
  {M.}~\bibnamefont {{Carrasco Kind}}}, \bibinfo {author} {\bibfnamefont
  {J.}~\bibnamefont {{Carretero}}}, \bibinfo {author} {\bibfnamefont {F.~J.}\
  \bibnamefont {{Castander}}}, \bibinfo {author} {\bibfnamefont
  {M.}~\bibnamefont {{Crocce}}}, \bibinfo {author} {\bibfnamefont {C.~E.}\
  \bibnamefont {{Cunha}}}, \bibinfo {author} {\bibfnamefont {C.~B.}\
  \bibnamefont {{D'Andrea}}}, \bibinfo {author} {\bibfnamefont {L.~N.}\
  \bibnamefont {{da Costa}}}, \bibinfo {author} {\bibfnamefont {D.~L.}\
  \bibnamefont {{DePoy}}}, \bibinfo {author} {\bibfnamefont {S.}~\bibnamefont
  {{Desai}}}, \bibinfo {author} {\bibfnamefont {H.~T.}\ \bibnamefont
  {{Diehl}}}, \bibinfo {author} {\bibfnamefont {P.}~\bibnamefont {{Doel}}},
  \bibinfo {author} {\bibfnamefont {A.}~\bibnamefont {{Fausti Neto}}}, \bibinfo
  {author} {\bibfnamefont {B.}~\bibnamefont {{Flaugher}}}, \bibinfo {author}
  {\bibfnamefont {P.}~\bibnamefont {{Fosalba}}}, \bibinfo {author}
  {\bibfnamefont {J.}~\bibnamefont {{Frieman}}}, \bibinfo {author}
  {\bibfnamefont {E.}~\bibnamefont {{Gaztanaga}}}, \bibinfo {author}
  {\bibfnamefont {D.~W.}\ \bibnamefont {{Gerdes}}}, \bibinfo {author}
  {\bibfnamefont {R.~A.}\ \bibnamefont {{Gruendl}}}, \bibinfo {author}
  {\bibfnamefont {G.}~\bibnamefont {{Gutierrez}}}, \bibinfo {author}
  {\bibfnamefont {K.}~\bibnamefont {{Honscheid}}}, \bibinfo {author}
  {\bibfnamefont {D.~J.}\ \bibnamefont {{James}}}, \bibinfo {author}
  {\bibfnamefont {K.}~\bibnamefont {{Kuehn}}}, \bibinfo {author} {\bibfnamefont
  {N.}~\bibnamefont {{Kuropatkin}}}, \bibinfo {author} {\bibfnamefont
  {O.}~\bibnamefont {{Lahav}}}, \bibinfo {author} {\bibfnamefont {T.~S.}\
  \bibnamefont {{Li}}}, \bibinfo {author} {\bibfnamefont {M.}~\bibnamefont
  {{Lima}}}, \bibinfo {author} {\bibfnamefont {M.}~\bibnamefont {{March}}},
  \bibinfo {author} {\bibfnamefont {P.}~\bibnamefont {{Martini}}}, \bibinfo
  {author} {\bibfnamefont {R.}~\bibnamefont {{Miquel}}}, \bibinfo {author}
  {\bibfnamefont {J.~J.}\ \bibnamefont {{Mohr}}}, \bibinfo {author}
  {\bibfnamefont {E.}~\bibnamefont {{Neilsen}}}, \bibinfo {author}
  {\bibfnamefont {B.}~\bibnamefont {{Nord}}}, \bibinfo {author} {\bibfnamefont
  {R.}~\bibnamefont {{Ogando}}}, \bibinfo {author} {\bibfnamefont
  {K.}~\bibnamefont {{Reil}}}, \bibinfo {author} {\bibfnamefont {A.~K.}\
  \bibnamefont {{Romer}}}, \bibinfo {author} {\bibfnamefont {A.}~\bibnamefont
  {{Roodman}}}, \bibinfo {author} {\bibfnamefont {M.}~\bibnamefont {{Sako}}},
  \bibinfo {author} {\bibfnamefont {E.}~\bibnamefont {{Sanchez}}}, \bibinfo
  {author} {\bibfnamefont {V.}~\bibnamefont {{Scarpine}}}, \bibinfo {author}
  {\bibfnamefont {M.}~\bibnamefont {{Schubnell}}}, \bibinfo {author}
  {\bibfnamefont {I.}~\bibnamefont {{Sevilla-Noarbe}}}, \bibinfo {author}
  {\bibfnamefont {R.~C.}\ \bibnamefont {{Smith}}}, \bibinfo {author}
  {\bibfnamefont {M.}~\bibnamefont {{Soares-Santos}}}, \bibinfo {author}
  {\bibfnamefont {F.}~\bibnamefont {{Sobreira}}}, \bibinfo {author}
  {\bibfnamefont {M.~E.~C.}\ \bibnamefont {{Swanson}}}, \bibinfo {author}
  {\bibfnamefont {G.}~\bibnamefont {{Tarle}}}, \bibinfo {author} {\bibfnamefont
  {J.}~\bibnamefont {{Thaler}}}, \bibinfo {author} {\bibfnamefont
  {D.}~\bibnamefont {{Thomas}}}, \bibinfo {author} {\bibfnamefont {A.~R.}\
  \bibnamefont {{Walker}}}, \ and\ \bibinfo {author} {\bibfnamefont {R.~H.}\
  \bibnamefont {{Wechsler}}},\ }\href@noop {} {\bibfield  {journal} {\bibinfo
  {journal} {arXiv:astro-ph/1507.05603}\ } (\bibinfo {year} {2015})},\ \Eprint
  {http://arxiv.org/abs/1507.05603} {arXiv:1507.05603 [astro-ph.IM]}
  \BibitemShut {NoStop}%
\bibitem [{\citenamefont {{Bonnett}}\ \emph {et~al.}(2015)\citenamefont
  {{Bonnett}}, \citenamefont {{Troxel}}, \citenamefont {{Hartley}},
  \citenamefont {{Amara}}, \citenamefont {{Leistedt}}, \citenamefont
  {{Becker}}, \citenamefont {{Bernstein}}, \citenamefont {{Bridle}},
  \citenamefont {{Bruderer}}, \citenamefont {{Busha}}, \citenamefont {{Carrasco
  Kind}}, \citenamefont {{Childress}}, \citenamefont {{Castander}},
  \citenamefont {{Chang}}, \citenamefont {{Crocce}}, \citenamefont {{Davis}},
  \citenamefont {{Eifler}}, \citenamefont {{Frieman}}, \citenamefont
  {{Gangkofner}}, \citenamefont {{Gaztanaga}}, \citenamefont {{Glazebrook}},
  \citenamefont {{Gruen}}, \citenamefont {{Kacprzak}}, \citenamefont {{King}},
  \citenamefont {{Kwan}}, \citenamefont {{Lahav}}, \citenamefont {{Lewis}},
  \citenamefont {{Lidman}}, \citenamefont {{Lin}}, \citenamefont {{MacCrann}},
  \citenamefont {{Miquel}}, \citenamefont {{O'Neill}}, \citenamefont
  {{Palmese}}, \citenamefont {{Peiris}}, \citenamefont {{Refregier}},
  \citenamefont {{Rozo}}, \citenamefont {{Rykoff}}, \citenamefont {{Sadeh}},
  \citenamefont {{S{\'a}nchez}}, \citenamefont {{Sheldon}}, \citenamefont
  {{Uddin}}, \citenamefont {{Wechsler}}, \citenamefont {{Zuntz}}, \citenamefont
  {{Abbott}}, \citenamefont {{Abdalla}}, \citenamefont {{Allam}}, \citenamefont
  {{Armstrong}}, \citenamefont {{Banerji}}, \citenamefont {{Bauer}},
  \citenamefont {{Benoit-L{\'e}vy}}, \citenamefont {{Bertin}}, \citenamefont
  {{Brooks}}, \citenamefont {{Buckley-Geer}}, \citenamefont {{Burke}},
  \citenamefont {{Capozzi}}, \citenamefont {{Carnero Rosell}}, \citenamefont
  {{Carretero}}, \citenamefont {{Cunha}}, \citenamefont {{D'Andrea}},
  \citenamefont {{da Costa}}, \citenamefont {{DePoy}}, \citenamefont {{Desai}},
  \citenamefont {{Diehl}}, \citenamefont {{Dietrich}}, \citenamefont {{Doel}},
  \citenamefont {{Fausti Neto}}, \citenamefont {{Fernandez}}, \citenamefont
  {{Flaugher}}, \citenamefont {{Fosalba}}, \citenamefont {{Gerdes}},
  \citenamefont {{Gruendl}}, \citenamefont {{Honscheid}}, \citenamefont
  {{Jain}}, \citenamefont {{James}}, \citenamefont {{Jarvis}}, \citenamefont
  {{Kim}}, \citenamefont {{Kuehn}}, \citenamefont {{Kuropatkin}}, \citenamefont
  {{Li}}, \citenamefont {{Lima}}, \citenamefont {{Maia}}, \citenamefont
  {{March}}, \citenamefont {{Marshall}}, \citenamefont {{Martini}},
  \citenamefont {{Melchior}}, \citenamefont {{Miller}}, \citenamefont
  {{Neilsen}}, \citenamefont {{Nichol}}, \citenamefont {{Nord}}, \citenamefont
  {{Ogando}}, \citenamefont {{Plazas}}, \citenamefont {{Reil}}, \citenamefont
  {{Romer}}, \citenamefont {{Roodman}}, \citenamefont {{Sako}}, \citenamefont
  {{Sanchez}}, \citenamefont {{Santiago}}, \citenamefont {{Smith}},
  \citenamefont {{Soares-Santos}}, \citenamefont {{Sobreira}}, \citenamefont
  {{Suchyta}}, \citenamefont {{Swanson}}, \citenamefont {{Tarle}},
  \citenamefont {{Thaler}}, \citenamefont {{Thomas}}, \citenamefont
  {{Vikram}},\ and\ \citenamefont {{Walker}}}]{bonnett2015b}%
  \BibitemOpen
  \bibfield  {author} {\bibinfo {author} {\bibfnamefont {C.}~\bibnamefont
  {{Bonnett}}}, \bibinfo {author} {\bibfnamefont {M.~A.}\ \bibnamefont
  {{Troxel}}}, \bibinfo {author} {\bibfnamefont {W.}~\bibnamefont {{Hartley}}},
  \bibinfo {author} {\bibfnamefont {A.}~\bibnamefont {{Amara}}}, \bibinfo
  {author} {\bibfnamefont {B.}~\bibnamefont {{Leistedt}}}, \bibinfo {author}
  {\bibfnamefont {M.~R.}\ \bibnamefont {{Becker}}}, \bibinfo {author}
  {\bibfnamefont {G.~M.}\ \bibnamefont {{Bernstein}}}, \bibinfo {author}
  {\bibfnamefont {S.}~\bibnamefont {{Bridle}}}, \bibinfo {author}
  {\bibfnamefont {C.}~\bibnamefont {{Bruderer}}}, \bibinfo {author}
  {\bibfnamefont {M.~T.}\ \bibnamefont {{Busha}}}, \bibinfo {author}
  {\bibfnamefont {M.}~\bibnamefont {{Carrasco Kind}}}, \bibinfo {author}
  {\bibfnamefont {M.~J.}\ \bibnamefont {{Childress}}}, \bibinfo {author}
  {\bibfnamefont {F.~J.}\ \bibnamefont {{Castander}}}, \bibinfo {author}
  {\bibfnamefont {C.}~\bibnamefont {{Chang}}}, \bibinfo {author} {\bibfnamefont
  {M.}~\bibnamefont {{Crocce}}}, \bibinfo {author} {\bibfnamefont {T.~M.}\
  \bibnamefont {{Davis}}}, \bibinfo {author} {\bibfnamefont {T.~F.}\
  \bibnamefont {{Eifler}}}, \bibinfo {author} {\bibfnamefont {J.}~\bibnamefont
  {{Frieman}}}, \bibinfo {author} {\bibfnamefont {C.}~\bibnamefont
  {{Gangkofner}}}, \bibinfo {author} {\bibfnamefont {E.}~\bibnamefont
  {{Gaztanaga}}}, \bibinfo {author} {\bibfnamefont {K.}~\bibnamefont
  {{Glazebrook}}}, \bibinfo {author} {\bibfnamefont {D.}~\bibnamefont
  {{Gruen}}}, \bibinfo {author} {\bibfnamefont {T.}~\bibnamefont {{Kacprzak}}},
  \bibinfo {author} {\bibfnamefont {A.}~\bibnamefont {{King}}}, \bibinfo
  {author} {\bibfnamefont {J.}~\bibnamefont {{Kwan}}}, \bibinfo {author}
  {\bibfnamefont {O.}~\bibnamefont {{Lahav}}}, \bibinfo {author} {\bibfnamefont
  {G.}~\bibnamefont {{Lewis}}}, \bibinfo {author} {\bibfnamefont
  {C.}~\bibnamefont {{Lidman}}}, \bibinfo {author} {\bibfnamefont
  {H.}~\bibnamefont {{Lin}}}, \bibinfo {author} {\bibfnamefont
  {N.}~\bibnamefont {{MacCrann}}}, \bibinfo {author} {\bibfnamefont
  {R.}~\bibnamefont {{Miquel}}}, \bibinfo {author} {\bibfnamefont {C.~R.}\
  \bibnamefont {{O'Neill}}}, \bibinfo {author} {\bibfnamefont {A.}~\bibnamefont
  {{Palmese}}}, \bibinfo {author} {\bibfnamefont {H.~V.}\ \bibnamefont
  {{Peiris}}}, \bibinfo {author} {\bibfnamefont {A.}~\bibnamefont
  {{Refregier}}}, \bibinfo {author} {\bibfnamefont {E.}~\bibnamefont {{Rozo}}},
  \bibinfo {author} {\bibfnamefont {E.~S.}\ \bibnamefont {{Rykoff}}}, \bibinfo
  {author} {\bibfnamefont {I.}~\bibnamefont {{Sadeh}}}, \bibinfo {author}
  {\bibfnamefont {C.}~\bibnamefont {{S{\'a}nchez}}}, \bibinfo {author}
  {\bibfnamefont {E.}~\bibnamefont {{Sheldon}}}, \bibinfo {author}
  {\bibfnamefont {S.}~\bibnamefont {{Uddin}}}, \bibinfo {author} {\bibfnamefont
  {R.~H.}\ \bibnamefont {{Wechsler}}}, \bibinfo {author} {\bibfnamefont
  {J.}~\bibnamefont {{Zuntz}}}, \bibinfo {author} {\bibfnamefont
  {T.}~\bibnamefont {{Abbott}}}, \bibinfo {author} {\bibfnamefont {F.~B.}\
  \bibnamefont {{Abdalla}}}, \bibinfo {author} {\bibfnamefont {S.}~\bibnamefont
  {{Allam}}}, \bibinfo {author} {\bibfnamefont {R.}~\bibnamefont
  {{Armstrong}}}, \bibinfo {author} {\bibfnamefont {M.}~\bibnamefont
  {{Banerji}}}, \bibinfo {author} {\bibfnamefont {A.~H.}\ \bibnamefont
  {{Bauer}}}, \bibinfo {author} {\bibfnamefont {A.}~\bibnamefont
  {{Benoit-L{\'e}vy}}}, \bibinfo {author} {\bibfnamefont {E.}~\bibnamefont
  {{Bertin}}}, \bibinfo {author} {\bibfnamefont {D.}~\bibnamefont {{Brooks}}},
  \bibinfo {author} {\bibfnamefont {E.}~\bibnamefont {{Buckley-Geer}}},
  \bibinfo {author} {\bibfnamefont {D.~L.}\ \bibnamefont {{Burke}}}, \bibinfo
  {author} {\bibfnamefont {D.}~\bibnamefont {{Capozzi}}}, \bibinfo {author}
  {\bibfnamefont {A.}~\bibnamefont {{Carnero Rosell}}}, \bibinfo {author}
  {\bibfnamefont {J.}~\bibnamefont {{Carretero}}}, \bibinfo {author}
  {\bibfnamefont {C.~E.}\ \bibnamefont {{Cunha}}}, \bibinfo {author}
  {\bibfnamefont {C.~B.}\ \bibnamefont {{D'Andrea}}}, \bibinfo {author}
  {\bibfnamefont {L.~N.}\ \bibnamefont {{da Costa}}}, \bibinfo {author}
  {\bibfnamefont {D.~L.}\ \bibnamefont {{DePoy}}}, \bibinfo {author}
  {\bibfnamefont {S.}~\bibnamefont {{Desai}}}, \bibinfo {author} {\bibfnamefont
  {H.~T.}\ \bibnamefont {{Diehl}}}, \bibinfo {author} {\bibfnamefont {J.~P.}\
  \bibnamefont {{Dietrich}}}, \bibinfo {author} {\bibfnamefont
  {P.}~\bibnamefont {{Doel}}}, \bibinfo {author} {\bibfnamefont
  {A.}~\bibnamefont {{Fausti Neto}}}, \bibinfo {author} {\bibfnamefont
  {E.}~\bibnamefont {{Fernandez}}}, \bibinfo {author} {\bibfnamefont
  {B.}~\bibnamefont {{Flaugher}}}, \bibinfo {author} {\bibfnamefont
  {P.}~\bibnamefont {{Fosalba}}}, \bibinfo {author} {\bibfnamefont {D.~W.}\
  \bibnamefont {{Gerdes}}}, \bibinfo {author} {\bibfnamefont {R.~A.}\
  \bibnamefont {{Gruendl}}}, \bibinfo {author} {\bibfnamefont {K.}~\bibnamefont
  {{Honscheid}}}, \bibinfo {author} {\bibfnamefont {B.}~\bibnamefont {{Jain}}},
  \bibinfo {author} {\bibfnamefont {D.~J.}\ \bibnamefont {{James}}}, \bibinfo
  {author} {\bibfnamefont {M.}~\bibnamefont {{Jarvis}}}, \bibinfo {author}
  {\bibfnamefont {A.~G.}\ \bibnamefont {{Kim}}}, \bibinfo {author}
  {\bibfnamefont {K.}~\bibnamefont {{Kuehn}}}, \bibinfo {author} {\bibfnamefont
  {N.}~\bibnamefont {{Kuropatkin}}}, \bibinfo {author} {\bibfnamefont {T.~S.}\
  \bibnamefont {{Li}}}, \bibinfo {author} {\bibfnamefont {M.}~\bibnamefont
  {{Lima}}}, \bibinfo {author} {\bibfnamefont {M.~A.~G.}\ \bibnamefont
  {{Maia}}}, \bibinfo {author} {\bibfnamefont {M.}~\bibnamefont {{March}}},
  \bibinfo {author} {\bibfnamefont {J.~L.}\ \bibnamefont {{Marshall}}},
  \bibinfo {author} {\bibfnamefont {P.}~\bibnamefont {{Martini}}}, \bibinfo
  {author} {\bibfnamefont {P.}~\bibnamefont {{Melchior}}}, \bibinfo {author}
  {\bibfnamefont {C.~J.}\ \bibnamefont {{Miller}}}, \bibinfo {author}
  {\bibfnamefont {E.}~\bibnamefont {{Neilsen}}}, \bibinfo {author}
  {\bibfnamefont {R.~C.}\ \bibnamefont {{Nichol}}}, \bibinfo {author}
  {\bibfnamefont {B.}~\bibnamefont {{Nord}}}, \bibinfo {author} {\bibfnamefont
  {R.}~\bibnamefont {{Ogando}}}, \bibinfo {author} {\bibfnamefont {A.~A.}\
  \bibnamefont {{Plazas}}}, \bibinfo {author} {\bibfnamefont {K.}~\bibnamefont
  {{Reil}}}, \bibinfo {author} {\bibfnamefont {A.~K.}\ \bibnamefont {{Romer}}},
  \bibinfo {author} {\bibfnamefont {A.}~\bibnamefont {{Roodman}}}, \bibinfo
  {author} {\bibfnamefont {M.}~\bibnamefont {{Sako}}}, \bibinfo {author}
  {\bibfnamefont {E.}~\bibnamefont {{Sanchez}}}, \bibinfo {author}
  {\bibfnamefont {B.}~\bibnamefont {{Santiago}}}, \bibinfo {author}
  {\bibfnamefont {R.~C.}\ \bibnamefont {{Smith}}}, \bibinfo {author}
  {\bibfnamefont {M.}~\bibnamefont {{Soares-Santos}}}, \bibinfo {author}
  {\bibfnamefont {F.}~\bibnamefont {{Sobreira}}}, \bibinfo {author}
  {\bibfnamefont {E.}~\bibnamefont {{Suchyta}}}, \bibinfo {author}
  {\bibfnamefont {M.~E.~C.}\ \bibnamefont {{Swanson}}}, \bibinfo {author}
  {\bibfnamefont {G.}~\bibnamefont {{Tarle}}}, \bibinfo {author} {\bibfnamefont
  {J.}~\bibnamefont {{Thaler}}}, \bibinfo {author} {\bibfnamefont
  {D.}~\bibnamefont {{Thomas}}}, \bibinfo {author} {\bibfnamefont
  {V.}~\bibnamefont {{Vikram}}}, \ and\ \bibinfo {author} {\bibfnamefont
  {A.~R.}\ \bibnamefont {{Walker}}},\ }\href@noop {} {\bibfield  {journal}
  {\bibinfo  {journal} {arXiv:astro-ph/1507.05909}\ } (\bibinfo {year}
  {2015})},\ \Eprint {http://arxiv.org/abs/1507.05909} {arXiv:1507.05909}
  \BibitemShut {NoStop}%
\bibitem [{\citenamefont {{Diehl}}(2012)}]{DiehlDECam2012}%
  \BibitemOpen
  \bibfield  {author} {\bibinfo {author} {\bibfnamefont {H.~T.}\ \bibnamefont
  {{Diehl}}},\ }in\ \href {\doibase
  http://dx.doi.org/10.1016/j.phpro.2012.02.472} {\emph {\bibinfo {booktitle}
  {Proceedings of the 2nd International Conference on Technology and
  Instrumentation in Particle Physics (TIPP 2011)}}},\ \bibinfo {series}
  {Physics Procedia}, Vol.~\bibinfo {volume} {37}\ (\bibinfo {year} {2012})\
  pp.\ \bibinfo {pages} {1332 -- 1340},\ \bibinfo {note} {proceedings of the
  2nd International Conference on Technology and Instrumentation in Particle
  Physics (TIPP 2011)}\BibitemShut {NoStop}%
\bibitem [{\citenamefont {{Flaugher}}\ \emph {et~al.}(2012)\citenamefont
  {{Flaugher}}, \citenamefont {{Abbott}}, \citenamefont {{Angstadt}},
  \citenamefont {{Annis}}, \citenamefont {{Antonik}}, \citenamefont {{Bailey}},
  \citenamefont {{Ballester}}, \citenamefont {{Bernstein}}, \citenamefont
  {{Bernstein}}, \citenamefont {{Bonati}}, \citenamefont {{Bremer}},
  \citenamefont {{Briones}}, \citenamefont {{Brooks}}, \citenamefont
  {{Buckley-Geer}}, \citenamefont {{Campa}}, \citenamefont {{Cardiel-Sas}},
  \citenamefont {{Castander}}, \citenamefont {{Castilla}}, \citenamefont
  {{Cease}}, \citenamefont {{Chappa}}, \citenamefont {{Chi}}, \citenamefont
  {{da Costa}}, \citenamefont {{DePoy}}, \citenamefont {{Derylo}},
  \citenamefont {{de Vincente}}, \citenamefont {{Diehl}}, \citenamefont
  {{Doel}}, \citenamefont {{Estrada}}, \citenamefont {{Eiting}}, \citenamefont
  {{Elliott}}, \citenamefont {{Finley}}, \citenamefont {{Flores}},
  \citenamefont {{Frieman}}, \citenamefont {{Gaztanaga}}, \citenamefont
  {{Gerdes}}, \citenamefont {{Gladders}}, \citenamefont {{Guarino}},
  \citenamefont {{Gutierrez}}, \citenamefont {{Grudzinski}}, \citenamefont
  {{Hanlon}}, \citenamefont {{Hao}}, \citenamefont {{Holland}}, \citenamefont
  {{Honscheid}}, \citenamefont {{Huffman}}, \citenamefont {{Jackson}},
  \citenamefont {{Jonas}}, \citenamefont {{Karliner}}, \citenamefont {{Kau}},
  \citenamefont {{Kent}}, \citenamefont {{Kozlovsky}}, \citenamefont
  {{Krempetz}}, \citenamefont {{Krider}}, \citenamefont {{Kubik}},
  \citenamefont {{Kuehn}}, \citenamefont {{Kuhlmann}}, \citenamefont {{Kuk}},
  \citenamefont {{Lahav}}, \citenamefont {{Langellier}}, \citenamefont
  {{Lathrop}}, \citenamefont {{Lewis}}, \citenamefont {{Lin}}, \citenamefont
  {{Lorenzon}}, \citenamefont {{Martinez}}, \citenamefont {{McKay}},
  \citenamefont {{Merritt}}, \citenamefont {{Meyer}}, \citenamefont {{Miquel}},
  \citenamefont {{Morgan}}, \citenamefont {{Moore}}, \citenamefont {{Moore}},
  \citenamefont {{Neilsen}}, \citenamefont {{Nord}}, \citenamefont {{Ogando}},
  \citenamefont {{Olson}}, \citenamefont {{Patton}}, \citenamefont {{Peoples}},
  \citenamefont {{Plazas}}, \citenamefont {{Qian}}, \citenamefont {{Roe}},
  \citenamefont {{Roodman}}, \citenamefont {{Rossetto}}, \citenamefont
  {{Sanchez}}, \citenamefont {{Soares-Santos}}, \citenamefont {{Scarpine}},
  \citenamefont {{Schalk}}, \citenamefont {{Schindler}}, \citenamefont
  {{Schmidt}}, \citenamefont {{Schmitt}}, \citenamefont {{Schubnell}},
  \citenamefont {{Schultz}}, \citenamefont {{Selen}}, \citenamefont
  {{Serrano}}, \citenamefont {{Shaw}}, \citenamefont {{Simaitis}},
  \citenamefont {{Slaughter}}, \citenamefont {{Smith}}, \citenamefont
  {{Spinka}}, \citenamefont {{Stefanik}}, \citenamefont {{Stuermer}},
  \citenamefont {{Sypniewski}}, \citenamefont {{Talaga}}, \citenamefont
  {{Tarle}}, \citenamefont {{Thaler}}, \citenamefont {{Tucker}}, \citenamefont
  {{Walker}}, \citenamefont {{Weaverdyck}}, \citenamefont {{Wester}},
  \citenamefont {{Woods}}, \citenamefont {{Worswick}},\ and\ \citenamefont
  {{Zhao}}}]{flaugher2012}%
  \BibitemOpen
  \bibfield  {author} {\bibinfo {author} {\bibfnamefont {B.~L.}\ \bibnamefont
  {{Flaugher}}}, \bibinfo {author} {\bibfnamefont {T.~M.~C.}\ \bibnamefont
  {{Abbott}}}, \bibinfo {author} {\bibfnamefont {R.}~\bibnamefont
  {{Angstadt}}}, \bibinfo {author} {\bibfnamefont {J.}~\bibnamefont {{Annis}}},
  \bibinfo {author} {\bibfnamefont {M.~L.}\ \bibnamefont {{Antonik}}}, \bibinfo
  {author} {\bibfnamefont {J.}~\bibnamefont {{Bailey}}}, \bibinfo {author}
  {\bibfnamefont {O.}~\bibnamefont {{Ballester}}}, \bibinfo {author}
  {\bibfnamefont {J.~P.}\ \bibnamefont {{Bernstein}}}, \bibinfo {author}
  {\bibfnamefont {R.~A.}\ \bibnamefont {{Bernstein}}}, \bibinfo {author}
  {\bibfnamefont {M.}~\bibnamefont {{Bonati}}}, \bibinfo {author}
  {\bibfnamefont {G.}~\bibnamefont {{Bremer}}}, \bibinfo {author}
  {\bibfnamefont {J.}~\bibnamefont {{Briones}}}, \bibinfo {author}
  {\bibfnamefont {D.}~\bibnamefont {{Brooks}}}, \bibinfo {author}
  {\bibfnamefont {E.~J.}\ \bibnamefont {{Buckley-Geer}}}, \bibinfo {author}
  {\bibfnamefont {J.}~\bibnamefont {{Campa}}}, \bibinfo {author} {\bibfnamefont
  {L.}~\bibnamefont {{Cardiel-Sas}}}, \bibinfo {author} {\bibfnamefont
  {F.}~\bibnamefont {{Castander}}}, \bibinfo {author} {\bibfnamefont
  {J.}~\bibnamefont {{Castilla}}}, \bibinfo {author} {\bibfnamefont
  {H.}~\bibnamefont {{Cease}}}, \bibinfo {author} {\bibfnamefont
  {S.}~\bibnamefont {{Chappa}}}, \bibinfo {author} {\bibfnamefont {E.~C.}\
  \bibnamefont {{Chi}}}, \bibinfo {author} {\bibfnamefont {L.}~\bibnamefont
  {{da Costa}}}, \bibinfo {author} {\bibfnamefont {D.~L.}\ \bibnamefont
  {{DePoy}}}, \bibinfo {author} {\bibfnamefont {G.}~\bibnamefont {{Derylo}}},
  \bibinfo {author} {\bibfnamefont {J.}~\bibnamefont {{de Vincente}}}, \bibinfo
  {author} {\bibfnamefont {H.~T.}\ \bibnamefont {{Diehl}}}, \bibinfo {author}
  {\bibfnamefont {P.}~\bibnamefont {{Doel}}}, \bibinfo {author} {\bibfnamefont
  {J.}~\bibnamefont {{Estrada}}}, \bibinfo {author} {\bibfnamefont
  {J.}~\bibnamefont {{Eiting}}}, \bibinfo {author} {\bibfnamefont {A.~E.}\
  \bibnamefont {{Elliott}}}, \bibinfo {author} {\bibfnamefont {D.~A.}\
  \bibnamefont {{Finley}}}, \bibinfo {author} {\bibfnamefont {R.}~\bibnamefont
  {{Flores}}}, \bibinfo {author} {\bibfnamefont {J.}~\bibnamefont {{Frieman}}},
  \bibinfo {author} {\bibfnamefont {E.}~\bibnamefont {{Gaztanaga}}}, \bibinfo
  {author} {\bibfnamefont {D.}~\bibnamefont {{Gerdes}}}, \bibinfo {author}
  {\bibfnamefont {M.}~\bibnamefont {{Gladders}}}, \bibinfo {author}
  {\bibfnamefont {V.}~\bibnamefont {{Guarino}}}, \bibinfo {author}
  {\bibfnamefont {G.}~\bibnamefont {{Gutierrez}}}, \bibinfo {author}
  {\bibfnamefont {J.}~\bibnamefont {{Grudzinski}}}, \bibinfo {author}
  {\bibfnamefont {B.}~\bibnamefont {{Hanlon}}}, \bibinfo {author}
  {\bibfnamefont {J.}~\bibnamefont {{Hao}}}, \bibinfo {author} {\bibfnamefont
  {S.}~\bibnamefont {{Holland}}}, \bibinfo {author} {\bibfnamefont
  {K.}~\bibnamefont {{Honscheid}}}, \bibinfo {author} {\bibfnamefont
  {D.}~\bibnamefont {{Huffman}}}, \bibinfo {author} {\bibfnamefont
  {C.}~\bibnamefont {{Jackson}}}, \bibinfo {author} {\bibfnamefont
  {M.}~\bibnamefont {{Jonas}}}, \bibinfo {author} {\bibfnamefont
  {I.}~\bibnamefont {{Karliner}}}, \bibinfo {author} {\bibfnamefont
  {D.}~\bibnamefont {{Kau}}}, \bibinfo {author} {\bibfnamefont
  {S.}~\bibnamefont {{Kent}}}, \bibinfo {author} {\bibfnamefont
  {M.}~\bibnamefont {{Kozlovsky}}}, \bibinfo {author} {\bibfnamefont
  {K.}~\bibnamefont {{Krempetz}}}, \bibinfo {author} {\bibfnamefont
  {J.}~\bibnamefont {{Krider}}}, \bibinfo {author} {\bibfnamefont
  {D.}~\bibnamefont {{Kubik}}}, \bibinfo {author} {\bibfnamefont
  {K.}~\bibnamefont {{Kuehn}}}, \bibinfo {author} {\bibfnamefont {S.~E.}\
  \bibnamefont {{Kuhlmann}}}, \bibinfo {author} {\bibfnamefont
  {K.}~\bibnamefont {{Kuk}}}, \bibinfo {author} {\bibfnamefont
  {O.}~\bibnamefont {{Lahav}}}, \bibinfo {author} {\bibfnamefont
  {N.}~\bibnamefont {{Langellier}}}, \bibinfo {author} {\bibfnamefont
  {A.}~\bibnamefont {{Lathrop}}}, \bibinfo {author} {\bibfnamefont {P.~M.}\
  \bibnamefont {{Lewis}}}, \bibinfo {author} {\bibfnamefont {H.}~\bibnamefont
  {{Lin}}}, \bibinfo {author} {\bibfnamefont {W.}~\bibnamefont {{Lorenzon}}},
  \bibinfo {author} {\bibfnamefont {G.}~\bibnamefont {{Martinez}}}, \bibinfo
  {author} {\bibfnamefont {T.}~\bibnamefont {{McKay}}}, \bibinfo {author}
  {\bibfnamefont {W.}~\bibnamefont {{Merritt}}}, \bibinfo {author}
  {\bibfnamefont {M.}~\bibnamefont {{Meyer}}}, \bibinfo {author} {\bibfnamefont
  {R.}~\bibnamefont {{Miquel}}}, \bibinfo {author} {\bibfnamefont
  {J.}~\bibnamefont {{Morgan}}}, \bibinfo {author} {\bibfnamefont
  {P.}~\bibnamefont {{Moore}}}, \bibinfo {author} {\bibfnamefont
  {T.}~\bibnamefont {{Moore}}}, \bibinfo {author} {\bibfnamefont
  {E.}~\bibnamefont {{Neilsen}}}, \bibinfo {author} {\bibfnamefont
  {B.}~\bibnamefont {{Nord}}}, \bibinfo {author} {\bibfnamefont
  {R.}~\bibnamefont {{Ogando}}}, \bibinfo {author} {\bibfnamefont
  {J.}~\bibnamefont {{Olson}}}, \bibinfo {author} {\bibfnamefont
  {K.}~\bibnamefont {{Patton}}}, \bibinfo {author} {\bibfnamefont
  {J.}~\bibnamefont {{Peoples}}}, \bibinfo {author} {\bibfnamefont
  {A.}~\bibnamefont {{Plazas}}}, \bibinfo {author} {\bibfnamefont
  {T.}~\bibnamefont {{Qian}}}, \bibinfo {author} {\bibfnamefont
  {N.}~\bibnamefont {{Roe}}}, \bibinfo {author} {\bibfnamefont
  {A.}~\bibnamefont {{Roodman}}}, \bibinfo {author} {\bibfnamefont
  {B.}~\bibnamefont {{Rossetto}}}, \bibinfo {author} {\bibfnamefont
  {E.}~\bibnamefont {{Sanchez}}}, \bibinfo {author} {\bibfnamefont
  {M.}~\bibnamefont {{Soares-Santos}}}, \bibinfo {author} {\bibfnamefont
  {V.}~\bibnamefont {{Scarpine}}}, \bibinfo {author} {\bibfnamefont
  {T.}~\bibnamefont {{Schalk}}}, \bibinfo {author} {\bibfnamefont
  {R.}~\bibnamefont {{Schindler}}}, \bibinfo {author} {\bibfnamefont
  {R.}~\bibnamefont {{Schmidt}}}, \bibinfo {author} {\bibfnamefont
  {R.}~\bibnamefont {{Schmitt}}}, \bibinfo {author} {\bibfnamefont
  {M.}~\bibnamefont {{Schubnell}}}, \bibinfo {author} {\bibfnamefont
  {K.}~\bibnamefont {{Schultz}}}, \bibinfo {author} {\bibfnamefont
  {M.}~\bibnamefont {{Selen}}}, \bibinfo {author} {\bibfnamefont
  {S.}~\bibnamefont {{Serrano}}}, \bibinfo {author} {\bibfnamefont
  {T.}~\bibnamefont {{Shaw}}}, \bibinfo {author} {\bibfnamefont
  {V.}~\bibnamefont {{Simaitis}}}, \bibinfo {author} {\bibfnamefont
  {J.}~\bibnamefont {{Slaughter}}}, \bibinfo {author} {\bibfnamefont {R.~C.}\
  \bibnamefont {{Smith}}}, \bibinfo {author} {\bibfnamefont {H.}~\bibnamefont
  {{Spinka}}}, \bibinfo {author} {\bibfnamefont {A.}~\bibnamefont
  {{Stefanik}}}, \bibinfo {author} {\bibfnamefont {W.}~\bibnamefont
  {{Stuermer}}}, \bibinfo {author} {\bibfnamefont {A.}~\bibnamefont
  {{Sypniewski}}}, \bibinfo {author} {\bibfnamefont {R.}~\bibnamefont
  {{Talaga}}}, \bibinfo {author} {\bibfnamefont {G.}~\bibnamefont {{Tarle}}},
  \bibinfo {author} {\bibfnamefont {J.}~\bibnamefont {{Thaler}}}, \bibinfo
  {author} {\bibfnamefont {D.}~\bibnamefont {{Tucker}}}, \bibinfo {author}
  {\bibfnamefont {A.~R.}\ \bibnamefont {{Walker}}}, \bibinfo {author}
  {\bibfnamefont {C.}~\bibnamefont {{Weaverdyck}}}, \bibinfo {author}
  {\bibfnamefont {W.}~\bibnamefont {{Wester}}}, \bibinfo {author}
  {\bibfnamefont {R.~J.}\ \bibnamefont {{Woods}}}, \bibinfo {author}
  {\bibfnamefont {S.}~\bibnamefont {{Worswick}}}, \ and\ \bibinfo {author}
  {\bibfnamefont {A.}~\bibnamefont {{Zhao}}},\ }in\ \href {\doibase
  10.1117/12.926216} {\emph {\bibinfo {booktitle} {Society of Photo-Optical
  Instrumentation Engineers (SPIE) Conference Series}}},\ \bibinfo {series}
  {Society of Photo-Optical Instrumentation Engineers (SPIE) Conference
  Series}, Vol.\ \bibinfo {volume} {8446}\ (\bibinfo {year} {2012})\
  p.~\bibinfo {pages} {11}\BibitemShut {NoStop}%
\bibitem [{\citenamefont {{Honscheid}}\ \emph {et~al.}(2012)\citenamefont
  {{Honscheid}}, \citenamefont {{Elliott}}, \citenamefont {{Annis}},
  \citenamefont {{Bonati}}, \citenamefont {{Buckley-Geer}}, \citenamefont
  {{Castander}}, \citenamefont {{daCosta}}, \citenamefont {{Fausti}},
  \citenamefont {{Karliner}}, \citenamefont {{Kuhlmann}}, \citenamefont
  {{Neilsen}}, \citenamefont {{Patton}}, \citenamefont {{Reil}}, \citenamefont
  {{Roodman}}, \citenamefont {{Thaler}}, \citenamefont {{Serrano}},
  \citenamefont {{Soares Santos}},\ and\ \citenamefont
  {{Suchyta}}}]{HonscheidControl2012}%
  \BibitemOpen
  \bibfield  {author} {\bibinfo {author} {\bibfnamefont {K.}~\bibnamefont
  {{Honscheid}}}, \bibinfo {author} {\bibfnamefont {A.}~\bibnamefont
  {{Elliott}}}, \bibinfo {author} {\bibfnamefont {J.}~\bibnamefont {{Annis}}},
  \bibinfo {author} {\bibfnamefont {M.}~\bibnamefont {{Bonati}}}, \bibinfo
  {author} {\bibfnamefont {E.}~\bibnamefont {{Buckley-Geer}}}, \bibinfo
  {author} {\bibfnamefont {F.}~\bibnamefont {{Castander}}}, \bibinfo {author}
  {\bibfnamefont {L.}~\bibnamefont {{daCosta}}}, \bibinfo {author}
  {\bibfnamefont {A.}~\bibnamefont {{Fausti}}}, \bibinfo {author}
  {\bibfnamefont {I.}~\bibnamefont {{Karliner}}}, \bibinfo {author}
  {\bibfnamefont {S.}~\bibnamefont {{Kuhlmann}}}, \bibinfo {author}
  {\bibfnamefont {E.}~\bibnamefont {{Neilsen}}}, \bibinfo {author}
  {\bibfnamefont {K.}~\bibnamefont {{Patton}}}, \bibinfo {author}
  {\bibfnamefont {K.}~\bibnamefont {{Reil}}}, \bibinfo {author} {\bibfnamefont
  {A.}~\bibnamefont {{Roodman}}}, \bibinfo {author} {\bibfnamefont
  {J.}~\bibnamefont {{Thaler}}}, \bibinfo {author} {\bibfnamefont
  {S.}~\bibnamefont {{Serrano}}}, \bibinfo {author} {\bibfnamefont
  {M.}~\bibnamefont {{Soares Santos}}}, \ and\ \bibinfo {author} {\bibfnamefont
  {E.}~\bibnamefont {{Suchyta}}},\ }in\ \href {\doibase 10.1117/12.925717}
  {\emph {\bibinfo {booktitle} {Society of Photo-Optical Instrumentation
  Engineers (SPIE) Conference Series}}},\ \bibinfo {series} {Society of
  Photo-Optical Instrumentation Engineers (SPIE) Conference Series}, Vol.\
  \bibinfo {volume} {8451}\ (\bibinfo {year} {2012})\ p.~\bibinfo {pages}
  {12}\BibitemShut {NoStop}%
\bibitem [{\citenamefont {{Flaugher}}\ \emph {et~al.}(2015)\citenamefont
  {{Flaugher}}, \citenamefont {{Diehl}}, \citenamefont {{Honscheid}},
  \citenamefont {{Abbott}}, \citenamefont {{Alvarez}}, \citenamefont
  {{Angstadt}}, \citenamefont {{Annis}}, \citenamefont {{Antonik}},
  \citenamefont {{Ballester}}, \citenamefont {{Beaufore}}, \citenamefont
  {{Bernstein}}, \citenamefont {{Bernstein}}, \citenamefont {{Bigelow}},
  \citenamefont {{Bonati}}, \citenamefont {{Boprie}}, \citenamefont {{Brooks}},
  \citenamefont {{Buckley-Geer}}, \citenamefont {{Campa}}, \citenamefont
  {{Cardiel-Sas}}, \citenamefont {{Castander}}, \citenamefont {{Castilla}},
  \citenamefont {{Cease}}, \citenamefont {{Cela-Ruiz}}, \citenamefont
  {{Chappa}}, \citenamefont {{Chi}}, \citenamefont {{Cooper}}, \citenamefont
  {{da Costa}}, \citenamefont {{Dede}}, \citenamefont {{Derylo}}, \citenamefont
  {{DePoy}}, \citenamefont {{de Vicente}}, \citenamefont {{Doel}},
  \citenamefont {{Drlica-Wagner}}, \citenamefont {{Eiting}}, \citenamefont
  {{Elliott}}, \citenamefont {{Emes}}, \citenamefont {{Estrada}}, \citenamefont
  {{Fausti Neto}}, \citenamefont {{Finley}}, \citenamefont {{Flores}},
  \citenamefont {{Frieman}}, \citenamefont {{Gerdes}}, \citenamefont
  {{Gladders}}, \citenamefont {{Gregory}}, \citenamefont {{Gutierrez}},
  \citenamefont {{Hao}}, \citenamefont {{Holland}}, \citenamefont {{Holm}},
  \citenamefont {{Huffman}}, \citenamefont {{Jackson}}, \citenamefont
  {{James}}, \citenamefont {{Jonas}}, \citenamefont {{Karcher}}, \citenamefont
  {{Karliner}}, \citenamefont {{Kent}}, \citenamefont {{Kessler}},
  \citenamefont {{Kozlovsky}}, \citenamefont {{Kron}}, \citenamefont {{Kubik}},
  \citenamefont {{Kuehn}}, \citenamefont {{Kuhlmann}}, \citenamefont {{Kuk}},
  \citenamefont {{Lahav}}, \citenamefont {{Lathrop}}, \citenamefont {{Lee}},
  \citenamefont {{Levi}}, \citenamefont {{Lewis}}, \citenamefont {{Li}},
  \citenamefont {{Mandrichenko}}, \citenamefont {{Marshall}}, \citenamefont
  {{Martinez}}, \citenamefont {{Merritt}}, \citenamefont {{Miquel}},
  \citenamefont {{Mu{\~n}oz}}, \citenamefont {{Neilsen}}, \citenamefont
  {{Nichol}}, \citenamefont {{Nord}}, \citenamefont {{Ogando}}, \citenamefont
  {{Olsen}}, \citenamefont {{Palaio}}, \citenamefont {{Patton}}, \citenamefont
  {{Peoples}}, \citenamefont {{Plazas}}, \citenamefont {{Rauch}}, \citenamefont
  {{Reil}}, \citenamefont {{Rheault}}, \citenamefont {{Roe}}, \citenamefont
  {{Rogers}}, \citenamefont {{Roodman}}, \citenamefont {{Sanchez}},
  \citenamefont {{Scarpine}}, \citenamefont {{Schindler}}, \citenamefont
  {{Schmidt}}, \citenamefont {{Schmitt}}, \citenamefont {{Schubnell}},
  \citenamefont {{Schultz}}, \citenamefont {{Schurter}}, \citenamefont
  {{Scott}}, \citenamefont {{Serrano}}, \citenamefont {{Shaw}}, \citenamefont
  {{Smith}}, \citenamefont {{Soares-Santos}}, \citenamefont {{Stefanik}},
  \citenamefont {{Stuermer}}, \citenamefont {{Suchyta}}, \citenamefont
  {{Sypniewski}}, \citenamefont {{Tarle}}, \citenamefont {{Thaler}},
  \citenamefont {{Tighe}}, \citenamefont {{Tran}}, \citenamefont {{Tucker}},
  \citenamefont {{Walker}}, \citenamefont {{Wang}}, \citenamefont {{Watson}},
  \citenamefont {{Weaverdyck}}, \citenamefont {{Wester}}, \citenamefont
  {{Woods}}, \citenamefont {{Yanny}},\ and\ \citenamefont {{DES
  Collaboration}}}]{flaugher2015}%
  \BibitemOpen
  \bibfield  {author} {\bibinfo {author} {\bibfnamefont {B.}~\bibnamefont
  {{Flaugher}}}, \bibinfo {author} {\bibfnamefont {H.~T.}\ \bibnamefont
  {{Diehl}}}, \bibinfo {author} {\bibfnamefont {K.}~\bibnamefont
  {{Honscheid}}}, \bibinfo {author} {\bibfnamefont {T.~M.~C.}\ \bibnamefont
  {{Abbott}}}, \bibinfo {author} {\bibfnamefont {O.}~\bibnamefont {{Alvarez}}},
  \bibinfo {author} {\bibfnamefont {R.}~\bibnamefont {{Angstadt}}}, \bibinfo
  {author} {\bibfnamefont {J.~T.}\ \bibnamefont {{Annis}}}, \bibinfo {author}
  {\bibfnamefont {M.}~\bibnamefont {{Antonik}}}, \bibinfo {author}
  {\bibfnamefont {O.}~\bibnamefont {{Ballester}}}, \bibinfo {author}
  {\bibfnamefont {L.}~\bibnamefont {{Beaufore}}}, \bibinfo {author}
  {\bibfnamefont {G.~M.}\ \bibnamefont {{Bernstein}}}, \bibinfo {author}
  {\bibfnamefont {R.~A.}\ \bibnamefont {{Bernstein}}}, \bibinfo {author}
  {\bibfnamefont {B.}~\bibnamefont {{Bigelow}}}, \bibinfo {author}
  {\bibfnamefont {M.}~\bibnamefont {{Bonati}}}, \bibinfo {author}
  {\bibfnamefont {D.}~\bibnamefont {{Boprie}}}, \bibinfo {author}
  {\bibfnamefont {D.}~\bibnamefont {{Brooks}}}, \bibinfo {author}
  {\bibfnamefont {E.~J.}\ \bibnamefont {{Buckley-Geer}}}, \bibinfo {author}
  {\bibfnamefont {J.}~\bibnamefont {{Campa}}}, \bibinfo {author} {\bibfnamefont
  {L.}~\bibnamefont {{Cardiel-Sas}}}, \bibinfo {author} {\bibfnamefont {F.~J.}\
  \bibnamefont {{Castander}}}, \bibinfo {author} {\bibfnamefont
  {J.}~\bibnamefont {{Castilla}}}, \bibinfo {author} {\bibfnamefont
  {H.}~\bibnamefont {{Cease}}}, \bibinfo {author} {\bibfnamefont {J.~M.}\
  \bibnamefont {{Cela-Ruiz}}}, \bibinfo {author} {\bibfnamefont
  {S.}~\bibnamefont {{Chappa}}}, \bibinfo {author} {\bibfnamefont
  {E.}~\bibnamefont {{Chi}}}, \bibinfo {author} {\bibfnamefont
  {C.}~\bibnamefont {{Cooper}}}, \bibinfo {author} {\bibfnamefont {L.~N.}\
  \bibnamefont {{da Costa}}}, \bibinfo {author} {\bibfnamefont
  {E.}~\bibnamefont {{Dede}}}, \bibinfo {author} {\bibfnamefont
  {G.}~\bibnamefont {{Derylo}}}, \bibinfo {author} {\bibfnamefont {D.~L.}\
  \bibnamefont {{DePoy}}}, \bibinfo {author} {\bibfnamefont {J.}~\bibnamefont
  {{de Vicente}}}, \bibinfo {author} {\bibfnamefont {P.}~\bibnamefont
  {{Doel}}}, \bibinfo {author} {\bibfnamefont {A.}~\bibnamefont
  {{Drlica-Wagner}}}, \bibinfo {author} {\bibfnamefont {J.}~\bibnamefont
  {{Eiting}}}, \bibinfo {author} {\bibfnamefont {A.~E.}\ \bibnamefont
  {{Elliott}}}, \bibinfo {author} {\bibfnamefont {J.}~\bibnamefont {{Emes}}},
  \bibinfo {author} {\bibfnamefont {J.}~\bibnamefont {{Estrada}}}, \bibinfo
  {author} {\bibfnamefont {A.}~\bibnamefont {{Fausti Neto}}}, \bibinfo {author}
  {\bibfnamefont {D.~A.}\ \bibnamefont {{Finley}}}, \bibinfo {author}
  {\bibfnamefont {R.}~\bibnamefont {{Flores}}}, \bibinfo {author}
  {\bibfnamefont {J.}~\bibnamefont {{Frieman}}}, \bibinfo {author}
  {\bibfnamefont {D.}~\bibnamefont {{Gerdes}}}, \bibinfo {author}
  {\bibfnamefont {M.~D.}\ \bibnamefont {{Gladders}}}, \bibinfo {author}
  {\bibfnamefont {B.}~\bibnamefont {{Gregory}}}, \bibinfo {author}
  {\bibfnamefont {G.~R.}\ \bibnamefont {{Gutierrez}}}, \bibinfo {author}
  {\bibfnamefont {J.}~\bibnamefont {{Hao}}}, \bibinfo {author} {\bibfnamefont
  {S.~E.}\ \bibnamefont {{Holland}}}, \bibinfo {author} {\bibfnamefont
  {S.}~\bibnamefont {{Holm}}}, \bibinfo {author} {\bibfnamefont
  {D.}~\bibnamefont {{Huffman}}}, \bibinfo {author} {\bibfnamefont
  {C.}~\bibnamefont {{Jackson}}}, \bibinfo {author} {\bibfnamefont {D.~J.}\
  \bibnamefont {{James}}}, \bibinfo {author} {\bibfnamefont {M.}~\bibnamefont
  {{Jonas}}}, \bibinfo {author} {\bibfnamefont {A.}~\bibnamefont {{Karcher}}},
  \bibinfo {author} {\bibfnamefont {I.}~\bibnamefont {{Karliner}}}, \bibinfo
  {author} {\bibfnamefont {S.}~\bibnamefont {{Kent}}}, \bibinfo {author}
  {\bibfnamefont {R.}~\bibnamefont {{Kessler}}}, \bibinfo {author}
  {\bibfnamefont {M.}~\bibnamefont {{Kozlovsky}}}, \bibinfo {author}
  {\bibfnamefont {R.~G.}\ \bibnamefont {{Kron}}}, \bibinfo {author}
  {\bibfnamefont {D.}~\bibnamefont {{Kubik}}}, \bibinfo {author} {\bibfnamefont
  {K.}~\bibnamefont {{Kuehn}}}, \bibinfo {author} {\bibfnamefont
  {S.}~\bibnamefont {{Kuhlmann}}}, \bibinfo {author} {\bibfnamefont
  {K.}~\bibnamefont {{Kuk}}}, \bibinfo {author} {\bibfnamefont
  {O.}~\bibnamefont {{Lahav}}}, \bibinfo {author} {\bibfnamefont
  {A.}~\bibnamefont {{Lathrop}}}, \bibinfo {author} {\bibfnamefont
  {J.}~\bibnamefont {{Lee}}}, \bibinfo {author} {\bibfnamefont {M.~E.}\
  \bibnamefont {{Levi}}}, \bibinfo {author} {\bibfnamefont {P.}~\bibnamefont
  {{Lewis}}}, \bibinfo {author} {\bibfnamefont {T.~S.}\ \bibnamefont {{Li}}},
  \bibinfo {author} {\bibfnamefont {I.}~\bibnamefont {{Mandrichenko}}},
  \bibinfo {author} {\bibfnamefont {J.~L.}\ \bibnamefont {{Marshall}}},
  \bibinfo {author} {\bibfnamefont {G.}~\bibnamefont {{Martinez}}}, \bibinfo
  {author} {\bibfnamefont {K.~W.}\ \bibnamefont {{Merritt}}}, \bibinfo {author}
  {\bibfnamefont {R.}~\bibnamefont {{Miquel}}}, \bibinfo {author}
  {\bibfnamefont {F.}~\bibnamefont {{Mu{\~n}oz}}}, \bibinfo {author}
  {\bibfnamefont {E.~H.}\ \bibnamefont {{Neilsen}}}, \bibinfo {author}
  {\bibfnamefont {R.~C.}\ \bibnamefont {{Nichol}}}, \bibinfo {author}
  {\bibfnamefont {B.}~\bibnamefont {{Nord}}}, \bibinfo {author} {\bibfnamefont
  {R.}~\bibnamefont {{Ogando}}}, \bibinfo {author} {\bibfnamefont
  {J.}~\bibnamefont {{Olsen}}}, \bibinfo {author} {\bibfnamefont
  {N.}~\bibnamefont {{Palaio}}}, \bibinfo {author} {\bibfnamefont
  {K.}~\bibnamefont {{Patton}}}, \bibinfo {author} {\bibfnamefont
  {J.}~\bibnamefont {{Peoples}}}, \bibinfo {author} {\bibfnamefont {A.~A.}\
  \bibnamefont {{Plazas}}}, \bibinfo {author} {\bibfnamefont {J.}~\bibnamefont
  {{Rauch}}}, \bibinfo {author} {\bibfnamefont {K.}~\bibnamefont {{Reil}}},
  \bibinfo {author} {\bibfnamefont {J.-P.}\ \bibnamefont {{Rheault}}}, \bibinfo
  {author} {\bibfnamefont {N.~A.}\ \bibnamefont {{Roe}}}, \bibinfo {author}
  {\bibfnamefont {H.}~\bibnamefont {{Rogers}}}, \bibinfo {author}
  {\bibfnamefont {A.}~\bibnamefont {{Roodman}}}, \bibinfo {author}
  {\bibfnamefont {E.}~\bibnamefont {{Sanchez}}}, \bibinfo {author}
  {\bibfnamefont {V.}~\bibnamefont {{Scarpine}}}, \bibinfo {author}
  {\bibfnamefont {R.~H.}\ \bibnamefont {{Schindler}}}, \bibinfo {author}
  {\bibfnamefont {R.}~\bibnamefont {{Schmidt}}}, \bibinfo {author}
  {\bibfnamefont {R.}~\bibnamefont {{Schmitt}}}, \bibinfo {author}
  {\bibfnamefont {M.}~\bibnamefont {{Schubnell}}}, \bibinfo {author}
  {\bibfnamefont {K.}~\bibnamefont {{Schultz}}}, \bibinfo {author}
  {\bibfnamefont {P.}~\bibnamefont {{Schurter}}}, \bibinfo {author}
  {\bibfnamefont {L.}~\bibnamefont {{Scott}}}, \bibinfo {author} {\bibfnamefont
  {S.}~\bibnamefont {{Serrano}}}, \bibinfo {author} {\bibfnamefont {T.~M.}\
  \bibnamefont {{Shaw}}}, \bibinfo {author} {\bibfnamefont {R.~C.}\
  \bibnamefont {{Smith}}}, \bibinfo {author} {\bibfnamefont {M.}~\bibnamefont
  {{Soares-Santos}}}, \bibinfo {author} {\bibfnamefont {A.}~\bibnamefont
  {{Stefanik}}}, \bibinfo {author} {\bibfnamefont {W.}~\bibnamefont
  {{Stuermer}}}, \bibinfo {author} {\bibfnamefont {E.}~\bibnamefont
  {{Suchyta}}}, \bibinfo {author} {\bibfnamefont {A.}~\bibnamefont
  {{Sypniewski}}}, \bibinfo {author} {\bibfnamefont {G.}~\bibnamefont
  {{Tarle}}}, \bibinfo {author} {\bibfnamefont {J.}~\bibnamefont {{Thaler}}},
  \bibinfo {author} {\bibfnamefont {R.}~\bibnamefont {{Tighe}}}, \bibinfo
  {author} {\bibfnamefont {C.}~\bibnamefont {{Tran}}}, \bibinfo {author}
  {\bibfnamefont {D.}~\bibnamefont {{Tucker}}}, \bibinfo {author}
  {\bibfnamefont {A.~R.}\ \bibnamefont {{Walker}}}, \bibinfo {author}
  {\bibfnamefont {G.}~\bibnamefont {{Wang}}}, \bibinfo {author} {\bibfnamefont
  {M.}~\bibnamefont {{Watson}}}, \bibinfo {author} {\bibfnamefont
  {C.}~\bibnamefont {{Weaverdyck}}}, \bibinfo {author} {\bibfnamefont
  {W.}~\bibnamefont {{Wester}}}, \bibinfo {author} {\bibfnamefont
  {R.}~\bibnamefont {{Woods}}}, \bibinfo {author} {\bibfnamefont
  {B.}~\bibnamefont {{Yanny}}}, \ and\ \bibinfo {author} {\bibnamefont {{DES
  Collaboration}}},\ }\href {\doibase 10.1088/0004-6256/150/5/150} {\bibfield
  {journal} {\bibinfo  {journal} {\aj}\ }\textbf {\bibinfo {volume} {150}},\
  \bibinfo {eid} {150} (\bibinfo {year} {2015})},\ \Eprint
  {http://arxiv.org/abs/1504.02900} {arXiv:1504.02900 [astro-ph.IM]}
  \BibitemShut {NoStop}%
\bibitem [{\citenamefont {{Desai}}\ \emph {et~al.}(2012)\citenamefont
  {{Desai}}, \citenamefont {{Armstrong}}, \citenamefont {{Mohr}}, \citenamefont
  {{Semler}}, \citenamefont {{Liu}}, \citenamefont {{Bertin}}, \citenamefont
  {{Allam}}, \citenamefont {{Barkhouse}}, \citenamefont {{Bazin}},
  \citenamefont {{Buckley-Geer}}, \citenamefont {{Cooper}}, \citenamefont
  {{Hansen}}, \citenamefont {{High}}, \citenamefont {{Lin}}, \citenamefont
  {{Lin}}, \citenamefont {{Ngeow}}, \citenamefont {{Rest}}, \citenamefont
  {{Song}}, \citenamefont {{Tucker}},\ and\ \citenamefont
  {{Zenteno}}}]{desai2012}%
  \BibitemOpen
  \bibfield  {author} {\bibinfo {author} {\bibfnamefont {S.}~\bibnamefont
  {{Desai}}}, \bibinfo {author} {\bibfnamefont {R.}~\bibnamefont
  {{Armstrong}}}, \bibinfo {author} {\bibfnamefont {J.~J.}\ \bibnamefont
  {{Mohr}}}, \bibinfo {author} {\bibfnamefont {D.~R.}\ \bibnamefont
  {{Semler}}}, \bibinfo {author} {\bibfnamefont {J.}~\bibnamefont {{Liu}}},
  \bibinfo {author} {\bibfnamefont {E.}~\bibnamefont {{Bertin}}}, \bibinfo
  {author} {\bibfnamefont {S.~S.}\ \bibnamefont {{Allam}}}, \bibinfo {author}
  {\bibfnamefont {W.~A.}\ \bibnamefont {{Barkhouse}}}, \bibinfo {author}
  {\bibfnamefont {G.}~\bibnamefont {{Bazin}}}, \bibinfo {author} {\bibfnamefont
  {E.~J.}\ \bibnamefont {{Buckley-Geer}}}, \bibinfo {author} {\bibfnamefont
  {M.~C.}\ \bibnamefont {{Cooper}}}, \bibinfo {author} {\bibfnamefont {S.~M.}\
  \bibnamefont {{Hansen}}}, \bibinfo {author} {\bibfnamefont {F.~W.}\
  \bibnamefont {{High}}}, \bibinfo {author} {\bibfnamefont {H.}~\bibnamefont
  {{Lin}}}, \bibinfo {author} {\bibfnamefont {Y.-T.}\ \bibnamefont {{Lin}}},
  \bibinfo {author} {\bibfnamefont {C.-C.}\ \bibnamefont {{Ngeow}}}, \bibinfo
  {author} {\bibfnamefont {A.}~\bibnamefont {{Rest}}}, \bibinfo {author}
  {\bibfnamefont {J.}~\bibnamefont {{Song}}}, \bibinfo {author} {\bibfnamefont
  {D.}~\bibnamefont {{Tucker}}}, \ and\ \bibinfo {author} {\bibfnamefont
  {A.}~\bibnamefont {{Zenteno}}},\ }\href {\doibase 10.1088/0004-637X/757/1/83}
  {\bibfield  {journal} {\bibinfo  {journal} {\apj}\ }\textbf {\bibinfo
  {volume} {757}},\ \bibinfo {eid} {83} (\bibinfo {year} {2012})},\ \Eprint
  {http://arxiv.org/abs/1204.1210} {arXiv:1204.1210 [astro-ph.CO]} \BibitemShut
  {NoStop}%
\bibitem [{\citenamefont {{Schneider}}\ and\ \citenamefont
  {{Seitz}}(1995)}]{schneider1995}%
  \BibitemOpen
  \bibfield  {author} {\bibinfo {author} {\bibfnamefont {P.}~\bibnamefont
  {{Schneider}}}\ and\ \bibinfo {author} {\bibfnamefont {C.}~\bibnamefont
  {{Seitz}}},\ }\href@noop {} {\bibfield  {journal} {\bibinfo  {journal}
  {\aap}\ }\textbf {\bibinfo {volume} {294}},\ \bibinfo {pages} {411} (\bibinfo
  {year} {1995})},\ \Eprint {http://arxiv.org/abs/astro-ph/9407032}
  {astro-ph/9407032} \BibitemShut {NoStop}%
\bibitem [{\citenamefont {{Sheldon}}(2014)}]{sheldon2014}%
  \BibitemOpen
  \bibfield  {author} {\bibinfo {author} {\bibfnamefont {E.~S.}\ \bibnamefont
  {{Sheldon}}},\ }\href {\doibase 10.1093/mnrasl/slu104} {\bibfield  {journal}
  {\bibinfo  {journal} {\mnras}\ }\textbf {\bibinfo {volume} {444}},\ \bibinfo
  {pages} {L25} (\bibinfo {year} {2014})},\ \Eprint
  {http://arxiv.org/abs/1403.7669} {arXiv:1403.7669} \BibitemShut {NoStop}%
\bibitem [{\citenamefont {{Hogg}}\ and\ \citenamefont
  {{Lang}}(2013)}]{hogg2013}%
  \BibitemOpen
  \bibfield  {author} {\bibinfo {author} {\bibfnamefont {D.~W.}\ \bibnamefont
  {{Hogg}}}\ and\ \bibinfo {author} {\bibfnamefont {D.}~\bibnamefont
  {{Lang}}},\ }\href {\doibase 10.1086/671228} {\bibfield  {journal} {\bibinfo
  {journal} {\pasp}\ }\textbf {\bibinfo {volume} {125}},\ \bibinfo {pages}
  {719} (\bibinfo {year} {2013})},\ \Eprint {http://arxiv.org/abs/1210.6563}
  {arXiv:1210.6563 [astro-ph.IM]} \BibitemShut {NoStop}%
\bibitem [{\citenamefont {{Miller}}\ \emph {et~al.}(2007)\citenamefont
  {{Miller}}, \citenamefont {{Kitching}}, \citenamefont {{Heymans}},
  \citenamefont {{Heavens}},\ and\ \citenamefont {{van
  Waerbeke}}}]{miller2007}%
  \BibitemOpen
  \bibfield  {author} {\bibinfo {author} {\bibfnamefont {L.}~\bibnamefont
  {{Miller}}}, \bibinfo {author} {\bibfnamefont {T.~D.}\ \bibnamefont
  {{Kitching}}}, \bibinfo {author} {\bibfnamefont {C.}~\bibnamefont
  {{Heymans}}}, \bibinfo {author} {\bibfnamefont {A.~F.}\ \bibnamefont
  {{Heavens}}}, \ and\ \bibinfo {author} {\bibfnamefont {L.}~\bibnamefont {{van
  Waerbeke}}},\ }\href {\doibase 10.1111/j.1365-2966.2007.12363.x} {\bibfield
  {journal} {\bibinfo  {journal} {\mnras}\ }\textbf {\bibinfo {volume} {382}},\
  \bibinfo {pages} {315} (\bibinfo {year} {2007})},\ \Eprint
  {http://arxiv.org/abs/0708.2340} {arXiv:0708.2340} \BibitemShut {NoStop}%
\bibitem [{\citenamefont {{Mandelbaum}}\ \emph {et~al.}(2014)\citenamefont
  {{Mandelbaum}}, \citenamefont {{Rowe}}, \citenamefont {{Bosch}},
  \citenamefont {{Chang}}, \citenamefont {{Courbin}}, \citenamefont {{Gill}},
  \citenamefont {{Jarvis}}, \citenamefont {{Kannawadi}}, \citenamefont
  {{Kacprzak}}, \citenamefont {{Lackner}}, \citenamefont {{Leauthaud}},
  \citenamefont {{Miyatake}}, \citenamefont {{Nakajima}}, \citenamefont
  {{Rhodes}}, \citenamefont {{Simet}}, \citenamefont {{Zuntz}}, \citenamefont
  {{Armstrong}}, \citenamefont {{Bridle}}, \citenamefont {{Coupon}},
  \citenamefont {{Dietrich}}, \citenamefont {{Gentile}}, \citenamefont
  {{Heymans}}, \citenamefont {{Jurling}}, \citenamefont {{Kent}}, \citenamefont
  {{Kirkby}}, \citenamefont {{Margala}}, \citenamefont {{Massey}},
  \citenamefont {{Melchior}}, \citenamefont {{Peterson}}, \citenamefont
  {{Roodman}},\ and\ \citenamefont {{Schrabback}}}]{great3hb}%
  \BibitemOpen
  \bibfield  {author} {\bibinfo {author} {\bibfnamefont {R.}~\bibnamefont
  {{Mandelbaum}}}, \bibinfo {author} {\bibfnamefont {B.}~\bibnamefont
  {{Rowe}}}, \bibinfo {author} {\bibfnamefont {J.}~\bibnamefont {{Bosch}}},
  \bibinfo {author} {\bibfnamefont {C.}~\bibnamefont {{Chang}}}, \bibinfo
  {author} {\bibfnamefont {F.}~\bibnamefont {{Courbin}}}, \bibinfo {author}
  {\bibfnamefont {M.}~\bibnamefont {{Gill}}}, \bibinfo {author} {\bibfnamefont
  {M.}~\bibnamefont {{Jarvis}}}, \bibinfo {author} {\bibfnamefont
  {A.}~\bibnamefont {{Kannawadi}}}, \bibinfo {author} {\bibfnamefont
  {T.}~\bibnamefont {{Kacprzak}}}, \bibinfo {author} {\bibfnamefont
  {C.}~\bibnamefont {{Lackner}}}, \bibinfo {author} {\bibfnamefont
  {A.}~\bibnamefont {{Leauthaud}}}, \bibinfo {author} {\bibfnamefont
  {H.}~\bibnamefont {{Miyatake}}}, \bibinfo {author} {\bibfnamefont
  {R.}~\bibnamefont {{Nakajima}}}, \bibinfo {author} {\bibfnamefont
  {J.}~\bibnamefont {{Rhodes}}}, \bibinfo {author} {\bibfnamefont
  {M.}~\bibnamefont {{Simet}}}, \bibinfo {author} {\bibfnamefont
  {J.}~\bibnamefont {{Zuntz}}}, \bibinfo {author} {\bibfnamefont
  {B.}~\bibnamefont {{Armstrong}}}, \bibinfo {author} {\bibfnamefont
  {S.}~\bibnamefont {{Bridle}}}, \bibinfo {author} {\bibfnamefont
  {J.}~\bibnamefont {{Coupon}}}, \bibinfo {author} {\bibfnamefont {J.~P.}\
  \bibnamefont {{Dietrich}}}, \bibinfo {author} {\bibfnamefont
  {M.}~\bibnamefont {{Gentile}}}, \bibinfo {author} {\bibfnamefont
  {C.}~\bibnamefont {{Heymans}}}, \bibinfo {author} {\bibfnamefont {A.~S.}\
  \bibnamefont {{Jurling}}}, \bibinfo {author} {\bibfnamefont {S.~M.}\
  \bibnamefont {{Kent}}}, \bibinfo {author} {\bibfnamefont {D.}~\bibnamefont
  {{Kirkby}}}, \bibinfo {author} {\bibfnamefont {D.}~\bibnamefont {{Margala}}},
  \bibinfo {author} {\bibfnamefont {R.}~\bibnamefont {{Massey}}}, \bibinfo
  {author} {\bibfnamefont {P.}~\bibnamefont {{Melchior}}}, \bibinfo {author}
  {\bibfnamefont {J.}~\bibnamefont {{Peterson}}}, \bibinfo {author}
  {\bibfnamefont {A.}~\bibnamefont {{Roodman}}}, \ and\ \bibinfo {author}
  {\bibfnamefont {T.}~\bibnamefont {{Schrabback}}},\ }\href {\doibase
  10.1088/0067-0049/212/1/5} {\bibfield  {journal} {\bibinfo  {journal}
  {\apjs}\ }\textbf {\bibinfo {volume} {212}},\ \bibinfo {eid} {5} (\bibinfo
  {year} {2014})},\ \Eprint {http://arxiv.org/abs/1308.4982} {arXiv:1308.4982}
  \BibitemShut {NoStop}%
\bibitem [{\citenamefont {{Zuntz}}\ \emph {et~al.}(2013)\citenamefont
  {{Zuntz}}, \citenamefont {{Kacprzak}}, \citenamefont {{Voigt}}, \citenamefont
  {{Hirsch}}, \citenamefont {{Rowe}},\ and\ \citenamefont
  {{Bridle}}}]{zuntz2013}%
  \BibitemOpen
  \bibfield  {author} {\bibinfo {author} {\bibfnamefont {J.}~\bibnamefont
  {{Zuntz}}}, \bibinfo {author} {\bibfnamefont {T.}~\bibnamefont {{Kacprzak}}},
  \bibinfo {author} {\bibfnamefont {L.}~\bibnamefont {{Voigt}}}, \bibinfo
  {author} {\bibfnamefont {M.}~\bibnamefont {{Hirsch}}}, \bibinfo {author}
  {\bibfnamefont {B.}~\bibnamefont {{Rowe}}}, \ and\ \bibinfo {author}
  {\bibfnamefont {S.}~\bibnamefont {{Bridle}}},\ }\href {\doibase
  10.1093/mnras/stt1125} {\bibfield  {journal} {\bibinfo  {journal} {\mnras}\
  }\textbf {\bibinfo {volume} {434}},\ \bibinfo {pages} {1604} (\bibinfo {year}
  {2013})},\ \Eprint {http://arxiv.org/abs/1302.0183} {arXiv:1302.0183
  [astro-ph.CO]} \BibitemShut {NoStop}%
\bibitem [{\citenamefont {{Refregier}}\ \emph {et~al.}(2012)\citenamefont
  {{Refregier}}, \citenamefont {{Kacprzak}}, \citenamefont {{Amara}},
  \citenamefont {{Bridle}},\ and\ \citenamefont {{Rowe}}}]{refregier2012}%
  \BibitemOpen
  \bibfield  {author} {\bibinfo {author} {\bibfnamefont {A.}~\bibnamefont
  {{Refregier}}}, \bibinfo {author} {\bibfnamefont {T.}~\bibnamefont
  {{Kacprzak}}}, \bibinfo {author} {\bibfnamefont {A.}~\bibnamefont {{Amara}}},
  \bibinfo {author} {\bibfnamefont {S.}~\bibnamefont {{Bridle}}}, \ and\
  \bibinfo {author} {\bibfnamefont {B.}~\bibnamefont {{Rowe}}},\ }\href
  {\doibase 10.1111/j.1365-2966.2012.21483.x} {\bibfield  {journal} {\bibinfo
  {journal} {\mnras}\ }\textbf {\bibinfo {volume} {425}},\ \bibinfo {pages}
  {1951} (\bibinfo {year} {2012})},\ \Eprint {http://arxiv.org/abs/1203.5050}
  {arXiv:1203.5050} \BibitemShut {NoStop}%
\bibitem [{\citenamefont {{Kacprzak}}\ \emph {et~al.}(2012)\citenamefont
  {{Kacprzak}}, \citenamefont {{Zuntz}}, \citenamefont {{Rowe}}, \citenamefont
  {{Bridle}}, \citenamefont {{Refregier}}, \citenamefont {{Amara}},
  \citenamefont {{Voigt}},\ and\ \citenamefont {{Hirsch}}}]{kacprzak2013}%
  \BibitemOpen
  \bibfield  {author} {\bibinfo {author} {\bibfnamefont {T.}~\bibnamefont
  {{Kacprzak}}}, \bibinfo {author} {\bibfnamefont {J.}~\bibnamefont {{Zuntz}}},
  \bibinfo {author} {\bibfnamefont {B.}~\bibnamefont {{Rowe}}}, \bibinfo
  {author} {\bibfnamefont {S.}~\bibnamefont {{Bridle}}}, \bibinfo {author}
  {\bibfnamefont {A.}~\bibnamefont {{Refregier}}}, \bibinfo {author}
  {\bibfnamefont {A.}~\bibnamefont {{Amara}}}, \bibinfo {author} {\bibfnamefont
  {L.}~\bibnamefont {{Voigt}}}, \ and\ \bibinfo {author} {\bibfnamefont
  {M.}~\bibnamefont {{Hirsch}}},\ }\href {\doibase
  10.1111/j.1365-2966.2012.21622.x} {\bibfield  {journal} {\bibinfo  {journal}
  {\mnras}\ }\textbf {\bibinfo {volume} {427}},\ \bibinfo {pages} {2711}
  (\bibinfo {year} {2012})},\ \Eprint {http://arxiv.org/abs/1203.5049}
  {arXiv:1203.5049 [astro-ph.CO]} \BibitemShut {NoStop}%
\bibitem [{\citenamefont {{S{\'a}nchez}}\ \emph {et~al.}(2014)\citenamefont
  {{S{\'a}nchez}}, \citenamefont {{Carrasco Kind}}, \citenamefont {{Lin}},
  \citenamefont {{Miquel}}, \citenamefont {{Abdalla}}, \citenamefont {{Amara}},
  \citenamefont {{Banerji}}, \citenamefont {{Bonnett}}, \citenamefont
  {{Brunner}}, \citenamefont {{Capozzi}}, \citenamefont {{Carnero}},
  \citenamefont {{Castander}}, \citenamefont {{da Costa}}, \citenamefont
  {{Cunha}}, \citenamefont {{Fausti}}, \citenamefont {{Gerdes}}, \citenamefont
  {{Greisel}}, \citenamefont {{Gschwend}}, \citenamefont {{Hartley}},
  \citenamefont {{Jouvel}}, \citenamefont {{Lahav}}, \citenamefont {{Lima}},
  \citenamefont {{Maia}}, \citenamefont {{Mart{\'{\i}}}}, \citenamefont
  {{Ogando}}, \citenamefont {{Ostrovski}}, \citenamefont {{Pellegrini}},
  \citenamefont {{Rau}}, \citenamefont {{Sadeh}}, \citenamefont {{Seitz}},
  \citenamefont {{Sevilla-Noarbe}}, \citenamefont {{Sypniewski}}, \citenamefont
  {{de Vicente}}, \citenamefont {{Abbot}}, \citenamefont {{Allam}},
  \citenamefont {{Atlee}}, \citenamefont {{Bernstein}}, \citenamefont
  {{Bernstein}}, \citenamefont {{Buckley-Geer}}, \citenamefont {{Burke}},
  \citenamefont {{Childress}}, \citenamefont {{Davis}}, \citenamefont
  {{DePoy}}, \citenamefont {{Dey}}, \citenamefont {{Desai}}, \citenamefont
  {{Diehl}}, \citenamefont {{Doel}}, \citenamefont {{Estrada}}, \citenamefont
  {{Evrard}}, \citenamefont {{Fern{\'a}ndez}}, \citenamefont {{Finley}},
  \citenamefont {{Flaugher}}, \citenamefont {{Frieman}}, \citenamefont
  {{Gaztanaga}}, \citenamefont {{Glazebrook}}, \citenamefont {{Honscheid}},
  \citenamefont {{Kim}}, \citenamefont {{Kuehn}}, \citenamefont {{Kuropatkin}},
  \citenamefont {{Lidman}}, \citenamefont {{Makler}}, \citenamefont
  {{Marshall}}, \citenamefont {{Nichol}}, \citenamefont {{Roodman}},
  \citenamefont {{S{\'a}nchez}}, \citenamefont {{Santiago}}, \citenamefont
  {{Sako}}, \citenamefont {{Scalzo}}, \citenamefont {{Smith}}, \citenamefont
  {{Swanson}}, \citenamefont {{Tarle}}, \citenamefont {{Thomas}}, \citenamefont
  {{Tucker}}, \citenamefont {{Uddin}}, \citenamefont {{Vald{\'e}s}},
  \citenamefont {{Walker}}, \citenamefont {{Yuan}},\ and\ \citenamefont
  {{Zuntz}}}]{sanchez2014}%
  \BibitemOpen
  \bibfield  {author} {\bibinfo {author} {\bibfnamefont {C.}~\bibnamefont
  {{S{\'a}nchez}}}, \bibinfo {author} {\bibfnamefont {M.}~\bibnamefont
  {{Carrasco Kind}}}, \bibinfo {author} {\bibfnamefont {H.}~\bibnamefont
  {{Lin}}}, \bibinfo {author} {\bibfnamefont {R.}~\bibnamefont {{Miquel}}},
  \bibinfo {author} {\bibfnamefont {F.~B.}\ \bibnamefont {{Abdalla}}}, \bibinfo
  {author} {\bibfnamefont {A.}~\bibnamefont {{Amara}}}, \bibinfo {author}
  {\bibfnamefont {M.}~\bibnamefont {{Banerji}}}, \bibinfo {author}
  {\bibfnamefont {C.}~\bibnamefont {{Bonnett}}}, \bibinfo {author}
  {\bibfnamefont {R.}~\bibnamefont {{Brunner}}}, \bibinfo {author}
  {\bibfnamefont {D.}~\bibnamefont {{Capozzi}}}, \bibinfo {author}
  {\bibfnamefont {A.}~\bibnamefont {{Carnero}}}, \bibinfo {author}
  {\bibfnamefont {F.~J.}\ \bibnamefont {{Castander}}}, \bibinfo {author}
  {\bibfnamefont {L.~A.~N.}\ \bibnamefont {{da Costa}}}, \bibinfo {author}
  {\bibfnamefont {C.}~\bibnamefont {{Cunha}}}, \bibinfo {author} {\bibfnamefont
  {A.}~\bibnamefont {{Fausti}}}, \bibinfo {author} {\bibfnamefont
  {D.}~\bibnamefont {{Gerdes}}}, \bibinfo {author} {\bibfnamefont
  {N.}~\bibnamefont {{Greisel}}}, \bibinfo {author} {\bibfnamefont
  {J.}~\bibnamefont {{Gschwend}}}, \bibinfo {author} {\bibfnamefont
  {W.}~\bibnamefont {{Hartley}}}, \bibinfo {author} {\bibfnamefont
  {S.}~\bibnamefont {{Jouvel}}}, \bibinfo {author} {\bibfnamefont
  {O.}~\bibnamefont {{Lahav}}}, \bibinfo {author} {\bibfnamefont
  {M.}~\bibnamefont {{Lima}}}, \bibinfo {author} {\bibfnamefont {M.~A.~G.}\
  \bibnamefont {{Maia}}}, \bibinfo {author} {\bibfnamefont {P.}~\bibnamefont
  {{Mart{\'{\i}}}}}, \bibinfo {author} {\bibfnamefont {R.~L.~C.}\ \bibnamefont
  {{Ogando}}}, \bibinfo {author} {\bibfnamefont {F.}~\bibnamefont
  {{Ostrovski}}}, \bibinfo {author} {\bibfnamefont {P.}~\bibnamefont
  {{Pellegrini}}}, \bibinfo {author} {\bibfnamefont {M.~M.}\ \bibnamefont
  {{Rau}}}, \bibinfo {author} {\bibfnamefont {I.}~\bibnamefont {{Sadeh}}},
  \bibinfo {author} {\bibfnamefont {S.}~\bibnamefont {{Seitz}}}, \bibinfo
  {author} {\bibfnamefont {I.}~\bibnamefont {{Sevilla-Noarbe}}}, \bibinfo
  {author} {\bibfnamefont {A.}~\bibnamefont {{Sypniewski}}}, \bibinfo {author}
  {\bibfnamefont {J.}~\bibnamefont {{de Vicente}}}, \bibinfo {author}
  {\bibfnamefont {T.}~\bibnamefont {{Abbot}}}, \bibinfo {author} {\bibfnamefont
  {S.~S.}\ \bibnamefont {{Allam}}}, \bibinfo {author} {\bibfnamefont
  {D.}~\bibnamefont {{Atlee}}}, \bibinfo {author} {\bibfnamefont
  {G.}~\bibnamefont {{Bernstein}}}, \bibinfo {author} {\bibfnamefont {J.~P.}\
  \bibnamefont {{Bernstein}}}, \bibinfo {author} {\bibfnamefont
  {E.}~\bibnamefont {{Buckley-Geer}}}, \bibinfo {author} {\bibfnamefont
  {D.}~\bibnamefont {{Burke}}}, \bibinfo {author} {\bibfnamefont {M.~J.}\
  \bibnamefont {{Childress}}}, \bibinfo {author} {\bibfnamefont
  {T.}~\bibnamefont {{Davis}}}, \bibinfo {author} {\bibfnamefont {D.~L.}\
  \bibnamefont {{DePoy}}}, \bibinfo {author} {\bibfnamefont {A.}~\bibnamefont
  {{Dey}}}, \bibinfo {author} {\bibfnamefont {S.}~\bibnamefont {{Desai}}},
  \bibinfo {author} {\bibfnamefont {H.~T.}\ \bibnamefont {{Diehl}}}, \bibinfo
  {author} {\bibfnamefont {P.}~\bibnamefont {{Doel}}}, \bibinfo {author}
  {\bibfnamefont {J.}~\bibnamefont {{Estrada}}}, \bibinfo {author}
  {\bibfnamefont {A.}~\bibnamefont {{Evrard}}}, \bibinfo {author}
  {\bibfnamefont {E.}~\bibnamefont {{Fern{\'a}ndez}}}, \bibinfo {author}
  {\bibfnamefont {D.}~\bibnamefont {{Finley}}}, \bibinfo {author}
  {\bibfnamefont {B.}~\bibnamefont {{Flaugher}}}, \bibinfo {author}
  {\bibfnamefont {J.}~\bibnamefont {{Frieman}}}, \bibinfo {author}
  {\bibfnamefont {E.}~\bibnamefont {{Gaztanaga}}}, \bibinfo {author}
  {\bibfnamefont {K.}~\bibnamefont {{Glazebrook}}}, \bibinfo {author}
  {\bibfnamefont {K.}~\bibnamefont {{Honscheid}}}, \bibinfo {author}
  {\bibfnamefont {A.}~\bibnamefont {{Kim}}}, \bibinfo {author} {\bibfnamefont
  {K.}~\bibnamefont {{Kuehn}}}, \bibinfo {author} {\bibfnamefont
  {N.}~\bibnamefont {{Kuropatkin}}}, \bibinfo {author} {\bibfnamefont
  {C.}~\bibnamefont {{Lidman}}}, \bibinfo {author} {\bibfnamefont
  {M.}~\bibnamefont {{Makler}}}, \bibinfo {author} {\bibfnamefont {J.~L.}\
  \bibnamefont {{Marshall}}}, \bibinfo {author} {\bibfnamefont {R.~C.}\
  \bibnamefont {{Nichol}}}, \bibinfo {author} {\bibfnamefont {A.}~\bibnamefont
  {{Roodman}}}, \bibinfo {author} {\bibfnamefont {E.}~\bibnamefont
  {{S{\'a}nchez}}}, \bibinfo {author} {\bibfnamefont {B.~X.}\ \bibnamefont
  {{Santiago}}}, \bibinfo {author} {\bibfnamefont {M.}~\bibnamefont {{Sako}}},
  \bibinfo {author} {\bibfnamefont {R.}~\bibnamefont {{Scalzo}}}, \bibinfo
  {author} {\bibfnamefont {R.~C.}\ \bibnamefont {{Smith}}}, \bibinfo {author}
  {\bibfnamefont {M.~E.~C.}\ \bibnamefont {{Swanson}}}, \bibinfo {author}
  {\bibfnamefont {G.}~\bibnamefont {{Tarle}}}, \bibinfo {author} {\bibfnamefont
  {D.}~\bibnamefont {{Thomas}}}, \bibinfo {author} {\bibfnamefont {D.~L.}\
  \bibnamefont {{Tucker}}}, \bibinfo {author} {\bibfnamefont {S.~A.}\
  \bibnamefont {{Uddin}}}, \bibinfo {author} {\bibfnamefont {F.}~\bibnamefont
  {{Vald{\'e}s}}}, \bibinfo {author} {\bibfnamefont {A.}~\bibnamefont
  {{Walker}}}, \bibinfo {author} {\bibfnamefont {F.}~\bibnamefont {{Yuan}}}, \
  and\ \bibinfo {author} {\bibfnamefont {J.}~\bibnamefont {{Zuntz}}},\ }\href
  {\doibase 10.1093/mnras/stu1836} {\bibfield  {journal} {\bibinfo  {journal}
  {\mnras}\ }\textbf {\bibinfo {volume} {445}},\ \bibinfo {pages} {1482}
  (\bibinfo {year} {2014})},\ \Eprint {http://arxiv.org/abs/1406.4407}
  {arXiv:1406.4407 [astro-ph.IM]} \BibitemShut {NoStop}%
\bibitem [{\citenamefont {Bonnett}(2015)}]{Bonnett2015}%
  \BibitemOpen
  \bibfield  {author} {\bibinfo {author} {\bibfnamefont {C.}~\bibnamefont
  {Bonnett}},\ }\href {\doibase 10.1093/mnras/stv230} {\bibfield  {journal}
  {\bibinfo  {journal} {Monthly Notices of the Royal Astronomical Society}\
  }\textbf {\bibinfo {volume} {449}},\ \bibinfo {pages} {1043} (\bibinfo {year}
  {2015})},\ \Eprint
  {http://arxiv.org/abs/http://mnras.oxfordjournals.org/content/449/1/1043.full.pdf+html}
  {http://mnras.oxfordjournals.org/content/449/1/1043.full.pdf+html}
  \BibitemShut {NoStop}%
\bibitem [{\citenamefont {{Graff}}\ and\ \citenamefont
  {{Feroz}}(2013)}]{Graff2013}%
  \BibitemOpen
  \bibfield  {author} {\bibinfo {author} {\bibfnamefont {P.}~\bibnamefont
  {{Graff}}}\ and\ \bibinfo {author} {\bibfnamefont {F.}~\bibnamefont
  {{Feroz}}},\ }\href@noop {} {\enquote {\bibinfo {title} {{SkyNet: Neural
  network training tool for machine learning in astronomy}},}\ }\bibinfo
  {howpublished} {Astrophysics Source Code Library} (\bibinfo {year} {2013}),\
  \Eprint {http://arxiv.org/abs/1312.007} {ascl:1312.007} \BibitemShut
  {NoStop}%
\bibitem [{\citenamefont {{Crocce}}\ \emph {et~al.}(2006)\citenamefont
  {{Crocce}}, \citenamefont {{Pueblas}},\ and\ \citenamefont
  {{Scoccimarro}}}]{crocce2006}%
  \BibitemOpen
  \bibfield  {author} {\bibinfo {author} {\bibfnamefont {M.}~\bibnamefont
  {{Crocce}}}, \bibinfo {author} {\bibfnamefont {S.}~\bibnamefont {{Pueblas}}},
  \ and\ \bibinfo {author} {\bibfnamefont {R.}~\bibnamefont {{Scoccimarro}}},\
  }\href {\doibase 10.1111/j.1365-2966.2006.11040.x} {\bibfield  {journal}
  {\bibinfo  {journal} {\mnras}\ }\textbf {\bibinfo {volume} {373}},\ \bibinfo
  {pages} {369} (\bibinfo {year} {2006})},\ \Eprint
  {http://arxiv.org/abs/astro-ph/0606505} {astro-ph/0606505} \BibitemShut
  {NoStop}%
\bibitem [{\citenamefont {Lewis}\ and\ \citenamefont
  {Bridle}(2002)}]{lewis2002}%
  \BibitemOpen
  \bibfield  {author} {\bibinfo {author} {\bibfnamefont {A.}~\bibnamefont
  {Lewis}}\ and\ \bibinfo {author} {\bibfnamefont {S.}~\bibnamefont {Bridle}},\
  }\href@noop {} {\bibfield  {journal} {\bibinfo  {journal} {Phys. Rev.}\
  }\textbf {\bibinfo {volume} {D66}},\ \bibinfo {pages} {103511} (\bibinfo
  {year} {2002})},\ \Eprint {http://arxiv.org/abs/astro-ph/0205436}
  {astro-ph/0205436} \BibitemShut {NoStop}%
\bibitem [{\citenamefont {{Springel}}(2005)}]{springel2005}%
  \BibitemOpen
  \bibfield  {author} {\bibinfo {author} {\bibfnamefont {V.}~\bibnamefont
  {{Springel}}},\ }\href {\doibase 10.1111/j.1365-2966.2005.09655.x} {\bibfield
   {journal} {\bibinfo  {journal} {\mnras}\ }\textbf {\bibinfo {volume}
  {364}},\ \bibinfo {pages} {1105} (\bibinfo {year} {2005})},\ \Eprint
  {http://arxiv.org/abs/arXiv:astro-ph/0505010} {arXiv:astro-ph/0505010}
  \BibitemShut {NoStop}%
\bibitem [{\citenamefont {{Becker}}(2013)}]{becker2013}%
  \BibitemOpen
  \bibfield  {author} {\bibinfo {author} {\bibfnamefont {M.~R.}\ \bibnamefont
  {{Becker}}},\ }\href {\doibase 10.1093/mnras/stt1396} {\bibfield  {journal}
  {\bibinfo  {journal} {\mnras}\ }\textbf {\bibinfo {volume} {435}},\ \bibinfo
  {pages} {1547} (\bibinfo {year} {2013})},\ \Eprint
  {http://arxiv.org/abs/1208.0068} {arXiv:1208.0068} \BibitemShut {NoStop}%
\bibitem [{\citenamefont {{Miller}}\ \emph {et~al.}(2013)\citenamefont
  {{Miller}}, \citenamefont {{Heymans}}, \citenamefont {{Kitching}},
  \citenamefont {{van Waerbeke}}, \citenamefont {{Erben}}, \citenamefont
  {{Hildebrandt}}, \citenamefont {{Hoekstra}}, \citenamefont {{Mellier}},
  \citenamefont {{Rowe}}, \citenamefont {{Coupon}}, \citenamefont {{Dietrich}},
  \citenamefont {{Fu}}, \citenamefont {{Harnois-D{\'e}raps}}, \citenamefont
  {{Hudson}}, \citenamefont {{Kilbinger}}, \citenamefont {{Kuijken}},
  \citenamefont {{Schrabback}}, \citenamefont {{Semboloni}}, \citenamefont
  {{Vafaei}},\ and\ \citenamefont {{Velander}}}]{miller2013}%
  \BibitemOpen
  \bibfield  {author} {\bibinfo {author} {\bibfnamefont {L.}~\bibnamefont
  {{Miller}}}, \bibinfo {author} {\bibfnamefont {C.}~\bibnamefont {{Heymans}}},
  \bibinfo {author} {\bibfnamefont {T.~D.}\ \bibnamefont {{Kitching}}},
  \bibinfo {author} {\bibfnamefont {L.}~\bibnamefont {{van Waerbeke}}},
  \bibinfo {author} {\bibfnamefont {T.}~\bibnamefont {{Erben}}}, \bibinfo
  {author} {\bibfnamefont {H.}~\bibnamefont {{Hildebrandt}}}, \bibinfo {author}
  {\bibfnamefont {H.}~\bibnamefont {{Hoekstra}}}, \bibinfo {author}
  {\bibfnamefont {Y.}~\bibnamefont {{Mellier}}}, \bibinfo {author}
  {\bibfnamefont {B.~T.~P.}\ \bibnamefont {{Rowe}}}, \bibinfo {author}
  {\bibfnamefont {J.}~\bibnamefont {{Coupon}}}, \bibinfo {author}
  {\bibfnamefont {J.~P.}\ \bibnamefont {{Dietrich}}}, \bibinfo {author}
  {\bibfnamefont {L.}~\bibnamefont {{Fu}}}, \bibinfo {author} {\bibfnamefont
  {J.}~\bibnamefont {{Harnois-D{\'e}raps}}}, \bibinfo {author} {\bibfnamefont
  {M.~J.}\ \bibnamefont {{Hudson}}}, \bibinfo {author} {\bibfnamefont
  {M.}~\bibnamefont {{Kilbinger}}}, \bibinfo {author} {\bibfnamefont
  {K.}~\bibnamefont {{Kuijken}}}, \bibinfo {author} {\bibfnamefont
  {T.}~\bibnamefont {{Schrabback}}}, \bibinfo {author} {\bibfnamefont
  {E.}~\bibnamefont {{Semboloni}}}, \bibinfo {author} {\bibfnamefont
  {S.}~\bibnamefont {{Vafaei}}}, \ and\ \bibinfo {author} {\bibfnamefont
  {M.}~\bibnamefont {{Velander}}},\ }\href {\doibase 10.1093/mnras/sts454}
  {\bibfield  {journal} {\bibinfo  {journal} {\mnras}\ }\textbf {\bibinfo
  {volume} {429}},\ \bibinfo {pages} {2858} (\bibinfo {year} {2013})},\ \Eprint
  {http://arxiv.org/abs/1210.8201} {arXiv:1210.8201 [astro-ph.CO]} \BibitemShut
  {NoStop}%
\bibitem [{\citenamefont {{Jarvis}}\ \emph {et~al.}(2004)\citenamefont
  {{Jarvis}}, \citenamefont {{Bernstein}},\ and\ \citenamefont
  {{Jain}}}]{jarvis2004}%
  \BibitemOpen
  \bibfield  {author} {\bibinfo {author} {\bibfnamefont {M.}~\bibnamefont
  {{Jarvis}}}, \bibinfo {author} {\bibfnamefont {G.}~\bibnamefont
  {{Bernstein}}}, \ and\ \bibinfo {author} {\bibfnamefont {B.}~\bibnamefont
  {{Jain}}},\ }\href {\doibase 10.1111/j.1365-2966.2004.07926.x} {\bibfield
  {journal} {\bibinfo  {journal} {\mnras}\ }\textbf {\bibinfo {volume} {352}},\
  \bibinfo {pages} {338} (\bibinfo {year} {2004})},\ \Eprint
  {http://arxiv.org/abs/astro-ph/0307393} {astro-ph/0307393} \BibitemShut
  {NoStop}%
\bibitem [{\citenamefont {{Hartlap}}\ \emph {et~al.}(2007)\citenamefont
  {{Hartlap}}, \citenamefont {{Simon}},\ and\ \citenamefont
  {{Schneider}}}]{hartlap2007}%
  \BibitemOpen
  \bibfield  {author} {\bibinfo {author} {\bibfnamefont {J.}~\bibnamefont
  {{Hartlap}}}, \bibinfo {author} {\bibfnamefont {P.}~\bibnamefont {{Simon}}},
  \ and\ \bibinfo {author} {\bibfnamefont {P.}~\bibnamefont {{Schneider}}},\
  }\href {\doibase 10.1051/0004-6361:20066170} {\bibfield  {journal} {\bibinfo
  {journal} {\aap}\ }\textbf {\bibinfo {volume} {464}},\ \bibinfo {pages} {399}
  (\bibinfo {year} {2007})},\ \Eprint {http://arxiv.org/abs/astro-ph/0608064}
  {astro-ph/0608064} \BibitemShut {NoStop}%
\bibitem [{\citenamefont {{Zuntz}}\ \emph {et~al.}(2015)\citenamefont
  {{Zuntz}}, \citenamefont {{Paterno}}, \citenamefont {{Jennings}},
  \citenamefont {{Rudd}}, \citenamefont {{Manzotti}}, \citenamefont
  {{Dodelson}}, \citenamefont {{Bridle}}, \citenamefont {{Sehrish}},\ and\
  \citenamefont {{Kowalkowski}}}]{cosmosis}%
  \BibitemOpen
  \bibfield  {author} {\bibinfo {author} {\bibfnamefont {J.}~\bibnamefont
  {{Zuntz}}}, \bibinfo {author} {\bibfnamefont {M.}~\bibnamefont {{Paterno}}},
  \bibinfo {author} {\bibfnamefont {E.}~\bibnamefont {{Jennings}}}, \bibinfo
  {author} {\bibfnamefont {D.}~\bibnamefont {{Rudd}}}, \bibinfo {author}
  {\bibfnamefont {A.}~\bibnamefont {{Manzotti}}}, \bibinfo {author}
  {\bibfnamefont {S.}~\bibnamefont {{Dodelson}}}, \bibinfo {author}
  {\bibfnamefont {S.}~\bibnamefont {{Bridle}}}, \bibinfo {author}
  {\bibfnamefont {S.}~\bibnamefont {{Sehrish}}}, \ and\ \bibinfo {author}
  {\bibfnamefont {J.}~\bibnamefont {{Kowalkowski}}},\ }\href {\doibase
  10.1016/j.ascom.2015.05.005} {\bibfield  {journal} {\bibinfo  {journal}
  {Astronomy and Computing}\ }\textbf {\bibinfo {volume} {12}},\ \bibinfo
  {pages} {45} (\bibinfo {year} {2015})},\ \Eprint
  {http://arxiv.org/abs/1409.3409} {arXiv:1409.3409} \BibitemShut {NoStop}%
\bibitem [{\citenamefont {{Takahashi}}\ \emph {et~al.}(2012)\citenamefont
  {{Takahashi}}, \citenamefont {{Sato}}, \citenamefont {{Nishimichi}},
  \citenamefont {{Taruya}},\ and\ \citenamefont {{Oguri}}}]{takahashi2012}%
  \BibitemOpen
  \bibfield  {author} {\bibinfo {author} {\bibfnamefont {R.}~\bibnamefont
  {{Takahashi}}}, \bibinfo {author} {\bibfnamefont {M.}~\bibnamefont {{Sato}}},
  \bibinfo {author} {\bibfnamefont {T.}~\bibnamefont {{Nishimichi}}}, \bibinfo
  {author} {\bibfnamefont {A.}~\bibnamefont {{Taruya}}}, \ and\ \bibinfo
  {author} {\bibfnamefont {M.}~\bibnamefont {{Oguri}}},\ }\href {\doibase
  10.1088/0004-637X/761/2/152} {\bibfield  {journal} {\bibinfo  {journal}
  {\apj}\ }\textbf {\bibinfo {volume} {761}},\ \bibinfo {eid} {152} (\bibinfo
  {year} {2012})},\ \Eprint {http://arxiv.org/abs/1208.2701} {arXiv:1208.2701
  [astro-ph.CO]} \BibitemShut {NoStop}%
\bibitem [{\citenamefont {{Becker}}\ and\ \citenamefont
  {{Rozo}}(2016)}]{becker2014}%
  \BibitemOpen
  \bibfield  {author} {\bibinfo {author} {\bibfnamefont {M.~R.}\ \bibnamefont
  {{Becker}}}\ and\ \bibinfo {author} {\bibfnamefont {E.}~\bibnamefont
  {{Rozo}}},\ }\href@noop {} {\bibfield  {journal} {\bibinfo  {journal}
  {\mnras}\ }\textbf {\bibinfo {volume} {457}},\ \bibinfo {pages} {304}
  (\bibinfo {year} {2016})},\ \Eprint {http://arxiv.org/abs/1412.3851}
  {arXiv:1412.3851} \BibitemShut {NoStop}%
\bibitem [{\citenamefont {{Szapudi}}\ \emph {et~al.}(2000)\citenamefont
  {{Szapudi}}, \citenamefont {{Prunet}}, \citenamefont {{Pogosyan}},
  \citenamefont {{Szalay}},\ and\ \citenamefont {{Bond}}}]{szapudi2001}%
  \BibitemOpen
  \bibfield  {author} {\bibinfo {author} {\bibfnamefont {I.}~\bibnamefont
  {{Szapudi}}}, \bibinfo {author} {\bibfnamefont {S.}~\bibnamefont {{Prunet}}},
  \bibinfo {author} {\bibfnamefont {D.}~\bibnamefont {{Pogosyan}}}, \bibinfo
  {author} {\bibfnamefont {A.~S.}\ \bibnamefont {{Szalay}}}, \ and\ \bibinfo
  {author} {\bibfnamefont {J.~R.}\ \bibnamefont {{Bond}}},\ }\href@noop {}
  {\bibfield  {journal} {\bibinfo  {journal} {arXiv:astro-ph/0010256}\ }
  (\bibinfo {year} {2000})},\ \Eprint {http://arxiv.org/abs/astro-ph/0010256}
  {astro-ph/0010256} \BibitemShut {NoStop}%
\bibitem [{\citenamefont {{Chon}}\ \emph {et~al.}(2004)\citenamefont {{Chon}},
  \citenamefont {{Challinor}}, \citenamefont {{Prunet}}, \citenamefont
  {{Hivon}},\ and\ \citenamefont {{Szapudi}}}]{chon2004}%
  \BibitemOpen
  \bibfield  {author} {\bibinfo {author} {\bibfnamefont {G.}~\bibnamefont
  {{Chon}}}, \bibinfo {author} {\bibfnamefont {A.}~\bibnamefont {{Challinor}}},
  \bibinfo {author} {\bibfnamefont {S.}~\bibnamefont {{Prunet}}}, \bibinfo
  {author} {\bibfnamefont {E.}~\bibnamefont {{Hivon}}}, \ and\ \bibinfo
  {author} {\bibfnamefont {I.}~\bibnamefont {{Szapudi}}},\ }\href {\doibase
  10.1111/j.1365-2966.2004.07737.x} {\bibfield  {journal} {\bibinfo  {journal}
  {\mnras}\ }\textbf {\bibinfo {volume} {350}},\ \bibinfo {pages} {914}
  (\bibinfo {year} {2004})},\ \Eprint {http://arxiv.org/abs/astro-ph/0303414}
  {astro-ph/0303414} \BibitemShut {NoStop}%
\bibitem [{\citenamefont {{Eifler}}\ \emph {et~al.}(2014)\citenamefont
  {{Eifler}}, \citenamefont {{Krause}}, \citenamefont {{Schneider}},\ and\
  \citenamefont {{Honscheid}}}]{eifler2014}%
  \BibitemOpen
  \bibfield  {author} {\bibinfo {author} {\bibfnamefont {T.}~\bibnamefont
  {{Eifler}}}, \bibinfo {author} {\bibfnamefont {E.}~\bibnamefont {{Krause}}},
  \bibinfo {author} {\bibfnamefont {P.}~\bibnamefont {{Schneider}}}, \ and\
  \bibinfo {author} {\bibfnamefont {K.}~\bibnamefont {{Honscheid}}},\ }\href
  {\doibase 10.1093/mnras/stu251} {\bibfield  {journal} {\bibinfo  {journal}
  {\mnras}\ }\textbf {\bibinfo {volume} {440}},\ \bibinfo {pages} {1379}
  (\bibinfo {year} {2014})},\ \Eprint {http://arxiv.org/abs/1302.2401}
  {arXiv:1302.2401} \BibitemShut {NoStop}%
\bibitem [{\citenamefont {{Krause}}\ \emph {et~al.}(2016)\citenamefont
  {{Krause}}, \citenamefont {{Eifler}},\ and\ \citenamefont
  {{Blazek}}}]{krause2015}%
  \BibitemOpen
  \bibfield  {author} {\bibinfo {author} {\bibfnamefont {E.}~\bibnamefont
  {{Krause}}}, \bibinfo {author} {\bibfnamefont {T.}~\bibnamefont {{Eifler}}},
  \ and\ \bibinfo {author} {\bibfnamefont {J.}~\bibnamefont {{Blazek}}},\
  }\href {\doibase 10.1093/mnras/stv2615} {\bibfield  {journal} {\bibinfo
  {journal} {\mnras}\ }\textbf {\bibinfo {volume} {456}},\ \bibinfo {pages}
  {207} (\bibinfo {year} {2016})},\ \Eprint {http://arxiv.org/abs/1506.08730}
  {arXiv:1506.08730} \BibitemShut {NoStop}%
\bibitem [{\citenamefont {{Takada}}\ and\ \citenamefont
  {{Hu}}(2013)}]{takada2013}%
  \BibitemOpen
  \bibfield  {author} {\bibinfo {author} {\bibfnamefont {M.}~\bibnamefont
  {{Takada}}}\ and\ \bibinfo {author} {\bibfnamefont {W.}~\bibnamefont
  {{Hu}}},\ }\href {\doibase 10.1103/PhysRevD.87.123504} {\bibfield  {journal}
  {\bibinfo  {journal} {\prd}\ }\textbf {\bibinfo {volume} {87}},\ \bibinfo
  {eid} {123504} (\bibinfo {year} {2013})},\ \Eprint
  {http://arxiv.org/abs/1302.6994} {arXiv:1302.6994 [astro-ph.CO]} \BibitemShut
  {NoStop}%
\bibitem [{\citenamefont {{Albrecht}}\ \emph {et~al.}(2009)\citenamefont
  {{Albrecht}}, \citenamefont {{Amendola}}, \citenamefont {{Bernstein}},
  \citenamefont {{Clowe}}, \citenamefont {{Eisenstein}}, \citenamefont
  {{Guzzo}}, \citenamefont {{Hirata}}, \citenamefont {{Huterer}}, \citenamefont
  {{Kirshner}}, \citenamefont {{Kolb}},\ and\ \citenamefont
  {{Nichol}}}]{albrecht2009}%
  \BibitemOpen
  \bibfield  {author} {\bibinfo {author} {\bibfnamefont {A.}~\bibnamefont
  {{Albrecht}}}, \bibinfo {author} {\bibfnamefont {L.}~\bibnamefont
  {{Amendola}}}, \bibinfo {author} {\bibfnamefont {G.}~\bibnamefont
  {{Bernstein}}}, \bibinfo {author} {\bibfnamefont {D.}~\bibnamefont
  {{Clowe}}}, \bibinfo {author} {\bibfnamefont {D.}~\bibnamefont
  {{Eisenstein}}}, \bibinfo {author} {\bibfnamefont {L.}~\bibnamefont
  {{Guzzo}}}, \bibinfo {author} {\bibfnamefont {C.}~\bibnamefont {{Hirata}}},
  \bibinfo {author} {\bibfnamefont {D.}~\bibnamefont {{Huterer}}}, \bibinfo
  {author} {\bibfnamefont {R.}~\bibnamefont {{Kirshner}}}, \bibinfo {author}
  {\bibfnamefont {E.}~\bibnamefont {{Kolb}}}, \ and\ \bibinfo {author}
  {\bibfnamefont {R.}~\bibnamefont {{Nichol}}},\ }\href@noop {} {\bibfield
  {journal} {\bibinfo  {journal} {arXiv:astro-ph/0901.0721}\ } (\bibinfo {year}
  {2009})},\ \Eprint {http://arxiv.org/abs/0901.0721} {arXiv:0901.0721
  [astro-ph.IM]} \BibitemShut {NoStop}%
\bibitem [{\citenamefont {{Dodelson}}\ and\ \citenamefont
  {{Schneider}}(2013)}]{dodelson2013}%
  \BibitemOpen
  \bibfield  {author} {\bibinfo {author} {\bibfnamefont {S.}~\bibnamefont
  {{Dodelson}}}\ and\ \bibinfo {author} {\bibfnamefont {M.~D.}\ \bibnamefont
  {{Schneider}}},\ }\href {\doibase 10.1103/PhysRevD.88.063537} {\bibfield
  {journal} {\bibinfo  {journal} {\prd}\ }\textbf {\bibinfo {volume} {88}},\
  \bibinfo {eid} {063537} (\bibinfo {year} {2013})},\ \Eprint
  {http://arxiv.org/abs/1304.2593} {arXiv:1304.2593 [astro-ph.CO]} \BibitemShut
  {NoStop}%
\bibitem [{\citenamefont {{Friedrich}}\ \emph {et~al.}(2016)\citenamefont
  {{Friedrich}}, \citenamefont {{Seitz}}, \citenamefont {{Eifler}},\ and\
  \citenamefont {{Gruen}}}]{friedrich2016}%
  \BibitemOpen
  \bibfield  {author} {\bibinfo {author} {\bibfnamefont {O.}~\bibnamefont
  {{Friedrich}}}, \bibinfo {author} {\bibfnamefont {S.}~\bibnamefont
  {{Seitz}}}, \bibinfo {author} {\bibfnamefont {T.~F.}\ \bibnamefont
  {{Eifler}}}, \ and\ \bibinfo {author} {\bibfnamefont {D.}~\bibnamefont
  {{Gruen}}},\ }\href {\doibase 10.1093/mnras/stv2833} {\bibfield  {journal}
  {\bibinfo  {journal} {\mnras}\ }\textbf {\bibinfo {volume} {456}},\ \bibinfo
  {pages} {2662} (\bibinfo {year} {2016})},\ \Eprint
  {http://arxiv.org/abs/1508.00895} {arXiv:1508.00895} \BibitemShut {NoStop}%
\bibitem [{\citenamefont {{Norberg}}\ \emph {et~al.}(2009)\citenamefont
  {{Norberg}}, \citenamefont {{Baugh}}, \citenamefont {{Gazta{\~n}aga}},\ and\
  \citenamefont {{Croton}}}]{2009MNRAS.396...19N}%
  \BibitemOpen
  \bibfield  {author} {\bibinfo {author} {\bibfnamefont {P.}~\bibnamefont
  {{Norberg}}}, \bibinfo {author} {\bibfnamefont {C.~M.}\ \bibnamefont
  {{Baugh}}}, \bibinfo {author} {\bibfnamefont {E.}~\bibnamefont
  {{Gazta{\~n}aga}}}, \ and\ \bibinfo {author} {\bibfnamefont {D.~J.}\
  \bibnamefont {{Croton}}},\ }\href {\doibase 10.1111/j.1365-2966.2009.14389.x}
  {\bibfield  {journal} {\bibinfo  {journal} {\mnras}\ }\textbf {\bibinfo
  {volume} {396}},\ \bibinfo {pages} {19} (\bibinfo {year} {2009})},\ \Eprint
  {http://arxiv.org/abs/0810.1885} {arXiv:0810.1885} \BibitemShut {NoStop}%
\bibitem [{\citenamefont {Bartelmann}(2010)}]{bartelmann2010}%
  \BibitemOpen
  \bibfield  {author} {\bibinfo {author} {\bibfnamefont {M.}~\bibnamefont
  {Bartelmann}},\ }\href {\doibase 10.1088/0264-9381/27/23/233001} {\bibfield
  {journal} {\bibinfo  {journal} {Class. Quant. Grav.}\ }\textbf {\bibinfo
  {volume} {27}},\ \bibinfo {pages} {233001} (\bibinfo {year} {2010})},\
  \Eprint {http://arxiv.org/abs/1010.3829} {arXiv:1010.3829 [astro-ph.CO]}
  \BibitemShut {NoStop}%
\bibitem [{\citenamefont {{Kaiser}}(1992)}]{kaiser1992}%
  \BibitemOpen
  \bibfield  {author} {\bibinfo {author} {\bibfnamefont {N.}~\bibnamefont
  {{Kaiser}}},\ }\href {\doibase 10.1086/171151} {\bibfield  {journal}
  {\bibinfo  {journal} {\apj}\ }\textbf {\bibinfo {volume} {388}},\ \bibinfo
  {pages} {272} (\bibinfo {year} {1992})}\BibitemShut {NoStop}%
\bibitem [{\citenamefont {{Hilbert}}\ \emph {et~al.}(2009)\citenamefont
  {{Hilbert}}, \citenamefont {{Hartlap}}, \citenamefont {{White}},\ and\
  \citenamefont {{Schneider}}}]{hilbert2009}%
  \BibitemOpen
  \bibfield  {author} {\bibinfo {author} {\bibfnamefont {S.}~\bibnamefont
  {{Hilbert}}}, \bibinfo {author} {\bibfnamefont {J.}~\bibnamefont
  {{Hartlap}}}, \bibinfo {author} {\bibfnamefont {S.~D.~M.}\ \bibnamefont
  {{White}}}, \ and\ \bibinfo {author} {\bibfnamefont {P.}~\bibnamefont
  {{Schneider}}},\ }\href {\doibase 10.1051/0004-6361/200811054} {\bibfield
  {journal} {\bibinfo  {journal} {\aap}\ }\textbf {\bibinfo {volume} {499}},\
  \bibinfo {pages} {31} (\bibinfo {year} {2009})},\ \Eprint
  {http://arxiv.org/abs/0809.5035} {arXiv:0809.5035} \BibitemShut {NoStop}%
\bibitem [{\citenamefont {{Krause}}\ and\ \citenamefont
  {{Hirata}}(2010)}]{krause2010}%
  \BibitemOpen
  \bibfield  {author} {\bibinfo {author} {\bibfnamefont {E.}~\bibnamefont
  {{Krause}}}\ and\ \bibinfo {author} {\bibfnamefont {C.~M.}\ \bibnamefont
  {{Hirata}}},\ }\href {\doibase 10.1051/0004-6361/200913524} {\bibfield
  {journal} {\bibinfo  {journal} {\aap}\ }\textbf {\bibinfo {volume} {523}},\
  \bibinfo {eid} {A28} (\bibinfo {year} {2010})},\ \Eprint
  {http://arxiv.org/abs/0910.3786} {arXiv:0910.3786 [astro-ph.CO]} \BibitemShut
  {NoStop}%
\bibitem [{\citenamefont {{Schneider}}\ \emph {et~al.}(1998)\citenamefont
  {{Schneider}}, \citenamefont {{van Waerbeke}}, \citenamefont {{Jain}},\ and\
  \citenamefont {{Kruse}}}]{schneider1998}%
  \BibitemOpen
  \bibfield  {author} {\bibinfo {author} {\bibfnamefont {P.}~\bibnamefont
  {{Schneider}}}, \bibinfo {author} {\bibfnamefont {L.}~\bibnamefont {{van
  Waerbeke}}}, \bibinfo {author} {\bibfnamefont {B.}~\bibnamefont {{Jain}}}, \
  and\ \bibinfo {author} {\bibfnamefont {G.}~\bibnamefont {{Kruse}}},\ }\href
  {\doibase 10.1046/j.1365-8711.1998.01422.x} {\bibfield  {journal} {\bibinfo
  {journal} {\mnras}\ }\textbf {\bibinfo {volume} {296}},\ \bibinfo {pages}
  {873} (\bibinfo {year} {1998})},\ \Eprint
  {http://arxiv.org/abs/arXiv:astro-ph/9708143} {arXiv:astro-ph/9708143}
  \BibitemShut {NoStop}%
\bibitem [{\citenamefont {{Seljak}}(1998)}]{seljak1998}%
  \BibitemOpen
  \bibfield  {author} {\bibinfo {author} {\bibfnamefont {U.}~\bibnamefont
  {{Seljak}}},\ }\href {\doibase 10.1086/306225} {\bibfield  {journal}
  {\bibinfo  {journal} {\apj}\ }\textbf {\bibinfo {volume} {506}},\ \bibinfo
  {pages} {64} (\bibinfo {year} {1998})},\ \Eprint
  {http://arxiv.org/abs/arXiv:astro-ph/9711124} {arXiv:astro-ph/9711124}
  \BibitemShut {NoStop}%
\bibitem [{\citenamefont {{Hu}}\ and\ \citenamefont {{White}}(2001)}]{hu2001b}%
  \BibitemOpen
  \bibfield  {author} {\bibinfo {author} {\bibfnamefont {W.}~\bibnamefont
  {{Hu}}}\ and\ \bibinfo {author} {\bibfnamefont {M.}~\bibnamefont {{White}}},\
  }\href {\doibase 10.1086/321380} {\bibfield  {journal} {\bibinfo  {journal}
  {\apj}\ }\textbf {\bibinfo {volume} {554}},\ \bibinfo {pages} {67} (\bibinfo
  {year} {2001})},\ \Eprint {http://arxiv.org/abs/arXiv:astro-ph/0010352}
  {arXiv:astro-ph/0010352} \BibitemShut {NoStop}%
\bibitem [{\citenamefont {{Crittenden}}\ \emph {et~al.}(2002)\citenamefont
  {{Crittenden}}, \citenamefont {{Natarajan}}, \citenamefont {{Pen}},\ and\
  \citenamefont {{Theuns}}}]{2002ApJ...568...20C}%
  \BibitemOpen
  \bibfield  {author} {\bibinfo {author} {\bibfnamefont {R.~G.}\ \bibnamefont
  {{Crittenden}}}, \bibinfo {author} {\bibfnamefont {P.}~\bibnamefont
  {{Natarajan}}}, \bibinfo {author} {\bibfnamefont {U.-L.}\ \bibnamefont
  {{Pen}}}, \ and\ \bibinfo {author} {\bibfnamefont {T.}~\bibnamefont
  {{Theuns}}},\ }\href {\doibase 10.1086/338838} {\bibfield  {journal}
  {\bibinfo  {journal} {\apj}\ }\textbf {\bibinfo {volume} {568}},\ \bibinfo
  {pages} {20} (\bibinfo {year} {2002})},\ \Eprint
  {http://arxiv.org/abs/astro-ph/0012336} {astro-ph/0012336} \BibitemShut
  {NoStop}%
\bibitem [{\citenamefont {{Schneider}}\ and\ \citenamefont
  {{Kilbinger}}(2007)}]{schneider2007}%
  \BibitemOpen
  \bibfield  {author} {\bibinfo {author} {\bibfnamefont {P.}~\bibnamefont
  {{Schneider}}}\ and\ \bibinfo {author} {\bibfnamefont {M.}~\bibnamefont
  {{Kilbinger}}},\ }\href {\doibase 10.1051/0004-6361:20065532} {\bibfield
  {journal} {\bibinfo  {journal} {\aap}\ }\textbf {\bibinfo {volume} {462}},\
  \bibinfo {pages} {841} (\bibinfo {year} {2007})},\ \Eprint
  {http://arxiv.org/abs/arXiv:astro-ph/0605084} {arXiv:astro-ph/0605084}
  \BibitemShut {NoStop}%
\bibitem [{\citenamefont {{Schneider}}\ \emph {et~al.}(2010)\citenamefont
  {{Schneider}}, \citenamefont {{Eifler}},\ and\ \citenamefont
  {{Krause}}}]{schneider2010}%
  \BibitemOpen
  \bibfield  {author} {\bibinfo {author} {\bibfnamefont {P.}~\bibnamefont
  {{Schneider}}}, \bibinfo {author} {\bibfnamefont {T.}~\bibnamefont
  {{Eifler}}}, \ and\ \bibinfo {author} {\bibfnamefont {E.}~\bibnamefont
  {{Krause}}},\ }\href {\doibase 10.1051/0004-6361/201014235} {\bibfield
  {journal} {\bibinfo  {journal} {\aap}\ }\textbf {\bibinfo {volume} {520}},\
  \bibinfo {eid} {A116} (\bibinfo {year} {2010})},\ \Eprint
  {http://arxiv.org/abs/1002.2136} {arXiv:1002.2136 [astro-ph.CO]} \BibitemShut
  {NoStop}%
\bibitem [{\citenamefont {{Leistedt}}\ \emph {et~al.}(2015)\citenamefont
  {{Leistedt}}, \citenamefont {{Peiris}}, \citenamefont {{Elsner}},
  \citenamefont {{Benoit-L{\'e}vy}}, \citenamefont {{Amara}}, \citenamefont
  {{Bauer}}, \citenamefont {{Becker}}, \citenamefont {{Bonnett}}, \citenamefont
  {{Bruderer}}, \citenamefont {{Busha}}, \citenamefont {{Carrasco Kind}},
  \citenamefont {{Chang}}, \citenamefont {{Crocce}}, \citenamefont {{da
  Costa}}, \citenamefont {{Gaztanaga}}, \citenamefont {{Huff}}, \citenamefont
  {{Lahav}}, \citenamefont {{Palmese}}, \citenamefont {{Percival}},
  \citenamefont {{Refregier}}, \citenamefont {{Ross}}, \citenamefont {{Rozo}},
  \citenamefont {{Rykoff}}, \citenamefont {{S{\'a}nchez}}, \citenamefont
  {{Sadeh}}, \citenamefont {{Sevilla-Noarbe}}, \citenamefont {{Sobreira}},
  \citenamefont {{Suchyta}}, \citenamefont {{Swanson}}, \citenamefont
  {{Wechsler}}, \citenamefont {{Abdalla}}, \citenamefont {{Allam}},
  \citenamefont {{Banerji}}, \citenamefont {{Bernstein}}, \citenamefont
  {{Bernstein}}, \citenamefont {{Bertin}}, \citenamefont {{Bridle}},
  \citenamefont {{Brooks}}, \citenamefont {{Buckley-Geer}}, \citenamefont
  {{Burke}}, \citenamefont {{Capozzi}}, \citenamefont {{Carnero Rosell}},
  \citenamefont {{Carretero}}, \citenamefont {{Cunha}}, \citenamefont
  {{D'Andrea}}, \citenamefont {{DePoy}}, \citenamefont {{Desai}}, \citenamefont
  {{Diehl}}, \citenamefont {{Doel}}, \citenamefont {{Eifler}}, \citenamefont
  {{Evrard}}, \citenamefont {{Fausti Neto}}, \citenamefont {{Flaugher}},
  \citenamefont {{Fosalba}}, \citenamefont {{Frieman}}, \citenamefont
  {{Gerdes}}, \citenamefont {{Gruen}}, \citenamefont {{Gruendl}}, \citenamefont
  {{Gutierrez}}, \citenamefont {{Honscheid}}, \citenamefont {{James}},
  \citenamefont {{Jarvis}}, \citenamefont {{Kent}}, \citenamefont {{Kuehn}},
  \citenamefont {{Kuropatkin}}, \citenamefont {{Li}}, \citenamefont {{Lima}},
  \citenamefont {{Maia}}, \citenamefont {{March}}, \citenamefont {{Marshall}},
  \citenamefont {{Martini}}, \citenamefont {{Melchior}}, \citenamefont
  {{Miller}}, \citenamefont {{Miquel}}, \citenamefont {{Nichol}}, \citenamefont
  {{Nord}}, \citenamefont {{Ogando}}, \citenamefont {{Plazas}}, \citenamefont
  {{Reil}}, \citenamefont {{Romer}}, \citenamefont {{Roodman}}, \citenamefont
  {{Sanchez}}, \citenamefont {{Santiago}}, \citenamefont {{Scarpine}},
  \citenamefont {{Schubnell}}, \citenamefont {{Smith}}, \citenamefont
  {{Soares-Santos}}, \citenamefont {{Tarle}}, \citenamefont {{Thaler}},
  \citenamefont {{Thomas}}, \citenamefont {{Vikram}}, \citenamefont {{Walker}},
  \citenamefont {{Wester}}, \citenamefont {{Zhang}},\ and\ \citenamefont
  {{Zuntz}}}]{leistedt2015}%
  \BibitemOpen
  \bibfield  {author} {\bibinfo {author} {\bibfnamefont {B.}~\bibnamefont
  {{Leistedt}}}, \bibinfo {author} {\bibfnamefont {H.~V.}\ \bibnamefont
  {{Peiris}}}, \bibinfo {author} {\bibfnamefont {F.}~\bibnamefont {{Elsner}}},
  \bibinfo {author} {\bibfnamefont {A.}~\bibnamefont {{Benoit-L{\'e}vy}}},
  \bibinfo {author} {\bibfnamefont {A.}~\bibnamefont {{Amara}}}, \bibinfo
  {author} {\bibfnamefont {A.~H.}\ \bibnamefont {{Bauer}}}, \bibinfo {author}
  {\bibfnamefont {M.~R.}\ \bibnamefont {{Becker}}}, \bibinfo {author}
  {\bibfnamefont {C.}~\bibnamefont {{Bonnett}}}, \bibinfo {author}
  {\bibfnamefont {C.}~\bibnamefont {{Bruderer}}}, \bibinfo {author}
  {\bibfnamefont {M.~T.}\ \bibnamefont {{Busha}}}, \bibinfo {author}
  {\bibfnamefont {M.}~\bibnamefont {{Carrasco Kind}}}, \bibinfo {author}
  {\bibfnamefont {C.}~\bibnamefont {{Chang}}}, \bibinfo {author} {\bibfnamefont
  {M.}~\bibnamefont {{Crocce}}}, \bibinfo {author} {\bibfnamefont {L.~N.}\
  \bibnamefont {{da Costa}}}, \bibinfo {author} {\bibfnamefont
  {E.}~\bibnamefont {{Gaztanaga}}}, \bibinfo {author} {\bibfnamefont {E.~M.}\
  \bibnamefont {{Huff}}}, \bibinfo {author} {\bibfnamefont {O.}~\bibnamefont
  {{Lahav}}}, \bibinfo {author} {\bibfnamefont {A.}~\bibnamefont {{Palmese}}},
  \bibinfo {author} {\bibfnamefont {W.~J.}\ \bibnamefont {{Percival}}},
  \bibinfo {author} {\bibfnamefont {A.}~\bibnamefont {{Refregier}}}, \bibinfo
  {author} {\bibfnamefont {A.~J.}\ \bibnamefont {{Ross}}}, \bibinfo {author}
  {\bibfnamefont {E.}~\bibnamefont {{Rozo}}}, \bibinfo {author} {\bibfnamefont
  {E.~S.}\ \bibnamefont {{Rykoff}}}, \bibinfo {author} {\bibfnamefont
  {C.}~\bibnamefont {{S{\'a}nchez}}}, \bibinfo {author} {\bibfnamefont
  {I.}~\bibnamefont {{Sadeh}}}, \bibinfo {author} {\bibfnamefont
  {I.}~\bibnamefont {{Sevilla-Noarbe}}}, \bibinfo {author} {\bibfnamefont
  {F.}~\bibnamefont {{Sobreira}}}, \bibinfo {author} {\bibfnamefont
  {E.}~\bibnamefont {{Suchyta}}}, \bibinfo {author} {\bibfnamefont {M.~E.~C.}\
  \bibnamefont {{Swanson}}}, \bibinfo {author} {\bibfnamefont {R.~H.}\
  \bibnamefont {{Wechsler}}}, \bibinfo {author} {\bibfnamefont {F.~B.}\
  \bibnamefont {{Abdalla}}}, \bibinfo {author} {\bibfnamefont {S.}~\bibnamefont
  {{Allam}}}, \bibinfo {author} {\bibfnamefont {M.}~\bibnamefont {{Banerji}}},
  \bibinfo {author} {\bibfnamefont {G.~M.}\ \bibnamefont {{Bernstein}}},
  \bibinfo {author} {\bibfnamefont {R.~A.}\ \bibnamefont {{Bernstein}}},
  \bibinfo {author} {\bibfnamefont {E.}~\bibnamefont {{Bertin}}}, \bibinfo
  {author} {\bibfnamefont {S.~L.}\ \bibnamefont {{Bridle}}}, \bibinfo {author}
  {\bibfnamefont {D.}~\bibnamefont {{Brooks}}}, \bibinfo {author}
  {\bibfnamefont {E.}~\bibnamefont {{Buckley-Geer}}}, \bibinfo {author}
  {\bibfnamefont {D.~L.}\ \bibnamefont {{Burke}}}, \bibinfo {author}
  {\bibfnamefont {D.}~\bibnamefont {{Capozzi}}}, \bibinfo {author}
  {\bibfnamefont {A.}~\bibnamefont {{Carnero Rosell}}}, \bibinfo {author}
  {\bibfnamefont {J.}~\bibnamefont {{Carretero}}}, \bibinfo {author}
  {\bibfnamefont {C.~E.}\ \bibnamefont {{Cunha}}}, \bibinfo {author}
  {\bibfnamefont {C.~B.}\ \bibnamefont {{D'Andrea}}}, \bibinfo {author}
  {\bibfnamefont {D.~L.}\ \bibnamefont {{DePoy}}}, \bibinfo {author}
  {\bibfnamefont {S.}~\bibnamefont {{Desai}}}, \bibinfo {author} {\bibfnamefont
  {H.~T.}\ \bibnamefont {{Diehl}}}, \bibinfo {author} {\bibfnamefont
  {P.}~\bibnamefont {{Doel}}}, \bibinfo {author} {\bibfnamefont {T.~F.}\
  \bibnamefont {{Eifler}}}, \bibinfo {author} {\bibfnamefont {A.~E.}\
  \bibnamefont {{Evrard}}}, \bibinfo {author} {\bibfnamefont {A.}~\bibnamefont
  {{Fausti Neto}}}, \bibinfo {author} {\bibfnamefont {B.}~\bibnamefont
  {{Flaugher}}}, \bibinfo {author} {\bibfnamefont {P.}~\bibnamefont
  {{Fosalba}}}, \bibinfo {author} {\bibfnamefont {J.}~\bibnamefont
  {{Frieman}}}, \bibinfo {author} {\bibfnamefont {D.~W.}\ \bibnamefont
  {{Gerdes}}}, \bibinfo {author} {\bibfnamefont {D.}~\bibnamefont {{Gruen}}},
  \bibinfo {author} {\bibfnamefont {R.~A.}\ \bibnamefont {{Gruendl}}}, \bibinfo
  {author} {\bibfnamefont {G.}~\bibnamefont {{Gutierrez}}}, \bibinfo {author}
  {\bibfnamefont {K.}~\bibnamefont {{Honscheid}}}, \bibinfo {author}
  {\bibfnamefont {D.~J.}\ \bibnamefont {{James}}}, \bibinfo {author}
  {\bibfnamefont {M.}~\bibnamefont {{Jarvis}}}, \bibinfo {author}
  {\bibfnamefont {S.}~\bibnamefont {{Kent}}}, \bibinfo {author} {\bibfnamefont
  {K.}~\bibnamefont {{Kuehn}}}, \bibinfo {author} {\bibfnamefont
  {N.}~\bibnamefont {{Kuropatkin}}}, \bibinfo {author} {\bibfnamefont {T.~S.}\
  \bibnamefont {{Li}}}, \bibinfo {author} {\bibfnamefont {M.}~\bibnamefont
  {{Lima}}}, \bibinfo {author} {\bibfnamefont {M.~A.~G.}\ \bibnamefont
  {{Maia}}}, \bibinfo {author} {\bibfnamefont {M.}~\bibnamefont {{March}}},
  \bibinfo {author} {\bibfnamefont {J.~L.}\ \bibnamefont {{Marshall}}},
  \bibinfo {author} {\bibfnamefont {P.}~\bibnamefont {{Martini}}}, \bibinfo
  {author} {\bibfnamefont {P.}~\bibnamefont {{Melchior}}}, \bibinfo {author}
  {\bibfnamefont {C.~J.}\ \bibnamefont {{Miller}}}, \bibinfo {author}
  {\bibfnamefont {R.}~\bibnamefont {{Miquel}}}, \bibinfo {author}
  {\bibfnamefont {R.~C.}\ \bibnamefont {{Nichol}}}, \bibinfo {author}
  {\bibfnamefont {B.}~\bibnamefont {{Nord}}}, \bibinfo {author} {\bibfnamefont
  {R.}~\bibnamefont {{Ogando}}}, \bibinfo {author} {\bibfnamefont {A.~A.}\
  \bibnamefont {{Plazas}}}, \bibinfo {author} {\bibfnamefont {K.}~\bibnamefont
  {{Reil}}}, \bibinfo {author} {\bibfnamefont {A.~K.}\ \bibnamefont {{Romer}}},
  \bibinfo {author} {\bibfnamefont {A.}~\bibnamefont {{Roodman}}}, \bibinfo
  {author} {\bibfnamefont {E.}~\bibnamefont {{Sanchez}}}, \bibinfo {author}
  {\bibfnamefont {B.}~\bibnamefont {{Santiago}}}, \bibinfo {author}
  {\bibfnamefont {V.}~\bibnamefont {{Scarpine}}}, \bibinfo {author}
  {\bibfnamefont {M.}~\bibnamefont {{Schubnell}}}, \bibinfo {author}
  {\bibfnamefont {R.~C.}\ \bibnamefont {{Smith}}}, \bibinfo {author}
  {\bibfnamefont {M.}~\bibnamefont {{Soares-Santos}}}, \bibinfo {author}
  {\bibfnamefont {G.}~\bibnamefont {{Tarle}}}, \bibinfo {author} {\bibfnamefont
  {J.}~\bibnamefont {{Thaler}}}, \bibinfo {author} {\bibfnamefont
  {D.}~\bibnamefont {{Thomas}}}, \bibinfo {author} {\bibfnamefont
  {V.}~\bibnamefont {{Vikram}}}, \bibinfo {author} {\bibfnamefont {A.~R.}\
  \bibnamefont {{Walker}}}, \bibinfo {author} {\bibfnamefont {W.}~\bibnamefont
  {{Wester}}}, \bibinfo {author} {\bibfnamefont {Y.}~\bibnamefont {{Zhang}}}, \
  and\ \bibinfo {author} {\bibfnamefont {J.}~\bibnamefont {{Zuntz}}},\
  }\href@noop {} {\bibfield  {journal} {\bibinfo  {journal}
  {arXiv:astro-ph/1507.05647}\ } (\bibinfo {year} {2015})},\ \Eprint
  {http://arxiv.org/abs/1507.05647} {arXiv:1507.05647} \BibitemShut {NoStop}%
\bibitem [{\citenamefont {Pedregosa}\ \emph {et~al.}(2011)\citenamefont
  {Pedregosa}, \citenamefont {Varoquaux}, \citenamefont {Gramfort},
  \citenamefont {Michel}, \citenamefont {Thirion}, \citenamefont {Grisel},
  \citenamefont {Blondel}, \citenamefont {Prettenhofer}, \citenamefont {Weiss},
  \citenamefont {Dubourg}, \citenamefont {Vanderplas}, \citenamefont {Passos},
  \citenamefont {Cournapeau}, \citenamefont {Brucher}, \citenamefont {Perrot},\
  and\ \citenamefont {Duchesnay}}]{scikit-learn}%
  \BibitemOpen
  \bibfield  {author} {\bibinfo {author} {\bibfnamefont {F.}~\bibnamefont
  {Pedregosa}}, \bibinfo {author} {\bibfnamefont {G.}~\bibnamefont
  {Varoquaux}}, \bibinfo {author} {\bibfnamefont {A.}~\bibnamefont {Gramfort}},
  \bibinfo {author} {\bibfnamefont {V.}~\bibnamefont {Michel}}, \bibinfo
  {author} {\bibfnamefont {B.}~\bibnamefont {Thirion}}, \bibinfo {author}
  {\bibfnamefont {O.}~\bibnamefont {Grisel}}, \bibinfo {author} {\bibfnamefont
  {M.}~\bibnamefont {Blondel}}, \bibinfo {author} {\bibfnamefont
  {P.}~\bibnamefont {Prettenhofer}}, \bibinfo {author} {\bibfnamefont
  {R.}~\bibnamefont {Weiss}}, \bibinfo {author} {\bibfnamefont
  {V.}~\bibnamefont {Dubourg}}, \bibinfo {author} {\bibfnamefont
  {J.}~\bibnamefont {Vanderplas}}, \bibinfo {author} {\bibfnamefont
  {A.}~\bibnamefont {Passos}}, \bibinfo {author} {\bibfnamefont
  {D.}~\bibnamefont {Cournapeau}}, \bibinfo {author} {\bibfnamefont
  {M.}~\bibnamefont {Brucher}}, \bibinfo {author} {\bibfnamefont
  {M.}~\bibnamefont {Perrot}}, \ and\ \bibinfo {author} {\bibfnamefont
  {E.}~\bibnamefont {Duchesnay}},\ }\href@noop {} {\bibfield  {journal}
  {\bibinfo  {journal} {Journal of Machine Learning Research}\ }\textbf
  {\bibinfo {volume} {12}},\ \bibinfo {pages} {2825} (\bibinfo {year}
  {2011})}\BibitemShut {NoStop}%
\bibitem [{\citenamefont {{G{\'o}rski}}\ \emph {et~al.}(2005)\citenamefont
  {{G{\'o}rski}}, \citenamefont {{Hivon}}, \citenamefont {{Banday}},
  \citenamefont {{Wandelt}}, \citenamefont {{Hansen}}, \citenamefont
  {{Reinecke}},\ and\ \citenamefont {{Bartelmann}}}]{gorski2005}%
  \BibitemOpen
  \bibfield  {author} {\bibinfo {author} {\bibfnamefont {K.~M.}\ \bibnamefont
  {{G{\'o}rski}}}, \bibinfo {author} {\bibfnamefont {E.}~\bibnamefont
  {{Hivon}}}, \bibinfo {author} {\bibfnamefont {A.~J.}\ \bibnamefont
  {{Banday}}}, \bibinfo {author} {\bibfnamefont {B.~D.}\ \bibnamefont
  {{Wandelt}}}, \bibinfo {author} {\bibfnamefont {F.~K.}\ \bibnamefont
  {{Hansen}}}, \bibinfo {author} {\bibfnamefont {M.}~\bibnamefont
  {{Reinecke}}}, \ and\ \bibinfo {author} {\bibfnamefont {M.}~\bibnamefont
  {{Bartelmann}}},\ }\href {\doibase 10.1086/427976} {\bibfield  {journal}
  {\bibinfo  {journal} {\apj}\ }\textbf {\bibinfo {volume} {622}},\ \bibinfo
  {pages} {759} (\bibinfo {year} {2005})},\ \Eprint
  {http://arxiv.org/abs/arXiv:astro-ph/0409513} {arXiv:astro-ph/0409513}
  \BibitemShut {NoStop}%
\bibitem [{\citenamefont {{Chiang}}\ \emph {et~al.}(2010)\citenamefont
  {{Chiang}}, \citenamefont {{Ade}}, \citenamefont {{Barkats}}, \citenamefont
  {{Battle}}, \citenamefont {{Bierman}}, \citenamefont {{Bock}}, \citenamefont
  {{Dowell}}, \citenamefont {{Duband}}, \citenamefont {{Hivon}}, \citenamefont
  {{Holzapfel}}, \citenamefont {{Hristov}}, \citenamefont {{Jones}},
  \citenamefont {{Keating}}, \citenamefont {{Kovac}}, \citenamefont {{Kuo}},
  \citenamefont {{Lange}}, \citenamefont {{Leitch}}, \citenamefont {{Mason}},
  \citenamefont {{Matsumura}}, \citenamefont {{Nguyen}}, \citenamefont
  {{Ponthieu}}, \citenamefont {{Pryke}}, \citenamefont {{Richter}},
  \citenamefont {{Rocha}}, \citenamefont {{Sheehy}}, \citenamefont
  {{Takahashi}}, \citenamefont {{Tolan}},\ and\ \citenamefont
  {{Yoon}}}]{chiang2010}%
  \BibitemOpen
  \bibfield  {author} {\bibinfo {author} {\bibfnamefont {H.~C.}\ \bibnamefont
  {{Chiang}}}, \bibinfo {author} {\bibfnamefont {P.~A.~R.}\ \bibnamefont
  {{Ade}}}, \bibinfo {author} {\bibfnamefont {D.}~\bibnamefont {{Barkats}}},
  \bibinfo {author} {\bibfnamefont {J.~O.}\ \bibnamefont {{Battle}}}, \bibinfo
  {author} {\bibfnamefont {E.~M.}\ \bibnamefont {{Bierman}}}, \bibinfo {author}
  {\bibfnamefont {J.~J.}\ \bibnamefont {{Bock}}}, \bibinfo {author}
  {\bibfnamefont {C.~D.}\ \bibnamefont {{Dowell}}}, \bibinfo {author}
  {\bibfnamefont {L.}~\bibnamefont {{Duband}}}, \bibinfo {author}
  {\bibfnamefont {E.~F.}\ \bibnamefont {{Hivon}}}, \bibinfo {author}
  {\bibfnamefont {W.~L.}\ \bibnamefont {{Holzapfel}}}, \bibinfo {author}
  {\bibfnamefont {V.~V.}\ \bibnamefont {{Hristov}}}, \bibinfo {author}
  {\bibfnamefont {W.~C.}\ \bibnamefont {{Jones}}}, \bibinfo {author}
  {\bibfnamefont {B.~G.}\ \bibnamefont {{Keating}}}, \bibinfo {author}
  {\bibfnamefont {J.~M.}\ \bibnamefont {{Kovac}}}, \bibinfo {author}
  {\bibfnamefont {C.~L.}\ \bibnamefont {{Kuo}}}, \bibinfo {author}
  {\bibfnamefont {A.~E.}\ \bibnamefont {{Lange}}}, \bibinfo {author}
  {\bibfnamefont {E.~M.}\ \bibnamefont {{Leitch}}}, \bibinfo {author}
  {\bibfnamefont {P.~V.}\ \bibnamefont {{Mason}}}, \bibinfo {author}
  {\bibfnamefont {T.}~\bibnamefont {{Matsumura}}}, \bibinfo {author}
  {\bibfnamefont {H.~T.}\ \bibnamefont {{Nguyen}}}, \bibinfo {author}
  {\bibfnamefont {N.}~\bibnamefont {{Ponthieu}}}, \bibinfo {author}
  {\bibfnamefont {C.}~\bibnamefont {{Pryke}}}, \bibinfo {author} {\bibfnamefont
  {S.}~\bibnamefont {{Richter}}}, \bibinfo {author} {\bibfnamefont
  {G.}~\bibnamefont {{Rocha}}}, \bibinfo {author} {\bibfnamefont
  {C.}~\bibnamefont {{Sheehy}}}, \bibinfo {author} {\bibfnamefont {Y.~D.}\
  \bibnamefont {{Takahashi}}}, \bibinfo {author} {\bibfnamefont {J.~E.}\
  \bibnamefont {{Tolan}}}, \ and\ \bibinfo {author} {\bibfnamefont {K.~W.}\
  \bibnamefont {{Yoon}}},\ }\href {\doibase 10.1088/0004-637X/711/2/1123}
  {\bibfield  {journal} {\bibinfo  {journal} {\apj}\ }\textbf {\bibinfo
  {volume} {711}},\ \bibinfo {pages} {1123} (\bibinfo {year} {2010})},\ \Eprint
  {http://arxiv.org/abs/0906.1181} {arXiv:0906.1181 [astro-ph.CO]} \BibitemShut
  {NoStop}%
\bibitem [{\citenamefont {{Crocce}}\ \emph {et~al.}(2016)\citenamefont
  {{Crocce}}, \citenamefont {{Carretero}}, \citenamefont {{Bauer}},
  \citenamefont {{Ross}}, \citenamefont {{Sevilla-Noarbe}}, \citenamefont
  {{Giannantonio}}, \citenamefont {{Sobreira}}, \citenamefont {{Sanchez}},
  \citenamefont {{Gaztanaga}}, \citenamefont {{Kind}}, \citenamefont
  {{S{\'a}nchez}}, \citenamefont {{Bonnett}}, \citenamefont
  {{Benoit-L{\'e}vy}}, \citenamefont {{Brunner}}, \citenamefont {{Rosell}},
  \citenamefont {{Cawthon}}, \citenamefont {{Fosalba}}, \citenamefont
  {{Hartley}}, \citenamefont {{Kim}}, \citenamefont {{Leistedt}}, \citenamefont
  {{Miquel}}, \citenamefont {{Peiris}}, \citenamefont {{Percival}},
  \citenamefont {{Rosenfeld}}, \citenamefont {{Rykoff}}, \citenamefont
  {{S{\'a}nchez}}, \citenamefont {{Abbott}}, \citenamefont {{Abdalla}},
  \citenamefont {{Allam}}, \citenamefont {{Banerji}}, \citenamefont
  {{Bernstein}}, \citenamefont {{Bertin}}, \citenamefont {{Brooks}},
  \citenamefont {{Buckley-Geer}}, \citenamefont {{Burke}}, \citenamefont
  {{Capozzi}}, \citenamefont {{Castander}}, \citenamefont {{Cunha}},
  \citenamefont {{D'Andrea}}, \citenamefont {{da Costa}}, \citenamefont
  {{Desai}}, \citenamefont {{Diehl}}, \citenamefont {{Eifler}}, \citenamefont
  {{Evrard}}, \citenamefont {{Neto}}, \citenamefont {{Fernandez}},
  \citenamefont {{Finley}}, \citenamefont {{Flaugher}}, \citenamefont
  {{Frieman}}, \citenamefont {{Gerdes}}, \citenamefont {{Gruen}}, \citenamefont
  {{Gruendl}}, \citenamefont {{Gutierrez}}, \citenamefont {{Honscheid}},
  \citenamefont {{James}}, \citenamefont {{Kuehn}}, \citenamefont
  {{Kuropatkin}}, \citenamefont {{Lahav}}, \citenamefont {{Li}}, \citenamefont
  {{Lima}}, \citenamefont {{Maia}}, \citenamefont {{March}}, \citenamefont
  {{Marshall}}, \citenamefont {{Martini}}, \citenamefont {{Melchior}},
  \citenamefont {{Miller}}, \citenamefont {{Neilsen}}, \citenamefont
  {{Nichol}}, \citenamefont {{Nord}}, \citenamefont {{Ogando}}, \citenamefont
  {{Plazas}}, \citenamefont {{Romer}}, \citenamefont {{Sako}}, \citenamefont
  {{Santiago}}, \citenamefont {{Schubnell}}, \citenamefont {{Smith}},
  \citenamefont {{Soares-Santos}}, \citenamefont {{Suchyta}}, \citenamefont
  {{Swanson}}, \citenamefont {{Tarle}}, \citenamefont {{Thaler}}, \citenamefont
  {{Thomas}}, \citenamefont {{Vikram}}, \citenamefont {{Walker}}, \citenamefont
  {{Wechsler}}, \citenamefont {{Weller}}, \citenamefont {{Zuntz}},\ and\
  \citenamefont {{DES Collaboration}}}]{crocce2016}%
  \BibitemOpen
  \bibfield  {author} {\bibinfo {author} {\bibfnamefont {M.}~\bibnamefont
  {{Crocce}}}, \bibinfo {author} {\bibfnamefont {J.}~\bibnamefont
  {{Carretero}}}, \bibinfo {author} {\bibfnamefont {A.~H.}\ \bibnamefont
  {{Bauer}}}, \bibinfo {author} {\bibfnamefont {A.~J.}\ \bibnamefont {{Ross}}},
  \bibinfo {author} {\bibfnamefont {I.}~\bibnamefont {{Sevilla-Noarbe}}},
  \bibinfo {author} {\bibfnamefont {T.}~\bibnamefont {{Giannantonio}}},
  \bibinfo {author} {\bibfnamefont {F.}~\bibnamefont {{Sobreira}}}, \bibinfo
  {author} {\bibfnamefont {J.}~\bibnamefont {{Sanchez}}}, \bibinfo {author}
  {\bibfnamefont {E.}~\bibnamefont {{Gaztanaga}}}, \bibinfo {author}
  {\bibfnamefont {M.~C.}\ \bibnamefont {{Kind}}}, \bibinfo {author}
  {\bibfnamefont {C.}~\bibnamefont {{S{\'a}nchez}}}, \bibinfo {author}
  {\bibfnamefont {C.}~\bibnamefont {{Bonnett}}}, \bibinfo {author}
  {\bibfnamefont {A.}~\bibnamefont {{Benoit-L{\'e}vy}}}, \bibinfo {author}
  {\bibfnamefont {R.~J.}\ \bibnamefont {{Brunner}}}, \bibinfo {author}
  {\bibfnamefont {A.~C.}\ \bibnamefont {{Rosell}}}, \bibinfo {author}
  {\bibfnamefont {R.}~\bibnamefont {{Cawthon}}}, \bibinfo {author}
  {\bibfnamefont {P.}~\bibnamefont {{Fosalba}}}, \bibinfo {author}
  {\bibfnamefont {W.}~\bibnamefont {{Hartley}}}, \bibinfo {author}
  {\bibfnamefont {E.~J.}\ \bibnamefont {{Kim}}}, \bibinfo {author}
  {\bibfnamefont {B.}~\bibnamefont {{Leistedt}}}, \bibinfo {author}
  {\bibfnamefont {R.}~\bibnamefont {{Miquel}}}, \bibinfo {author}
  {\bibfnamefont {H.~V.}\ \bibnamefont {{Peiris}}}, \bibinfo {author}
  {\bibfnamefont {W.~J.}\ \bibnamefont {{Percival}}}, \bibinfo {author}
  {\bibfnamefont {R.}~\bibnamefont {{Rosenfeld}}}, \bibinfo {author}
  {\bibfnamefont {E.~S.}\ \bibnamefont {{Rykoff}}}, \bibinfo {author}
  {\bibfnamefont {E.}~\bibnamefont {{S{\'a}nchez}}}, \bibinfo {author}
  {\bibfnamefont {T.}~\bibnamefont {{Abbott}}}, \bibinfo {author}
  {\bibfnamefont {F.~B.}\ \bibnamefont {{Abdalla}}}, \bibinfo {author}
  {\bibfnamefont {S.}~\bibnamefont {{Allam}}}, \bibinfo {author} {\bibfnamefont
  {M.}~\bibnamefont {{Banerji}}}, \bibinfo {author} {\bibfnamefont {G.~M.}\
  \bibnamefont {{Bernstein}}}, \bibinfo {author} {\bibfnamefont
  {E.}~\bibnamefont {{Bertin}}}, \bibinfo {author} {\bibfnamefont
  {D.}~\bibnamefont {{Brooks}}}, \bibinfo {author} {\bibfnamefont
  {E.}~\bibnamefont {{Buckley-Geer}}}, \bibinfo {author} {\bibfnamefont
  {D.~L.}\ \bibnamefont {{Burke}}}, \bibinfo {author} {\bibfnamefont
  {D.}~\bibnamefont {{Capozzi}}}, \bibinfo {author} {\bibfnamefont {F.~J.}\
  \bibnamefont {{Castander}}}, \bibinfo {author} {\bibfnamefont {C.~E.}\
  \bibnamefont {{Cunha}}}, \bibinfo {author} {\bibfnamefont {C.~B.}\
  \bibnamefont {{D'Andrea}}}, \bibinfo {author} {\bibfnamefont {L.~N.}\
  \bibnamefont {{da Costa}}}, \bibinfo {author} {\bibfnamefont
  {S.}~\bibnamefont {{Desai}}}, \bibinfo {author} {\bibfnamefont {H.~T.}\
  \bibnamefont {{Diehl}}}, \bibinfo {author} {\bibfnamefont {T.~F.}\
  \bibnamefont {{Eifler}}}, \bibinfo {author} {\bibfnamefont {A.~E.}\
  \bibnamefont {{Evrard}}}, \bibinfo {author} {\bibfnamefont {A.~F.}\
  \bibnamefont {{Neto}}}, \bibinfo {author} {\bibfnamefont {E.}~\bibnamefont
  {{Fernandez}}}, \bibinfo {author} {\bibfnamefont {D.~A.}\ \bibnamefont
  {{Finley}}}, \bibinfo {author} {\bibfnamefont {B.}~\bibnamefont
  {{Flaugher}}}, \bibinfo {author} {\bibfnamefont {J.}~\bibnamefont
  {{Frieman}}}, \bibinfo {author} {\bibfnamefont {D.~W.}\ \bibnamefont
  {{Gerdes}}}, \bibinfo {author} {\bibfnamefont {D.}~\bibnamefont {{Gruen}}},
  \bibinfo {author} {\bibfnamefont {R.~A.}\ \bibnamefont {{Gruendl}}}, \bibinfo
  {author} {\bibfnamefont {G.}~\bibnamefont {{Gutierrez}}}, \bibinfo {author}
  {\bibfnamefont {K.}~\bibnamefont {{Honscheid}}}, \bibinfo {author}
  {\bibfnamefont {D.~J.}\ \bibnamefont {{James}}}, \bibinfo {author}
  {\bibfnamefont {K.}~\bibnamefont {{Kuehn}}}, \bibinfo {author} {\bibfnamefont
  {N.}~\bibnamefont {{Kuropatkin}}}, \bibinfo {author} {\bibfnamefont
  {O.}~\bibnamefont {{Lahav}}}, \bibinfo {author} {\bibfnamefont {T.~S.}\
  \bibnamefont {{Li}}}, \bibinfo {author} {\bibfnamefont {M.}~\bibnamefont
  {{Lima}}}, \bibinfo {author} {\bibfnamefont {M.~A.~G.}\ \bibnamefont
  {{Maia}}}, \bibinfo {author} {\bibfnamefont {M.}~\bibnamefont {{March}}},
  \bibinfo {author} {\bibfnamefont {J.~L.}\ \bibnamefont {{Marshall}}},
  \bibinfo {author} {\bibfnamefont {P.}~\bibnamefont {{Martini}}}, \bibinfo
  {author} {\bibfnamefont {P.}~\bibnamefont {{Melchior}}}, \bibinfo {author}
  {\bibfnamefont {C.~J.}\ \bibnamefont {{Miller}}}, \bibinfo {author}
  {\bibfnamefont {E.}~\bibnamefont {{Neilsen}}}, \bibinfo {author}
  {\bibfnamefont {R.~C.}\ \bibnamefont {{Nichol}}}, \bibinfo {author}
  {\bibfnamefont {B.}~\bibnamefont {{Nord}}}, \bibinfo {author} {\bibfnamefont
  {R.}~\bibnamefont {{Ogando}}}, \bibinfo {author} {\bibfnamefont {A.~A.}\
  \bibnamefont {{Plazas}}}, \bibinfo {author} {\bibfnamefont {A.~K.}\
  \bibnamefont {{Romer}}}, \bibinfo {author} {\bibfnamefont {M.}~\bibnamefont
  {{Sako}}}, \bibinfo {author} {\bibfnamefont {B.}~\bibnamefont {{Santiago}}},
  \bibinfo {author} {\bibfnamefont {M.}~\bibnamefont {{Schubnell}}}, \bibinfo
  {author} {\bibfnamefont {R.~C.}\ \bibnamefont {{Smith}}}, \bibinfo {author}
  {\bibfnamefont {M.}~\bibnamefont {{Soares-Santos}}}, \bibinfo {author}
  {\bibfnamefont {E.}~\bibnamefont {{Suchyta}}}, \bibinfo {author}
  {\bibfnamefont {M.~E.~C.}\ \bibnamefont {{Swanson}}}, \bibinfo {author}
  {\bibfnamefont {G.}~\bibnamefont {{Tarle}}}, \bibinfo {author} {\bibfnamefont
  {J.}~\bibnamefont {{Thaler}}}, \bibinfo {author} {\bibfnamefont
  {D.}~\bibnamefont {{Thomas}}}, \bibinfo {author} {\bibfnamefont
  {V.}~\bibnamefont {{Vikram}}}, \bibinfo {author} {\bibfnamefont {A.~R.}\
  \bibnamefont {{Walker}}}, \bibinfo {author} {\bibfnamefont {R.~H.}\
  \bibnamefont {{Wechsler}}}, \bibinfo {author} {\bibfnamefont
  {J.}~\bibnamefont {{Weller}}}, \bibinfo {author} {\bibfnamefont
  {J.}~\bibnamefont {{Zuntz}}}, \ and\ \bibinfo {author} {\bibnamefont {{DES
  Collaboration}}},\ }\href {\doibase 10.1093/mnras/stv2590} {\bibfield
  {journal} {\bibinfo  {journal} {\mnras}\ }\textbf {\bibinfo {volume} {455}},\
  \bibinfo {pages} {4301} (\bibinfo {year} {2016})},\ \Eprint
  {http://arxiv.org/abs/1507.05360} {arXiv:1507.05360} \BibitemShut {NoStop}%
\bibitem [{\citenamefont {{Cooray}}\ and\ \citenamefont
  {{Hu}}(2001)}]{ch01cov}%
  \BibitemOpen
  \bibfield  {author} {\bibinfo {author} {\bibfnamefont {A.}~\bibnamefont
  {{Cooray}}}\ and\ \bibinfo {author} {\bibfnamefont {W.}~\bibnamefont
  {{Hu}}},\ }\href {\doibase 10.1086/321376} {\bibfield  {journal} {\bibinfo
  {journal} {\apj}\ }\textbf {\bibinfo {volume} {554}},\ \bibinfo {pages} {56}
  (\bibinfo {year} {2001})},\ \Eprint {http://arxiv.org/abs/astro-ph/0012087}
  {astro-ph/0012087} \BibitemShut {NoStop}%
\bibitem [{\citenamefont {{Hu}}\ and\ \citenamefont {{Jain}}(2004)}]{huj04}%
  \BibitemOpen
  \bibfield  {author} {\bibinfo {author} {\bibfnamefont {W.}~\bibnamefont
  {{Hu}}}\ and\ \bibinfo {author} {\bibfnamefont {B.}~\bibnamefont {{Jain}}},\
  }\href {\doibase 10.1103/PhysRevD.70.043009} {\bibfield  {journal} {\bibinfo
  {journal} {\prd}\ }\textbf {\bibinfo {volume} {70}},\ \bibinfo {eid} {043009}
  (\bibinfo {year} {2004})},\ \Eprint {http://arxiv.org/abs/astro-ph/0312395}
  {astro-ph/0312395} \BibitemShut {NoStop}%
\bibitem [{\citenamefont {{Sato}}\ \emph {et~al.}(2009)\citenamefont {{Sato}},
  \citenamefont {{Hamana}}, \citenamefont {{Takahashi}}, \citenamefont
  {{Takada}}, \citenamefont {{Yoshida}}, \citenamefont {{Matsubara}},\ and\
  \citenamefont {{Sugiyama}}}]{sht09}%
  \BibitemOpen
  \bibfield  {author} {\bibinfo {author} {\bibfnamefont {M.}~\bibnamefont
  {{Sato}}}, \bibinfo {author} {\bibfnamefont {T.}~\bibnamefont {{Hamana}}},
  \bibinfo {author} {\bibfnamefont {R.}~\bibnamefont {{Takahashi}}}, \bibinfo
  {author} {\bibfnamefont {M.}~\bibnamefont {{Takada}}}, \bibinfo {author}
  {\bibfnamefont {N.}~\bibnamefont {{Yoshida}}}, \bibinfo {author}
  {\bibfnamefont {T.}~\bibnamefont {{Matsubara}}}, \ and\ \bibinfo {author}
  {\bibfnamefont {N.}~\bibnamefont {{Sugiyama}}},\ }\href {\doibase
  10.1088/0004-637X/701/2/945} {\bibfield  {journal} {\bibinfo  {journal}
  {\apj}\ }\textbf {\bibinfo {volume} {701}},\ \bibinfo {pages} {945} (\bibinfo
  {year} {2009})},\ \Eprint {http://arxiv.org/abs/0906.2237} {arXiv:0906.2237
  [astro-ph.CO]} \BibitemShut {NoStop}%
\bibitem [{\citenamefont {{Eifler}}\ \emph {et~al.}(2015)\citenamefont
  {{Eifler}}, \citenamefont {{Krause}}, \citenamefont {{Dodelson}},
  \citenamefont {{Zentner}}, \citenamefont {{Hearin}},\ and\ \citenamefont
  {{Gnedin}}}]{ekd14}%
  \BibitemOpen
  \bibfield  {author} {\bibinfo {author} {\bibfnamefont {T.}~\bibnamefont
  {{Eifler}}}, \bibinfo {author} {\bibfnamefont {E.}~\bibnamefont {{Krause}}},
  \bibinfo {author} {\bibfnamefont {S.}~\bibnamefont {{Dodelson}}}, \bibinfo
  {author} {\bibfnamefont {A.~R.}\ \bibnamefont {{Zentner}}}, \bibinfo {author}
  {\bibfnamefont {A.~P.}\ \bibnamefont {{Hearin}}}, \ and\ \bibinfo {author}
  {\bibfnamefont {N.~Y.}\ \bibnamefont {{Gnedin}}},\ }\href {\doibase
  10.1093/mnras/stv2000} {\bibfield  {journal} {\bibinfo  {journal} {\mnras}\
  }\textbf {\bibinfo {volume} {454}},\ \bibinfo {pages} {2451} (\bibinfo {year}
  {2015})},\ \Eprint {http://arxiv.org/abs/1405.7423} {arXiv:1405.7423}
  \BibitemShut {NoStop}%
\bibitem [{\citenamefont {{Sato}}\ \emph {et~al.}(2011)\citenamefont {{Sato}},
  \citenamefont {{Takada}}, \citenamefont {{Hamana}},\ and\ \citenamefont
  {{Matsubara}}}]{sato2011}%
  \BibitemOpen
  \bibfield  {author} {\bibinfo {author} {\bibfnamefont {M.}~\bibnamefont
  {{Sato}}}, \bibinfo {author} {\bibfnamefont {M.}~\bibnamefont {{Takada}}},
  \bibinfo {author} {\bibfnamefont {T.}~\bibnamefont {{Hamana}}}, \ and\
  \bibinfo {author} {\bibfnamefont {T.}~\bibnamefont {{Matsubara}}},\ }\href
  {\doibase 10.1088/0004-637X/734/2/76} {\bibfield  {journal} {\bibinfo
  {journal} {\apj}\ }\textbf {\bibinfo {volume} {734}},\ \bibinfo {eid} {76}
  (\bibinfo {year} {2011})},\ \Eprint {http://arxiv.org/abs/1009.2558}
  {arXiv:1009.2558} \BibitemShut {NoStop}%
\bibitem [{\citenamefont {{Takada}}\ and\ \citenamefont
  {{Jain}}(2009)}]{takada2009}%
  \BibitemOpen
  \bibfield  {author} {\bibinfo {author} {\bibfnamefont {M.}~\bibnamefont
  {{Takada}}}\ and\ \bibinfo {author} {\bibfnamefont {B.}~\bibnamefont
  {{Jain}}},\ }\href {\doibase 10.1111/j.1365-2966.2009.14504.x} {\bibfield
  {journal} {\bibinfo  {journal} {\mnras}\ }\textbf {\bibinfo {volume} {395}},\
  \bibinfo {pages} {2065} (\bibinfo {year} {2009})},\ \Eprint
  {http://arxiv.org/abs/0810.4170} {arXiv:0810.4170} \BibitemShut {NoStop}%
\bibitem [{\citenamefont {{Li}}\ \emph {et~al.}(2014)\citenamefont {{Li}},
  \citenamefont {{Hu}},\ and\ \citenamefont {{Takada}}}]{li2014}%
  \BibitemOpen
  \bibfield  {author} {\bibinfo {author} {\bibfnamefont {Y.}~\bibnamefont
  {{Li}}}, \bibinfo {author} {\bibfnamefont {W.}~\bibnamefont {{Hu}}}, \ and\
  \bibinfo {author} {\bibfnamefont {M.}~\bibnamefont {{Takada}}},\ }\href
  {\doibase 10.1103/PhysRevD.89.083519} {\bibfield  {journal} {\bibinfo
  {journal} {\prd}\ }\textbf {\bibinfo {volume} {89}},\ \bibinfo {eid} {083519}
  (\bibinfo {year} {2014})},\ \Eprint {http://arxiv.org/abs/1401.0385}
  {arXiv:1401.0385} \BibitemShut {NoStop}%
\bibitem [{\citenamefont {{Joachimi}}\ \emph {et~al.}(2008)\citenamefont
  {{Joachimi}}, \citenamefont {{Schneider}},\ and\ \citenamefont
  {{Eifler}}}]{JSE08}%
  \BibitemOpen
  \bibfield  {author} {\bibinfo {author} {\bibfnamefont {B.}~\bibnamefont
  {{Joachimi}}}, \bibinfo {author} {\bibfnamefont {P.}~\bibnamefont
  {{Schneider}}}, \ and\ \bibinfo {author} {\bibfnamefont {T.}~\bibnamefont
  {{Eifler}}},\ }\href {\doibase 10.1051/0004-6361:20078400} {\bibfield
  {journal} {\bibinfo  {journal} {\aap}\ }\textbf {\bibinfo {volume} {477}},\
  \bibinfo {pages} {43} (\bibinfo {year} {2008})},\ \Eprint
  {http://arxiv.org/abs/0708.0387} {arXiv:0708.0387} \BibitemShut {NoStop}%
\end{thebibliography}%

\label{lastpage}

\end{document}